\documentstyle[preprint,eqsecnum,aps]{revtex}

\begin{document}

\begin{center}
{\large {\bf Flocks, herds, and schools:  A
quantitative theory of flocking}}
\vskip .75cm
 {\bf  John Toner }
\vskip.5cm
 {Materials Science Institute, Institute of Theoretical Science\\ and
Department of Physics\\ University of Oregon, Eugene, OR
97403-5203}
\vskip .75cm
{\bf  Yuhai Tu}
\vskip.5cm
 {IBM T. J. Watson Research Center,\\ P. O. Box 218,
Yorktown Heights, NY 10598}
\end{center}

\begin{abstract} We present a quantitative continuum theory of
``flocking'':  the collective coherent motion of large numbers of
self-propelled organisms.
In agreement with everyday experience, our model predicts the
existence of an ``ordered phase'' of flocks, in which all members
of even an arbitrarily large flock move together with the same
mean velocity $\left< \vec{v}\right> \neq 0$.  This coherent
motion of the flock is an example of spontaneously
broken symmetry:  no preferred direction for the motion is
picked out a priori in the model; rather, each flock is allowed to,
and does, spontaneously pick out some completely arbitrary
direction to move in.  By analyzing our model we can make
detailed, quantitative predictions for the long-distance,
long-time behavior of this ``broken symmetry state''.  The
``Goldstone modes'' associated with this ``spontaneously broken
rotational symmetry'' are fluctuations in the direction of motion of a
large part of the flock away from the mean direction of motion of the
flock as a whole.  These ``Goldstone modes'' mix with modes associated
with conservation of bird number to produce propagating sound
modes.  These sound modes lead to enormous fluctuations of the
density of the flock, far larger, at long wavelengths, than those in,
e.\ g., an equilibrium gas.  Our model is similar in many ways to the
Navier-Stokes equations for a simple compressible fluid; in other ways, it
resembles a relaxational time dependent
Ginsburg-Landau theory for an $n = d$ component isotropic
ferromagnet.  In spatial dimensions $d > 4$, the long distance
behavior is correctly described by a linearized theory, and is
equivalent to that of an unusual but nonetheless equilibrium model
for spin systems.  For $d < 4$, non-linear fluctuation effects radically
alter the long distance behavior, making it different from that of any
known equilibrium model. In particular, we find that in
$d = 2$,
where we can calculate the scaling exponents 
\underline{exactly}, flocks exhibit a true, long-range ordered,
spontaneously broken symmetry state, in contrast to equilibrium
systems, which cannot spontaneously break a continuous symmetry
in $d = 2$ (the ``Mermin-Wagner'' theorem).  We make detailed
predictions for various correlation functions that could be measured
either in simulations, or by quantitative imaging of real flocks.  We
also consider an anisotropic model, in which the birds move
preferentially in an ``easy'' (e.\ g., horizontal) plane, and make
analogous, but quantitatively different, predictions for that model as
well.  For this anisotropic model, we obtain
\underline{exact} scaling exponents for all spatial dimensions,
including the physically relevant case $d = 3$.
\end{abstract}

\section{Introduction}

A wide variety of non-equilibrium dynamical systems with many degrees of
freedom have recently been studied using powerful techniques developed for
equilibrium condensed matter physics (e.g., scaling, the renormalization group,
etc).  One of the most familiar examples of a many degree of freedom,
non-equilibrium dynamical system is a large flock of birds.  Myriad other
examples of the collective, coherent motion of large numbers of self-propelled
organisms occur in biology:  schools of fish, swarms of insects, slime molds,
herds of wildebeest, etc.

Recently, a number of simulations of this phenomenon have been
performed.\cite{Vicsek,Partridge,reynolds}
Following Reynolds \cite{reynolds}, we will use the term
``boid" and bird interchangeably for the particles in these simulations.
All of these simulations have
several essential features in common:

\begin{enumerate}

\item  A large number (a ``flock'') of point particles (``boids'') each move
over
time through a space of dimension $d$ ($=2,3$,...),
{\it attempting} at all times to ``follow'' (i.e., move in the same
direction as)
its neighbors.

\item The interactions are purely short ranged:  each ``boid'' only responds to
its neighbors, defined as those ``boids'' within some fixed, finite distance
$R_0$, which is assumed to be much less than L, the size of the
``flock.''

\item The ``following'' is not perfect: the ``boids'' make errors at all times,
which
are modeled as a stochastic noise.  This noise is assumed to have only short
ranged spatio-temporal correlations.

\item The underlying model has complete rotational symmetry:  the flock is
equally likely, a priori, to move in any direction.

\end{enumerate}

The development of a non-zero mean center of mass
velocity $\left<\vec{v}\right>$ for the flock as a whole therefore requires
spontaneous breaking of a continuous symmetry (namely, rotational).

In an earlier paper\cite{TT}, we formulated a continuum model for
such dynamics of flocking, and obtained some exact results for
that model in spatial dimensions $d = 2$ (appropriate for the
description of the motion of land animals on the earth's surface).
Our most surprising result \cite{TT} was that two-dimensional
moving herds with strictly short-ranged interactions appear to
violate the Mermin--Wagner theorem\cite{MW}, in that they can
acquire long-ranged order, by picking out a consistent direction of
motion across an arbitrarily large herd, despite the fact that this
involves spontaneously breaking a continuous (rotational)
symmetry.

Of course, this result does not, in fact, violate the Mermin-Wagner theorem,
since flocks are a non-equilibrium dynamical system.  What is fascinating (at
least to us) about our result is that the non-equilibrium aspects of the flock
dynamics that make the long-distance, long-time behavior of the flock
different from that of otherwise analogous equilibrium systems are
fundamentally non-linear, strong-fluctuation effects.  Indeed, a ``breakdown of
linearized hydrodynamics,'' analogous to that long known to occur in
equilibrium fluids\cite{FNS} in spatial dimensions $d=2$, occurs in flocks
for all
$d < 4$.  This breakdown of linearized hydrodynamics is essential to the very
existence of the ordered state in $d=2$.  Furthermore, it has dramatic
consequences even for $d>2$.

The physics of this breakdown is very simple:  above
$d=4$, where the breakdown does {\it not} occur, information about what is
going on in one part of the flock can be transmitted to another part of the
flock only by being passed sequentially through the intervening neighbors via
the assumed short ranged interactions.  Below $d=4$, where the breakdown
occurs, this slow, diffusive transport of information is replaced by direct,
convective transport:  fluctuations in the local velocity of the flock
became so
large, in these lower dimensions, that the motion of one part of the flock
relative to another becomes the principle means of information transport,
because it becomes faster than diffusion.  There is a sort of ``negative
feedback,'' in that this improved transport actually suppresses the very
fluctuations that give rise to it, leading to long-ranged order in
 $d=2$.  The purpose of the present paper is to study the properties of the
``ordered state'' of the flock, i.e.,  the state in which all members of
the flock
are moving in the same average direction.  Specifically, we will:

\begin{enumerate}
\item give the details of the derivation of the results of reference
\cite{TT}, and give detailed predictions for numerous correlation
functions that can be measured in both experiments and simulations.
In particular, we will show that two propagating sound modes exist in
flocks, with unusually anisotropic speeds, whose detailed dependence
on the direction of propagation we predict, making possible extremely
stringent quantitative tests of our theory. We also calculate their
attenuations, which show highly anomalous, and strongly anisotropic,
scaling,

\item formulate and study the most complete generalization of the model of
reference \cite{TT} for spatial dimensions
$d>2$, and

\item include the effect of spatial anisotropy (e.g., the fact that birds
prefer to fly horizontally rather than vertically) on flock motion.

\end{enumerate}

We describe the flock with coarse grained density and velocity fields
$\rho(\vec{r},t)$ and $\vec{v}(\vec{r},t)$, respectively, giving the average
number density and velocity of the birds at time
$t$ within some coarse graining distance $\ell_0$ of a given position
$\vec{r}$ in space.  The coarse graining distance $\ell_0$ is chosen to be as
small as possible, consistent with being large enough that the averaging can be
done sensibly (in particular, $\ell_0$ must be greater than the mean interbird
distance).  Our description is then valid for distances large compared to
$\ell_0$, and for times $t$ much greater than some microscopic time $t_0$,
presumably of order
$\ell_0/v_T$, where $v_T$ is a typical speed of a bird.   Collective motion of
the flock as a whole then requires that
$\left<\vec{v}(\vec{r},t)\right> \neq 0$; where the averaging can be
considered an ensemble average, a time average, or a spatial average.
Equivalently, long ranged order must develop for the flock as a whole
to move; i.e., the equal-time velocity auto-correlation function:
\begin{eqnarray}
C(\vec{R})\equiv
\left< \vec{v}(\vec{R}+\vec{r},t)\cdot\vec{v}(\vec{r},t)\right>
\label{C real}
\end{eqnarray}
must approach a non-zero constant as the separation
$|\vec{R}| \rightarrow \infty$; specifically:
\begin{eqnarray}
  C(\vec{R} \rightarrow \infty)
\rightarrow |\left<\vec{v} \right>|^2
\label{C real infty}
\end{eqnarray}

Thus the average velocity $\left< \vec{v} \right>$ of the flock is precisely
analogous to the order parameter
$\left< \vec{s} \right>$ in a ferromagnetic system, where $\vec{s}$ is a
local spin.

Our most dramatic result is that an intrinsically non-equilibrium and
nonlinear feature of our model, namely, convection, suppresses fluctuations of
the velocity $\vec{v}$ at long wavelengths, making them much smaller than
the analogous $\vec{s}$ fluctuations found in ferromagnets, for all spatial
dimensions of the flock $d<4$.  Specifically, the connected piece
$C_C(\vec{R})$ of  the correlation function $C(\vec{R})$, defined as:
\begin{eqnarray}
C_C(\vec{R})=C\left(\vec{R}\right)-
\lim_{\left|\vec{R}^\prime\right|
\rightarrow \infty} C\left(\vec{R}^\prime \right) \quad ,
\label{C connect}
\end{eqnarray}
which is a measure of the fluctuations, decays to zero much more rapidly, as
$|\vec{R}| \rightarrow \infty$, than the analogous correlation function in
magnets.  Quantitatively, for points whose separation $\vec{R} \equiv
\vec{R}_\perp$ lies perpendicular to the mean direction of motion of the flock,
\begin{eqnarray}
 C_C(\vec{R}) \propto R^{2\chi}_\perp,
\label{chi def}
\end{eqnarray}
where the  universal ``roughness exponent''
\begin{eqnarray}
\chi = -{1 \over 5}
\label{chi d = 2}
\end{eqnarray}
{\it exactly}, in $d=2$, and is $<1- {d \over 2}$, its value in
magnetic systems, for all $d < 4$.  For $d > 4$, $\chi = 1- {d \over 2}$ for
flocks as well as for magnets.

The physical mechanism for this suppression of fluctuations is easy to
understand:  increased fluctuations in the direction of motion of different
parts of the flock actually {\it enhance} the exchange of information between
those different parts.  This exchange, in turn, suppresses those very
fluctuations, since the interactions between birds tend to make them all move
in the same direction.

These non-equilibrium effects also lead to a spatial anisotropy of
scaling between the direction along ($\parallel$) and those orthogonal
to ($\perp$) the mean velocity $\left< \vec{v} \right>$. The physical
origin of the anisotropy is also simple:  if birds make small errors
$\delta\theta$ in their direction of motion, their random motion
{\it perpendicular} to the mean direction of motion $\left<\vec{v}\right>$ is
much larger than that {\it along}
$\left<\vec{v}\right>$; the former being $\propto \delta \theta$, while the
later is proportional to $1-\cos\delta\theta\sim\delta\theta^2$. As a result,
{\it any} equal-time correlation function in the system of {\it any}
combination of fields crosses over from dependence purely on
$|\vec{R}_\perp|$ to dependence purely on $R_\parallel$ when
\begin{eqnarray}
 {R_\parallel \over \ell_0} \approx \left( {|\vec{R}_\perp
| \over \ell_0}\right)^\zeta \quad ,
\label{cross real}
\end{eqnarray}
where $\ell_0$ is the bird interaction range.

The universal anisotropy exponent
\begin{eqnarray}
\zeta = {3 \over 5}\quad ,
\label{zeta d = 2}
\end{eqnarray}
{\it exactly} in $d=2$, and is $<1$ for all $d<4$.

In particular, the connected, equal-time, velocity autocorrelation
function $C_C(\vec{R})$ obeys the scaling law
\begin{eqnarray}
 C_C\left(\vec{R}\right) = \left|\vec{R}_\perp\right|^{2 \chi}
f_v \left({\left({R_\parallel/\ell_0}\right) \over \left(\left|\vec{R}_\perp
\right|/\ell_0\right)^\zeta}\right) \quad ,
\label{C conn scale}
\end{eqnarray}
where $f_v(x)$ is a {\it universal} scaling function.  We have,
alas, been unable to calculate this scaling function, even in $d=2$
where we know the exponents exactly.  However, the scaling form (\ref{C
conn scale})
immediately implies that
\begin{eqnarray}
 C_C\left(\vec{R}\right) \propto R^{- {2\chi \over \zeta}}_\parallel \, \,
, \quad\;\;\; when \;\;\;
R_\parallel/\ell_0 \gg \left(\vec{R}_\perp/\ell_0\right)^\zeta \quad .
\label{C conn parallel}
\end{eqnarray}

So far, our discussion has focussed on velocity fluctuations.  The
density $\rho(\vec{R},t)$ shows huge fluctuations as well:  indeed, at
long wavelengths, the fluctuations of the density of birds in a flock
become {\it infinitely} bigger than those in a fluid or an ideal
gas.  This fact is obvious to the eye in a picture of a flock
(see Fig. \ref{F1}). Quantitatively, we predict that the spatially Fourier
transformed, equal time density-density correlation function
$ C_{\rho}(\vec{q}) \equiv \left< |\rho(\vec{q},t)|^2 \right> $ obeys
the scaling law:
\begin{eqnarray}
 C_{\rho}(\vec{q}) =
{q^{3-d-\zeta-2\chi}_{\perp} \over  q^2} f_{\rho}
\left({q_{\parallel}\ell_0 \over
\left( q_{\perp}\ell_0\right)^{\zeta}}\right) Y (\theta_{\vec{q}})
\propto \left\{
\begin{array}{ll}
q^{1-d-\zeta-2\chi}_{\perp},& q_{\parallel} \ll
q_{\perp} \\
q_{||}^{-2}q^{3-d-\zeta-2\chi}_{\perp},&\left(\ell_0
q_{\perp}\right)^{\zeta}  \gg
\ell_0 q_{\parallel} \gg q_\perp \ell_0\\
q^{-3+{{1-d-2\chi}\over\zeta}}_{\parallel} q_\perp^2 ,&(q_\perp
\ell_0)^\zeta \ll
q_\parallel
\ell_0
\end{array} \right.
\label{C rho scale0}
\end{eqnarray}
where $Y(\theta_{\vec{q}})$ is a finite, non-vanishing, $O(1)$ function of
the angle
$\theta_{\vec{q}}$ between the wavevector $\vec{q}$ and the direction of
mean flock
motion,
$q_{||}$ and
$\vec{q}_{\perp}$ are the wavevectors parallel and  perpendicular to the
broken symmetry
direction, and
$q_{\perp}=|\vec{q}_{\perp}|$.

In $d=2$, $\zeta={3\over 5}$ and $\chi=-{1\over5}$, so
\begin{eqnarray}
 C_{\rho}(\vec{q}) =
{q^{{4\over5}}_{\perp} \over  q^2} f_{\rho} \left({q_{\parallel}\ell_0 \over
\left( q_{\perp}\ell_0\right)^{{3\over5}}}\right) Y (\theta_{\vec{q}})
\propto \left\{
\begin{array}{ll}
q^{-{6\over5}}_{\perp},& q_{\parallel} \ll
q_{\perp}\\
q_{||}^{-2}q^{{4\over5}}_{\perp},&\left(\ell_0
q_{\perp}\right)^{3\over5}  \gg
\ell_0 q_{\parallel} \gg q_\perp \ell_0\\
q^{-4}_{\parallel} q_\perp^2 ,&(q_\perp \ell_0)^{{3\over5}} \ll q_\parallel
\ell_0
\end{array} \right.
\label{C rho scale}
\end{eqnarray}

The most important thing to note about $C_\rho (\vec{q})$ is that it
diverges as $|\vec{q}| \rightarrow 0$, unlike $C_\rho (\vec{q})$ for,
say, a simple fluid or gas, or, indeed, for any equilibrium
condensed
matter system, which goes to a finite constant (the compressibility) as
$|\vec{q}| \rightarrow 0$.

This correlation function should be extremely easy to measure in
simulations, and in experiments on real herds or flocks, in which, say,
video tape allows one to measure the positions $\vec{r}_i(t)$ of all
the birds (labeled by $i$) in the flock at a variety of times $t$.  The
recipe is simple:
\begin{enumerate}
\item Calculate the complex numbers
\begin{eqnarray}
\rho (\vec{q},t) = \sum_i e^{i \vec{q}\cdot \vec{r}_i(t)}
\label{rho q discrete}
\end{eqnarray}
 for a variety of $\vec{q}$'s.
\item Average the squared magnitude of this number over time.  The
result is $C_\rho(\vec{q})$.
\end{enumerate}

Time dependent correlation functions of $\rho$ and
$\vec{v}$ in flocks also show interesting anomalous scaling behavior.
However, it is not so simple to summarize as the equal time correlation
functions.  Indeed, time dependent correlation functions (or, equivalently,
their
spatio-temporal Fourier transforms) do not have a simple scaling form.  This is
because the collective normal modes of the flock consist of propagating,
damped longitudinal "sound" modes (i.e., density waves), as well as, in $d >
2$, shear modes. The sound modes exhibit two different types of scaling:  the
period
$T$ of a wave is proportional to its wavelength $\lambda$ (the constant of
proportionality being the inverse sound speed); while the {\it lifetime} $\tau$
of the mode is proportional to
$\lambda^z$, with $z$ being another universal exponent.  In most systems
(e.g., fluids, crystals)
exhibiting sound modes, $z
= 2$, corresponding to conventional diffusive or viscous
damping\cite{FNS}.  In flocks, however, we find
\begin{eqnarray}
z ={6 \over 5}; \quad d=2
\label{z d = 2}
\end{eqnarray}
and $z < 2$  for all $d < 4$, for sound modes propagating orthogonal to
the mean direction of flock motion.  That is, sound modes are much
more heavily damped at long wavelengths in flocks than in most
\cite{smec} equilibrium condensed matter systems.

The full dispersion relation for the sound modes is
\begin{eqnarray}
\omega_\pm = {c  _\pm}(\theta_{\vec{q}})q
-iq^z_{\perp} f_\pm
\left({q_{\parallel}\ell_0 \over \left(q_{\perp}\ell_0\right)^\zeta}\right)
\label{omega pm}
\end{eqnarray}
where $\theta_{\vec{q}}$ is the angle between $\vec{q}$ and
$\left< \vec{v}\right>$, and the direction-dependent sound speeds $c_{\pm}
\left(\theta_{\vec{q}} \right)$ are given by equation (\ref{N+9a}) of
section (IV), with $\gamma$ and $\sigma_1$
flock-dependent parameters and $\rho_0$ the mean number density of
``birds'' in the flock.  A polar plot of these sound speeds is given in
Fig. (\ref{F2}). The
exponent
$\zeta$ is the universal anisotropy exponent described earlier, and $f_\pm
(x)$ are
universal scaling functions which, alas, we have been unable to calculate.
However, we do know some of their limits:
\begin{eqnarray}
 f_\pm(x \rightarrow 0) \rightarrow \mbox{constant} > 0; \quad f_\pm(x
\rightarrow \infty) \propto x^{{z \over \zeta}} \quad .
\label{f pm}
\end{eqnarray}

Note that this last result implies that the lifetime $\tau$ of the wave
is $\propto{q}^{-{z \over \zeta}}_{\parallel}$ for
$q_{\parallel}\ell_0 \gg \left(q_{\perp}\ell_0\right)^\zeta$ and $|\vec{q}|
\rightarrow 0$; this only happens for directions of propagation very nearly
parallel to
$\left< \vec{v} \right>$.  Note also that in $d = 2$ where $z =
{6 \over 5}$ and $\zeta = {3 \over 5}, {z \over \zeta}=2$ and the
damping is {\it conventional} for sound modes propagating
parallel to the mean motion of the flock.  For all {\it other}
directions of propagation, however, it is unconventional, and
characterized by $z = {6 \over 5}$.

This behavior of the damping (i.e. $Im \omega)$, is summarized in
Fig. (\ref{F3}).  For $d>2$, the ``hydrodynamic'' mode structure also
includes $d-2$ ``hyper diffusive'' shear modes, with identical
dispersion relations
\begin{eqnarray}
\omega_s = \gamma q_{\parallel} - iq^z_{\perp}f_s
\left({q_{\parallel}\ell_0
\over
\left(q_{\perp} \ell_0\right)^\zeta}\right) \, .
\label{omega shear}
\end{eqnarray}

The dispersion relations for
$\omega_\pm$ and
$\omega_s$ can be directly probed by measuring the
spatio-temporally Fourier transformed density--density and
velocity--velocity auto--correlation functions
\begin{eqnarray}
C_{\rho}(\vec{q},\omega) &\equiv& \left<|\rho
(\vec{q},\omega)|^2
\right>\\ \nonumber
 C_{ij}(\vec{q},\omega) &\equiv& \left<
v_i(\vec{q},\omega)v_j(-\vec{q},-\omega)
\right>
\label{C q omega}
\end{eqnarray}
respectively.  Experimentally, or in simulations, $C_\rho\left(\vec{q},
\omega \right)$ can be calculated by temporally Fourier transforming
the spatially Fourier transformed density
(\ref{rho q discrete}):
\begin{eqnarray}
\rho_n \left(\vec{q}, \omega\right) = \sum^{(n+1) \tau}_{t=n \tau}\rho_n
\left(\vec{q}, t\right) e^{-i\omega t}, \quad n = 0, 1, 2, \dots
\label{rho omega}
\end{eqnarray}
over a set of long ``bins'' of time intervals of length $\tau \gg t_0$ (the
``microscopic'' time step), and then averaging the squared magnitude
$\left|
\rho
\left(\vec{q},
\omega\right)\right|^2$ over bins:
\begin{eqnarray}
C_\rho \left(\vec{q}, \omega\right) \equiv \sum^{n_{max}}_{n=0} {\left|
\rho \left(\vec{q},
\omega\right)\right|^2 \over n_{max}}
\label{C rho omega}
\end{eqnarray}

Closed form expressions for these correlation functions in terms of the
scaling functions $f_L$ and $f_s$ are given in
section V.  Although these expressions look quite
complicated, the behavior they predict is really quite simple, as
illustrated in Fig. (\ref{F4}), where $C_\rho$ is plotted as a function of
$\omega$ for fixed $q$.  As shown there, $C_\rho$ has two sharp
peaks at $\omega = c_{\pm} \left(\theta_{\vec{q}}\right)q$,
of width
$\propto q_{\perp}^z f_L \left({q_{\parallel}\ell_0
\over \left(q_{\perp}\ell_0\right)^\zeta}\right)$
and height
$\propto q_{\perp}^{-\left(2\chi+z +3\zeta + d -3\right)} g
\left({q_{\parallel}\ell_0 \over
\left(q_{\perp}\ell_0\right)^\zeta}\right)$.  Thus,
$c_\pm\left(\theta_{\vec{q}}\right)$ can be simply extracted from the
position of the peaks, while the exponents $\chi, z$ and $\zeta$ can be
determined by comparing their widths and heights for different
$\vec{q}$\,'s.

The scaling properties of the flock are completely summarized by the universal
exponents $z$, $\zeta$, and
$\chi$.  In $d=2$, our predictions for these exponents are:
\begin{eqnarray}
 z= {6 \over 5}\, , \quad
\zeta  = {3 \over 5}\, , \quad
\chi =-{1 \over 5} .
\label{}
\end{eqnarray}
These results are {\it exact} and universal for
all flocks with the simple symmetries we discussed at the outset.

For $d\ne 2$, the situation is less clear. We have
performed a one loop,
$4-\epsilon$ expansion to attempt to calculate these exponents, and
find that, to this order, the model appears to have a fixed
{\it line} with continuously varying exponents
$z$, $\zeta$, and $\chi$.  Whether this is an artifact of our one
loop calculation, or actually happens, is unclear.  A two loop
calculation might clarify matters, but would be extremely long and
tedious.  (One loop was hard enough.)

The origin of this complication is an additional convective
non--linearity\cite{Sethna} not discussed in Ref. \cite{TT}.  This new
term (whose coefficient is a parameter we call
$\lambda_2$) is unrenormalized at one loop order, leading to the
apparent fixed line at that order.  In two dimensions, this extra term
can be written as a total derivative, and can be absorbed into the
non--linear term considered in Ref. \cite{TT}.  Hence, in $d=2$, the
results of
\cite{TT} are sound.  In $d>2$, however, this new term has a
different structure, and could, if it does not renormalize to zero,
change the exponents $\chi$, $z$, and
$\zeta$.  Since $\lambda_2$ does {\it not} renormalize at one
loop order, all we can say at this point is that there are three
possibilities:

\begin{enumerate}

\item  At higher order, $\lambda_2$ renormalizes to zero.
{\it If} this is the case, we can show that
\begin{eqnarray}
  z = {2(d+1) \over 5}\, , \quad
\zeta = {d+1 \over 5}\, , \quad
\chi = {3-2d \over 5}\, ,
\label{canon exp}
\end{eqnarray}
exactly, for all $d$ in the range $2  \leq d \leq 4$.
Note that these results linearly interpolate between the equilibrium
results
$z =2$, $\zeta=1$, and $\chi=1-{d \over 2}$ in $d=4$, and our $2d$
results $z={6 \over 5}$, $\zeta =
{3 \over 5}$, and $\chi = -{1 \over 5}$ in $d=2$.

\item At higher
order, $\lambda_2$ grows upon renormalization and reaches a
non-zero fixed point value $\lambda^*_2$ at some new fixed point
that differs from the $\lambda_2=0$ fixed point we've studied
previously, at which Eqn.\ (\ref{canon exp}) holds.  The exponents $\chi$, $z$,
and $\zeta$ would still be universal (i.\ e., depend only on the
dimension of space
$d$) for all flocks in this case, but those universal values would be
different from Eqn.\ (\ref{canon exp}).

\item $\lambda_2$ is unrenormalized to all orders.  Should this
happen, $\lambda_2$ would parameterize a fixed
{\it line}, with continuously varying values of the exponents
$z$, $\chi$, and $\zeta$.
\end{enumerate}

We reiterate:  we do not know which of the above possibilities holds
for $d > 2$.  However, whichever holds  is {\it universal}; that is, only
{\it one} of the three possibilities above applies to {\it all} flocks.  We
don't, however, know which one that is.

We also study an anisotropic model for flocking, which incorporates
the possibility that birds are averse to flying in certain directions (i.e.,
straight up or straight down).  In particular, we consider the case in
which, for arbitrary spatial dimensions $d \geq 2$, there is an easy
{\it plane} for motion (i.e., a $d_e = 2$ dimensional subspace
of the full d--dimensional space).  In this case, the relevant pieces of
the $\lambda_2$ vertex become a total derivative, and can be
absorbed into the non-linear term considered in reference \cite{TT},
for {\it all} spatial dimensions $d$, not just $d=2$ as in the
isotropic model.  Hence, we are able to obtain exact exponents for this
problem for {\it all} spatial dimensions $d$, {\it not}
just $d=2$.

We again find anisotropic, anomalous scaling for $d < 4$.  The
anisotropy of scaling is between the direction in the easy plane (call it
$x$) perpendicular to the mean direction of motion (call it $y$; which,
of course, also lies in the easy plane), and all $d-1$ other directions,
{\it including}
$y$ (see figure 5). That is, the equal time, velocity-velocity
autocorrelation function obeys the scaling law:
\begin{eqnarray}
C_v(\vec{R}) = x^{2\chi} f_v\,\left({\left(y/\ell _0\right) \over
\left(x/\ell _0\right)^{\zeta}}\, ,\,{|\vec{r}_H|/\ell _0 \over
\left(x/\ell _0\right)^{\zeta}}\right)
\label{anis v}
\end{eqnarray}
where $\vec{r}_H$ denotes the $d-2$ components of $\vec{r}$ in the
``hard'' directions orthogonal to the easy plane, with the scaling
exponents
$\chi$ and
$\zeta$ given by
\begin{eqnarray}
\chi = {1-d \over 7-d }
\label{chi anis}
\end{eqnarray}
\begin{eqnarray}
\zeta = {3 \over 7-d}
\label{zeta anis}
\end{eqnarray}
{\it exactly}, for all spatial dimensions $d$ in the range $2\leq d\leq 4$. For
$d>4$,
$\chi=1-{d\over 2}$ and $\zeta=1$, as in the isotropic case, while in
$d=2$, where the model
becomes identical to the isotropic model (the easy plane of motion being
the entire space in that
case), we again recover the isotropic results, $\zeta = {3\over 5}$,
and $\chi = -{1\over 5}$.  For the physical case $d = 3$, we have
\begin{eqnarray}
\chi(d=3) = -{1 \over 2} \quad ; \quad\zeta(d = 3) = {3 \over 4}
\label{chi 3 d = 3}
\end{eqnarray}
The Fourier-transformed, equal time density ($\rho-\rho$) correlation function
$C_\rho (\vec{q})$ also obeys a scaling law
\begin{eqnarray}
C_\rho (\vec{q}) =
q^{1-2\chi-(d-1)\zeta}_x(q_x^2+q_y^2)^{-1}
f_\rho^{A} \left({q_y\ell_0 \over \left(q_x
\ell_0 \right)^\zeta},  {|\vec{q}_H|\ell_0 \over \left(q_x
\ell_0 \right)^\zeta} \right)
Y_a(\theta_{xy}),
\label{an0}
\end{eqnarray}
where $Y_a(\theta_{xy})$ is a finite, non-zero, $O(1)$ function of the
angle $\theta_{xy}=
\tan^{-1}(q_x/q_y)$, and the scaling function $f_{\rho}^{A}$ follows
\begin{eqnarray}
f_\rho^{A} \left({q_y\ell_0 \over \left(q_x
\ell_0 \right)^\zeta},  {|\vec{q}_H|\ell_0 \over \left(q_x
\ell_0 \right)^\zeta} \right)
\propto\left\{\begin{array}{ll}
constant , & \ell_0^2
\left(q^2_y +
\nu \left|\vec{q}_H
\right|^2\right) \ll (\ell_0q_x)^\zeta \\
\left({q_x^{2\zeta} \over q^2_y +
\nu\left|\vec{q}_H\right|^2}\right)^{1+2\chi+(d-1)\zeta \over 2\zeta }, &
\ell _0^2
\left(q^2_y +
\nu\left|\vec{q}_H
\right|^2\right) \gg (\ell_0q_x)^\zeta
\end{array} \right.
\end{eqnarray}
where $\ell _0$ is a ``microscopic'' length (of the order the interbird
distance)
and $\nu$ a
dimensionless nonuniversal constant of order unity.

Finally, the hydrodynamic mode structure of this anisotropic flock consists
of a pair of
propagating longitudinal sound modes, with dispersion relation given, in
the co-ordinate system
of figure (\ref{F5}), by
\begin{eqnarray}
\omega = c_{\pm}\left(\theta_{\vec{q}}, \phi_{\vec{q}}\right)q - iq^z_x
f_A
\left({q_y\ell_0
\over
\left(q_x\ell_0\right)^\zeta}\, , \,{\left|\vec{q}_H\right| \ell_0  \over
\left(q_x\ell_0\right)^\zeta}
\right)
\label{}
\end{eqnarray}
where $f_A$ is a universal scaling
function, $c_{\pm}\left(\theta_{\vec{q}}, \phi_{\vec{q}}\right)$ is
given by equation (\ref{canis}) of Chapter 6 and the dynamical exponent
\begin{eqnarray}
z = 2\zeta = {6\over {7-d}}={3 \over 2}
\label{}
\end{eqnarray}
where the last equality holds in $d = 3$. Note that this value of $z$
again reduces to that of the isotropic model in $d = 2$, and $d = 4$.

The spatio-temporally Fourier transformed density-density
correlation function
$\left<|\rho(\vec{q},\omega)|^2\right>$ has the same structure as that
illustrated for the isotropic problem in figure (4), with the
modification that
$q_\perp$ is replaced by $q_x$, and the scaling function
$f_L$ is replaced by $f_A$. The detailed expression for
$\left<|\rho(\vec{q},\omega)|^2\right>$ is given by equation (\ref{Aniso Corr
rhorho}) of Section VI.

The remainder of this paper is organized (and we use the term
loosely) as follows: In Section II, we formulate the isotropic model. In
Section III, we specialize this model to the ``broken symmetry'' state,
in which the flock is moving with a non-zero mean speed
$\left<\vec{v}\right>$. In Section IV, we linearize the broken
symmetry state model, and calculate the correlation functions and
scaling laws in this linear approximation. In Section V, we study the
anharmonic corrections in the broken symmetry state, show
that they diverge in spatial dimensions
$d<4$, derive the new scaling laws that result in that case,  calculate
the exact exponents in $d=2$ and discuss the difficulties that prevent
us from obtaining these exponents for $2 < d < 4$. In Section VI, we
repeat all of the above for the anisotropic model.  In Section VII, we
describe in some detail how our predictions might be tested
experimentally, both by observations of real flocks of living organisms,
and in simulations. And finally, in Section VIII, we discuss some of the
open questions remaining in this problem, and suggest some possible
directions for future research.




\section{The isotropic model}
In this section, we formulate our model for isotropic
flocks.  As discussed in the introduction, the system we
wish to model is any collection of a large number $N$ of
organisms (hereafter referred to as ``birds'') in a $d$-dimensional
space, with each organism seeking to move in the same direction
as its immediate neighbors.

We further assume that each organism has no ``compass;'' i.\ e.,
no intrinsically preferred direction in which it wishes to move.
Rather, it is equally happy to move in any direction picked by its
neighbors.  However, the navigation of each organism is not
perfect; it makes some errors in attempting to follow its
neighbors.  We consider the case in which these errors have zero
mean; e.\ g., in two dimensions, a given bird is no more likely to
err to the right than to the left of the direction picked by its
neighbors.  We also assume that these errors have no long
temporal correlations; e.\ g., a bird that has erred to the right at
time $t$ is equally likely to err either left or right at a time
$t^\prime$ much later than $t$.

Although the continuum model we propose here will describe the
long distance behavior of {\it any} flock satisfying the
symmetry conditions we shall specify in a moment, it is instructive
to first consider an explicit example:  the automaton studied by
Vicsek et al \cite{Vicsek}.  In this discrete time model, a number
of boids labeled by $i$ in a two--dimensional plane with positions
$\{\vec{r}_i(t)\}$ at integer time $t$, each chooses the
direction it will move on the next time step (taken to be of
duration $\Delta t = 1$) by averaging the directions of
motion of all of those birds within a circle of radius $R _0$ (in
the most convenient units of length $R _0 = 1$) on the previous
time step (updating is simultaneous).  The distance $R _0$ is
assumed to be $\ll L$, the size of the flock.  The direction the bird
actually moves on the next time step differs from the above
described direction by a random angle $\eta_i(t)$, with zero
mean and standard deviation $\Delta$.  The distribution of
$\eta_i(t)$ is identical for all birds, time independent, {\it and}
uncorrelated between different birds and different time steps.
Each bird then, on the next time step, moves in the direction so
chosen a distance $v_0\Delta t$, where the speed $v_0$ is the
same for all birds.

To summarize, the rule for bird motion is
\begin{eqnarray}
\theta _i (t + 1) = \left< \theta _j (t)\right> + \eta_i (t)
\label{sigma rule}
\end{eqnarray}
\begin{eqnarray}
\vec{r}_i \left(t + 1 \right) = \vec{r}_i (t) + v_0
\left(\cos
\theta{(t + 1)}, \sin \theta{(t + 1)}\right)
\label{more}
\end{eqnarray}
\begin{eqnarray}
\left< \eta _i (t) \right> = 0
\label{ave eta}
\end{eqnarray}
\begin{eqnarray}
\left< \eta _i (t)\eta _j (t^\prime) \right> = \Delta \delta_{ij}
\delta_{tt^\prime}
\label{ave eta2}
\end{eqnarray}
where the average in eqn (\ref{sigma rule}) is over all birds $j$
satisfying
\begin{eqnarray}
\left|\vec{r}_j(t) -  \vec{r}_i(t) \right| < R _0
\label{circle}
\end{eqnarray}
and $\theta_i(t)$ is the angle of the direction of motion of the $i$th bird
(relative to some fixed reference axis) on the time step that ends
at $t$.  The flock evolves through the iteration of this rule.  Note
that the ``neighbors'' of a given bird may change on each time
step, since birds do not, in general, move in exactly the same
direction as their neighbors.

This model, though simple to simulate, is quite difficult to
treat analytically.  Our goal in our previous work\cite{TT} and
this paper is to capture the essential physics of this model in a
continuum, ``hydrodynamic'' description of the flock.  Clearly,
some short-ranged details must be lost in such a description.
However, as in hydrodynamic descriptions of equilibrium
systems\cite{FNS}, as well as many recent treatments\cite{KPZ} of
non-equilibrium systems, our hope is that our continuum
approach can correctly reproduce the long-distance, long-time
properties of the class of systems we wish to study.  This hope is
justified by the notion of universality:  all ``microscopic models''
(in our case, different specifications for the exact laws of motion
for an individual bird) that have the same symmetries and
conservation laws should have the same long distance behavior.
This belief can be justified by our
renormalization group treatment of the continuum model.

So, given this lengthy preamble, what {\it are} the
symmetries and conservation laws of flocks?

The only symmetry of the model is rotation invariance:  since the
``birds'' lack a compass, all direction of space are equivalent to
other directions.  Thus, the ``hydrodynamic'' equation of motion
we write down cannot have built into it any special direction
picked ``a priori''; all directions must be spontaneously picked out
by the motion and spatial structure of the flock.  As we shall see,
this symmetry {\it severely} restricts the allowed terms
in the equation of motion.

Note that the model does {\it not} have Galilean
invariance:  changing the velocities of all the birds by some
constant boost $\vec{v}_b$ does {\it not} leave the model
invariant.  Indeed, such a boost is {\it impossible} in a
model that strictly obeys Vicsek's rules, since the
{\it speeds} of all the birds will not remain equal to $v_0$
after the boost.  One could image relaxing this constraint on the
speed, and allowing birds to occasionally speed up or slow down,
while tending an average to move at speed $v_0$.  Then the
boost just described would be possible, but clearly would change
the subsequent evolution of the flock.

Another way to say this is that birds move through a resistive
medium, which provides a special Galilean reference frame, in
which the dynamics are particularly simple, and different from
those in other reference frames.  Since real organisms in flocks
always move through such a medium (birds through the air, fish
through the sea, wildebeest through the arid dust of the
Serengeti), this is a very realistic feature of the
model.

As we shall see shortly, this {\it lack} of Galilean
invariance
{\it allows} terms in the hydrodynamic equations of birds
that are {\it not} present in, e.\ g., the Navier-Stokes
equations for a simple fluid, which {\it must} be Galilean
invariant, due to the absence of a luminiferous ether.

The sole conservation law for flocks is conservation of birds:  we
do not allow birds to be born or die ``on the wing''.

In contrast to the Navier-Stokes equation, there is no
conservation of momentum. This is, ultimately, a consequence of
the absence of Galilean invariance.

Having established the symmetries and conservation laws
constraining our model, we need now to identify the
hydrodynamic variables.  They are:  the coarse grained bird
velocity field $\vec{v}(\vec{r},t)$, and the coarse grained bird
density $\rho(\vec{r},t)$.  The field $\vec{v}(\vec{r},t)$, which is
defined for all $\vec{r}$, is a suitable weighted average of the
velocities of the individual birds in some volume centered on
$\vec{r}$.  This volume is big enough to contain enough birds to
make the average well-behaved, but should have a spatial linear
extent of no more than a few ``microscopic'' lengths (i.\ e., the
interbird distance, or by a few times the interaction range
$R_0$).  By suitable weighting, we seek to make
$\vec{v}(\vec{r},t)$ fairly smoothly varying in space.

The density $\rho(\vec{r},t)$ is similarly defined, being just the
number of particles in a coarse graining volume, divided by that
volume.

The exact prescription for the coarse graining should be
unimportant, so long as $\rho(\vec{r},t)$ is normalized so as to
obey the ``sum rule'' that its integral over any
{\it macroscopic} volume (i.\ e., any volume compared
with the aforementioned microscopic lengths) be the total number
of birds in that volume.  Indeed, the coarse graining description
just outlined is the way that one imagines, in principle, going over from a
description of a simple fluid in terms of equations of motion for
the individual constituent molecules to the continuum
description of the Navier-Stokes equation.

We will also follow the historical precedent of the
Navier-Stokes\cite{FNS} equation by deriving our continuum, long
wavelength description of the flock {\it not} by explicitly
coarse graining the microscopic dynamics (a
{\it very} difficult procedure in practice), but, rather, by
writing down the most general continuum equations of motion for
$\vec{v}$ and $\rho$ consistent with the symmetries and
conservation laws of the problem.  This approach allows us to
bury our ignorance in a few phenomenological parameters, (e.\ g.,
the viscosity in the Navier-Stokes equation) whose numerical
values will depend on the detailed microscopic rules of individual
bird motion.  What terms can be present in the EOM's, however,
should depend only on symmetries and conservation laws, and
{\it not} on the microscopic rules.

To reduce the complexity of our equations of motion still further,
we will perform a spatial-temporal gradient expansion, and keep
only the lowest order terms in gradients and time derivatives of
$\vec{v}$ and $\rho$.  This is motivated and justified by our
desire to consider {\it only} the long distance, long time
properties of the flock.  Higher order terms in the gradient
expansion are ``irrelevant'':  they can lead to {\it finite}
``renormalization'' of the phenomenological parameters of the
long wavelength theory, but {\it cannot} change the type
of scaling of the allowed terms.

With this lengthy preamble in mind, we now write down the
equations of motion:
\begin{eqnarray}
\partial_{t}
\vec{v}+\lambda_1(\vec{v}\cdot\vec{\nabla})\vec{v}&+&
\lambda_2(\vec{\nabla}\cdot\vec{v})\vec{v}
+\lambda_3\vec{\nabla}(|\vec{v}|^2) = \nonumber \\
&&\alpha\vec{v}-\beta
|\vec{v}|^{2}\vec{v} -\vec{\nabla} P +D_{B} \vec{\nabla}
(\vec{\nabla}
\cdot \vec{v}) \nonumber \\
&&+ D_{T}\nabla^{2}\vec{v} +
D_{2}(\vec{v}\cdot\vec{\nabla})^{2}\vec{v}+\vec{f}
\label{EOM}
\end{eqnarray}
\begin{eqnarray}
P=P(\rho)=\sum_{n=1}^{\infty} \sigma_n
(\rho-\rho_0)^n
\label{P rho}
\end{eqnarray}
\begin{eqnarray}
{\partial\rho \over \partial
t}+\nabla\cdot(\vec{v}\rho)=0
\label{conservation}
\end{eqnarray}
where $\beta$, $D_{B}$, $D_{2}$ and $D_{T}$ are all
positive, and
$\alpha < 0$ in the disordered phase and $\alpha>0$ in
the ordered state (in mean field theory). The origin of
the various terms is as follows:  the $\lambda$ terms on the left
hand side of eq. (\ref{EOM}) are the analogs of the usual convective derivative
of the coarse-grained velocity field
$\vec{v}$ in the Navier-Stokes equation. Here the absence of Galilean
invariance allows all
{\it three} combinations of one spatial gradient and two
velocities that transform like vectors; if Galilean invariance
{\it did}  hold, it would force $\lambda_2=\lambda_3=0$ and
$\lambda_1=1$. However, Galilean invariance does {\it not}
hold, and so all three coefficients are non-zero phenomenological
parameters whose non-universal values
are determined by the microscopic rules.  The
$\alpha$ and
$\beta$ terms simply make the local
$\vec{v}$ have a non-zero magnitude
$(=\sqrt{\alpha/\beta})$ in the ordered phase, where
$\alpha>0$. $D_{L,1,2}$ are the diffusion constants (or
viscosities) reflecting the tendency of a localized fluctuation in the
velocities to spread out because of the coupling between
neighboring ``birds". The $\vec{f}$ term is a random
driving force representing the noise. We assume it is Gaussian with
white noise correlations:
\begin{eqnarray}
 <f_{i}(\vec{r},t)f_{j}(\vec{r'},t')>=\Delta
\delta_{ij}\delta^{d}(\vec{r}-\vec{r'})\delta(t-t')
\label{white noise}
\end{eqnarray}
 where $\Delta$ is a constant, and $i$ , $j$ denote
Cartesian components. Finally, P is the pressure, 
which tends to maintain the local number
density
$\rho(\vec{r})$ at its mean value $\rho_0$,
and $\delta \rho = \rho -
\rho_0$.

The final
equation (\ref{conservation}) is just conservation of bird number (we
don't allow our birds to reproduce or die ``on the wing").

Symmetry allows any of the phenomenological coefficients
$\lambda_i$, $\alpha$, $\sigma_n$, $\beta$, $D_i$ in equations (\ref{EOM})
and (\ref{P rho}) to be functions of the squared magnitude $\left|\vec{v}
\right|^2$ of the velocity, and of the density $\rho$ as well.




\section{The Broken Symmetry State}

We are mainly interested in the symmetry broken phase;
specifically in whether fluctuations around the symmetry broken
ground state destroy it (as in the analogous phase of the 2D XY
model). For $\alpha>0$, we can write the velocity field as
$\vec{v}=v_{0}\hat{x}_{\parallel}+\vec{\delta v}$, where
$v_{0}\hat{x}_{\parallel}=<\vec{v}>$ is the spontaneous average
value of
$\vec{v}$ in the ordered phase. We will
chose
$v_0=\sqrt{{\alpha} \over {\beta}}$ (which should be thought of as
an implicit condition on $v_0$, since $\alpha$ and
$\beta$ can, in general, depend on $\left| \vec{v}\right|^2$); with this
choice, the equation of motion for the fluctuation $\delta
v_\parallel$ of $v_\parallel$ is
\begin{eqnarray}
\partial_{t}\delta v_{\parallel} = - \sigma_1\partial_{\parallel}
\delta\rho - 2\alpha \delta v_{\parallel} + {\rm irrelevant \, terms}.
\label{v parallel elim}
\end{eqnarray}

Note now that if we are interested in ``hydrodynamic'' modes, by
which we mean modes for which frequency $\omega \rightarrow 0$
as wave vector $q \rightarrow 0$, we can, in the hydrodynamic
$(\omega, q \rightarrow 0)$ limit, neglect $\partial_t
\delta v_{\parallel}$ relative to $\alpha \delta v_{\parallel}$ in
(\ref{v parallel elim}).  The resultant equation can trivially be
solved for $\delta v_{\parallel}$:
\begin{eqnarray}
\delta v_{\parallel} = -D_{\rho} \partial_{\parallel} \delta\rho
\label{delta parallel elim}
\end{eqnarray}
where we've defined another diffusion constant
$D_{\rho} \equiv {\sigma_1 \over 2\alpha}$.  Inserting
(\ref{delta parallel  elim}) in the equations of motion for
$\vec{v}_{\perp}$ and
$\delta\rho$, we obtain, neglecting ``irrelevant'' terms:
\begin{eqnarray}
\partial_{t} \vec{v}_{\perp} &+& \gamma\partial_{\parallel} \vec{v}_{\perp}
+ \lambda_1 \left(\vec{v}_{\perp} \cdot
\vec{\nabla}_{\perp}\right) \vec{v}_{\perp} +
\lambda_2
\left(\vec{\nabla}_\perp \cdot \vec{v}_{\perp}\right)
\vec{v}_{\perp} \nonumber \\  &=& -\vec{\nabla}_{\perp} P +
D_B\vec{\nabla}_\perp\left(\vec{\nabla}_\perp\cdot\vec{v}_\perp\right)+
D_T\nabla^{2}_{\perp}\vec{v}_{\perp} +
D_{\parallel}\partial^{2}_{\parallel}\vec{v}_{\perp}+\vec{f}_{\perp}
\label{EOM broken}
\end{eqnarray}
\begin{eqnarray} {\partial\delta
\rho \over \partial t}+\rho_o\vec{\nabla}_\perp\cdot\vec{v}_\perp
+\vec{\nabla}_\perp\cdot(\vec{v}_\perp\delta\rho)+v_0\partial_{\parallel}\delta
\rho=D_
{\rho}\partial^2_{||}\delta\rho
\label{cons broken}
\end{eqnarray}
where $D_\rho$, $D_{B}$, $D_T$ and $D_{\parallel} \equiv
D_{T}+D_{2}v_0^2$ are the diffusion constants, and we've defined
\begin{eqnarray}
\gamma \equiv
\lambda_1v_0 \, .
\label{gamma def}
\end{eqnarray}
The pressure $P$ continues to be given, as it always will, by equation
(\ref{P rho}).

>From this point forward,  we will treat the phenomenological parameters
$\lambda_i$ ,
$\gamma$, and $D_i$ appearing in equations (\ref{EOM broken}) and
(\ref{cons broken}) as
constants, since they depend, in our original model (\ref{EOM}),only on the
scalar quantities $
\left|\vec{v}\right|^2$ and $\rho(\vec{r})$, whose fluctuations in the
broken symmetry state
away from their mean values $v_0^2$ and $\rho_0$ are small. Furthermore,
these fluctuations
lead only to ``irrelevant" terms in the equations of motion.

It should be emphasized here that, once non-linear fluctuation effects are
included, the $v_0$ in
equation (\ref{cons broken}) will $\it{not}$ be given by the ``mean''
velocity of the birds, in the
sense of
\begin{eqnarray}
\left< v \right> \equiv {\left| \sum_i
\vec{v}_i\right| \over N } \quad ,
\label{var v disc}
\end{eqnarray}
where $N$ is the number of birds. This is because, in our continuum
language,
\begin{eqnarray}
\left< v \right> = {\left|\left<\int\rho(\vec{r},t)\vec{v}(\vec{r} ,
t)d^dr\right>\right|
\over  \left<\int\rho(\vec{r} , t)d^dr\right>}
 = {{\left|\left<\rho\vec{v}\right>\right| \over \left<\rho\right>}}
\label{bar v cont}
\end{eqnarray}
while $v_0$ in equation (\ref{gamma def}) is
\begin{eqnarray}
v_0 = \left| \left<
\vec{v} \left(\vec{r}, t \right)\right>\right|
\label{v zero}
\end{eqnarray}
Once $\rho$ fluctuates, so that $\rho = <\rho> + \delta \rho$, the
``mean'' velocity of the birds
\begin{eqnarray}
\left< v \right> = \left|{\left<\rho\vec{v} \right>\over \left<\rho
\right>}\right| =
\left|{\left<\rho\right>\left<\vec{v} \right>\over \left<\rho
\right>} + {\left<\delta\rho\vec{v} \right>\over \left<\rho
\right>}\right|
\label{bar v 2}
\end{eqnarray}
which only $= v_0 \equiv \left|\left< \vec{v}\right> \right|$ if the
correlation function $\left< \delta \rho\vec{v}\right> = 0$, which it
will not, in general.  For instance, one could easily imagine that
denser regions of the flock might move faster; in which case  $\left<
\delta \rho\vec{v}\right>$ would be positive along $\left<
\vec{v}\right>$.  Thus, $\left< \vec{v}\right>$ measured in a
simulation by simply averaging the speed of all birds, as in equation
(\ref{var v disc}), will not be equal to $v_0$ in equation (\ref{gamma
def}).  Indeed, we can think of no simple way to measure $v_0$, and
so chose instead to think of it as an additional phenomenological
parameter in the broken symmetry state equations of motion
(\ref{EOM broken}).  It should, in simulations and experiments, be
determined by
fitting the correlation functions we will calculate in the next section.
One should {\it not} expect it to be given by $\left< v \right>$ as
defined in equation (\ref{bar v cont}).

Similar considerations apply to $\gamma$:  it should also be thought of
as an independent, phenomenological parameter, {\it not} necessarily
determined by the mean velocity and non-linear parameter
$\lambda_1$ through (\ref{gamma def}).




\section{Linearized Theory of the Broken Symmetry
State}

As a first step towards understanding the implications of these
equations of motion, we linearize them in
$\vec{v}_\perp$ and $\delta \rho \equiv \rho -
\rho_0$.  Doing this, and Fourier transforming in space and time,
we obtain the linear equations
\begin{eqnarray}
\left[-i\left(\omega - \gamma q_{\parallel} \right) +
\Gamma_T\left(\vec{q}\right)\right]
\vec{v}_T\left(\vec{q},\omega\right)  = \vec{f}_{T}\left(
\vec{q},\omega\right)
\label{vT lin}
\end{eqnarray}
\begin{eqnarray}
\left[-i\left(\omega - \gamma q_{\parallel} \right) +
\Gamma_L\left(\vec{q}\right)\right]
v_L + i\sigma_1 q_{\perp} \delta \rho = f_L\left(\vec{q},\omega\right)
\label{vL lin}
\end{eqnarray}
\begin{eqnarray}
\left[-i \left(\omega -v_0 q_{\parallel}\right)   + \Gamma_{\rho}(\vec{q})
\right]
\delta \rho + i\rho_0 q_\perp v_L = 0
\label{rho lin}
\end{eqnarray}
where
\begin{eqnarray}
v_L\left(\vec{q}, \omega\right) \equiv
{\vec{q}_\perp \cdot
\vec{v}_\perp \left(\vec{q}, \omega\right) \over q_\perp}
\label{N+3}
\end{eqnarray}
 and
\begin{eqnarray}
\vec{v}_T\left(\vec{q}, \omega \right) =
\vec{v}_\perp\left(\vec{q},
\omega\right) -{\vec{q}_\perp v_L \over q_\perp}
\label{N+4}
\end{eqnarray}
are the longitudinal and transverse (to
$\vec{q}_\perp$) pieces of the velocity, $\vec{f}_{T}
\left(\vec{q}, \omega\right)$ and $f_L \left(\vec{q},
\omega\right)$ are the analogous pieces of the Fourier
transformed random force
$\vec{f}
\left(\vec{q}, \omega\right)$, and we've defined wavevector
dependent transverse, longitudinal, and $\rho$ dampings
$\Gamma_{L,T,\rho}$:
\begin{eqnarray}
\Gamma_L \left(\vec{q}\right) \equiv D_L
 q_\perp^2 + D_{\parallel} q^2_{\parallel}
\label{N+5}
\end{eqnarray}
\begin{eqnarray}
\Gamma_T\left(\vec{q}\right)  = D_T q_\perp^2 + D_{\parallel}
q^2_{\parallel} \, ,
\label{N+6}
\end{eqnarray}
\begin{eqnarray}
\Gamma_{\rho}\left(\vec{q}\right) = D_\rho q^2_{\parallel} \, ,
\label{N+6a}
\end{eqnarray}
where we've defined $D_L \equiv D_T + D_B$, $q_{\perp}=|\vec{q}_{\perp}|$.

Note that in $d = 2$, the transverse velocity $\vec{v}_T$ does not
exist:  no vector can be perpendicular to {\it both} the
$x_{\parallel}$ axis {\it and} $\vec{q}_\perp$ in two dimensions.
This leads to many important simplifications in $d = 2$, as we will
see later; these simplifications make it (barely) possible to get {\it
exact} exponents in $d = 2$ for the full, non-linear problem.

The normal modes of these equations are $d-2$ purely diffusive
transverse modes associated with $\vec{v}_T$, all of which have
the same eigenfrequency
\begin{eqnarray}
\omega _T =  \gamma q_{\parallel}  -i\Gamma_T(\vec{q}) =
\gamma q_{\parallel} -i
\left(D_Tq_{\perp}^2 + D_{\parallel} q_{\parallel}^2  \right) \, ,
\label{N+7}
\end{eqnarray}
and a pair of damped, propagating sound modes
with complex (in {\it both} senses of the word) eigenfrequencies
\begin{eqnarray}
\omega _\pm &=&
c_\pm\left(\theta_{\vec{q}}
\right)q - i
\Gamma_L
 \left[{ v_{\pm} \left(\theta_{\vec{q}} \right) \over
2c_2 \left(\theta_{\vec{q}} \right)}\right] -
i \Gamma_{\rho}
 \left[{ v_{\mp} \left(\theta_{\vec{q}} \right) \over
2c_2 \left(\theta_{\vec{q}} \right)}\right]
\nonumber  \\
  &=&
c_\pm\left(\theta_{\vec{q}}
\right)q - i
(D_{L}q_{\parallel}^2+D_{\perp}q_{\perp}^2)
 \left[{ v_{\pm} \left(\theta_{\vec{q}} \right) \over
2c_2 \left(\theta_{\vec{q}} \right)}\right] -
i D_{\rho}q_{\parallel}^2
 \left[{ v_{\mp} \left(\theta_{\vec{q}} \right) \over
2c_2 \left(\theta_{\vec{q}} \right)}\right]
\label{Wplusminus}
\end{eqnarray}
where $\theta_{\vec{q}}$ is the angle between
$\vec{q}$ and the direction of flock motion (i.\ e., the
$x_\parallel$ axis),
\begin{eqnarray}
c_{\pm}\left(\theta_{\vec{q}} \right)  =
{\gamma + v_0 \over 2}\cos
\left(\theta_{\vec{q}}
\right)
\pm c_2\left(\theta_{\vec{q}} \right)
\label{N+9a}
\end{eqnarray}
\begin{eqnarray}
v_{\pm}\left(\theta_{\vec{q}} \right)  =
\pm {\gamma - v_0 \over 2}\cos
\left(\theta_{\vec{q}}
\right)
+ c_2\left(\theta_{\vec{q}} \right)
\label{vpm}
\end{eqnarray}

\begin{eqnarray}
c_2\left(\theta_{\vec{q}} \right) \equiv
\sqrt{{1 \over 4}\left(\gamma -v_0\right)^2 \cos^2
\left(\theta_{\vec{q}}
\right) + c^2_0
\sin^2 \left(\theta_{\vec{q}} \right)} \quad ,
\label{N+9b}
\end{eqnarray}
and $c_0 \equiv \sqrt{\sigma _1 \rho_0}$.  A
polar plot of this highly anisotropic sound speed is given in figure
2.  We remind the reader that here and hereafter, we only keep
the leading order terms in the long wave length limit, i. e., for
small
$q_{\parallel}$ and $q_{\perp}$.

The linear equations (\ref{vT lin}) - (\ref{rho lin}) are easily solved
for the fields
$\delta \rho$, $\vec{v}_T$ and $v_L$ in terms of the random
forces:
\begin{eqnarray}
\vec{v}_T \left(\vec{q},\omega \right) =
G_{TT}\left(\vec{q},\omega
\right)\vec{f}_T \left(\vec{q},\omega \right)
\label{N+10}
\end{eqnarray}
\begin{eqnarray}
v_L \left(\vec{q},\omega \right) = G_{LL} \left(\vec{q},\omega
\right) f_L\left(\vec{q},\omega
\right) + G_{L\rho}\left(\vec{q},\omega
\right) f_\rho \left(\vec{q},\omega\right)
\label{N+11}
\end{eqnarray}
\begin{eqnarray}
\delta \rho \left(\vec{q},\omega \right) = G_{\rho
L}\left(\vec{q},\omega
\right) f_L\left(\vec{q},\omega
\right) + G_{\rho \rho}\left(\vec{q},\omega
\right) f_\rho\left(\vec{q},\omega
\right)
\label{N+12}
\end{eqnarray}
where the propagators are
\begin{eqnarray}
G_{TT} = {1 \over -i\left(\omega - \gamma q_{\parallel} \right) +
\Gamma_T
\left(\vec{q}\right)}
\label{N+13}
\end{eqnarray}
\begin{eqnarray}
G_{LL} = {i\left(\omega - v_0 q_{\parallel}\right) -
\Gamma_{\rho}\left(\vec{q}\right)
\over
\left(\omega - c_+\left(\theta_{\vec{q}}\right) q
\right)\left(\omega - c_-\left(\theta_{\vec{q}}\right) q \right)+ i\omega
\left(\Gamma_L \left(\vec{q}\right)+
\Gamma_{\rho}\left(\vec{q}\right)\right)
-iq_{\parallel}\left(v_0\Gamma_L \left(\vec{q}\right) + \gamma
\Gamma_{\rho}\left(\vec{q}\right)
\right)}
\label{N+14}
\end{eqnarray}
\begin{eqnarray}
G_{L\rho} =
{i\sigma_1q_\perp \over
\left(\omega - c_+\left(\theta_{\vec{q}}\right) q
\right)\left(\omega - c_-\left(\theta_{\vec{q}}\right) q \right)+ i\omega
\left(\Gamma_L
\left(\vec{q}\right)+
\Gamma_{\rho}\left(\vec{q}\right)\right)  -iq_{\parallel}\left(v_0\Gamma_L
\left(\vec{q}\right)
+ \gamma \Gamma_\rho
\left(\vec{q}\right)
\right)}
\label{N+15}
\end{eqnarray}
\begin{eqnarray}
G_{\rho L} = {i\rho_0q_\perp \over \left(\omega -
c_+\left(\theta_{\vec{q}}\right) q
\right)\left(\omega - c_-\left(\theta_{\vec{q}}\right) q \right)+ i\omega
\left(\Gamma_L \left(\vec{q}\right)+
\Gamma_{\rho}\left(\vec{q}\right)\right) -iq_{\parallel}\left(v_0\Gamma_L
\left(\vec{q}\right)+
\gamma \Gamma_\rho
\left(\vec{q}\right)
\right)}
\label{N+16}
\end{eqnarray}
\begin{eqnarray}
G_{\rho \rho} = {i\left(\omega-\gamma q_\parallel
\right) - \Gamma _L
\left(\vec{q}\right)
\over
\left(\omega - c_+\left(\theta_{\vec{q}}\right) q
\right)\left(\omega - c_-\left(\theta_{\vec{q}}\right) q \right)+ i\omega
\left(\Gamma_L \left(\vec{q}\right)+
\Gamma_{\rho}\left(\vec{q}\right)\right) -iq_{\parallel}\left(v_0\Gamma_L
\left(\vec{q}\right)
+
\gamma \Gamma_\rho
\left(\vec{q}\right)
\right)}
\label{N+17}
\end{eqnarray}

In writing the definitions of the propagators
(\ref{N+10})-(\ref{N+12}), we have introduced a fictitious force
$f_\rho$ in the $\rho$ equation of motion (\ref{rho lin}).  Of
course, this force is, in fact, zero; but the propagators
$G_{\rho\rho}$ and
$G_{L \rho}$ nonetheless prove useful in the perturbative
treatment of the non-linear corrections to this linear theory, so we
have included $f_{\rho}$ here.

Given the expressions (\ref{N+10})-(\ref{N+17}) for the velocity
and density in terms of the random force $\vec{f}$, and the
autocorrelation (\ref{white noise}) of that random force, it is
straightforward to calculate the correlations of the densities and
velocities.  We find:
\begin{eqnarray}
C_{ij}\left(\vec{q},\omega
\right) &\equiv& \left<v^\perp_i\left(-\vec{q},-\omega
\right) \, v^\perp_j \left(\vec{q},\omega\right) \right> \nonumber
\\  &=& G_{TT}\left(\vec{q},\omega\right)
G_{TT}\left(-\vec{q},-\omega\right)\left<f_{T_i}\left(\vec{q},\omega
\right)f_{T_j}\left(-\vec{q},-\omega\right)\right>\nonumber \\
 &&+ \, G_{LL}\left(\vec{q},\omega\right)
G_{LL}\left(-\vec{q},-\omega\right){q^\perp_i  q^\perp_j\over
q^2_\perp}\left<f_L\left(\vec{q},\omega\right)f_L\left(-\vec{q},-\omega\right)
\right>\nonumber \\  &\equiv& C_{TT}\left(\vec{q},\omega\right)
P^\perp_{ij}\left(\vec{q}\right) + C_{LL}\left(\vec{q},\omega\right)
L^\perp_{ij}\left(\vec{q}\right)
\label{N+18}
\end{eqnarray}
where
\begin{eqnarray}
L^\perp_{ij}\left(\vec{q}\right) \equiv
{q^\perp_i  q^\perp_j\over q^2_\perp}
\label{N+19}
\end{eqnarray}
\begin{eqnarray}
P^\perp_{ij}\left(\vec{q}\right) \equiv
\delta^\perp_{ij} - L^\perp_{ij}\left(\vec{q}\right)
\label{N+20}
\end{eqnarray}
are longitudinal and transverse projection
operators that project any vector perpendicular to {\it both} the
flock motion
{\it and}
$\vec{q}_{\perp}$,
\begin{eqnarray}
C_{TT}\left(\vec{q},\omega\right) = {\Delta \over \left(\omega -
\gamma q_{\parallel} \right)^2 + \Gamma^2_T
\left(\vec{q}\right)}
\label{N+21}
\end{eqnarray}
and
\begin{eqnarray}
C_{LL}
\left(\vec{q},\omega\right) =
{\Delta
\left(\omega - v_0q_{\parallel}\right)^2
\over
\left(\omega-c_+\left(\theta_{\vec{q}}\right)q\right)^2 \,
\left(\omega-c_-\left(\theta_{\vec{q}}\right)q\right)^2\, +
\left(\omega\left(\Gamma_L
\left(\vec{q}\right) +
\Gamma_\rho
\left(\vec{q}\right)\right) -
q_{\parallel}
\left(v_0 \Gamma_L
\left(\vec{q}\right) +
\gamma\Gamma_{\rho}
\left(\vec{q}\right) \right)\right)^2}
\label{N+22}
\end{eqnarray}

The transverse and longitudinal correlation functions Eqn.
(\ref{N+21}) and (\ref{N+22}) are plotted as functions of
$\omega$ for fixed $\vec{q}$ in figure 6.  Note that they have
weight in entirely different regions of frequency:  $C_{TT}$ is
peaked at
$\omega = \gamma q_{\parallel}$, while
$C_{LL}$ has two peaks, at $\omega =
c_{\pm}\left(\theta_{\vec{q}}
\right)q$.  Since all three peaks have widths of order $q^2$, there
is little overlap between the transverse and the longitudinal peaks
as
$\left|\vec{q}\right| \rightarrow 0$.

The density-density correlation function
\begin{eqnarray}
 C_{\rho \rho}  = {\Delta \rho_0^2 q^2_\perp
\over
\left(\omega-c_+ \left(\theta_{\vec{q}}\right)q
\right)^2 \, \left(\omega-c_-\left(\theta_{\vec{q}}\right)q
\right)^2\, +
\left(\omega\left(\Gamma_L\left(\vec{q}\right) +
\Gamma_{\rho}\left(\vec{q}\right)\right) -
 q_{\parallel} \left(v_0 \Gamma_L\left(\vec{q}\right) +
\gamma\Gamma_{\rho}\left(\vec{q}\right)\right)\right)^2}
\label{N+23}
\end{eqnarray}
looks almost identical to $C_{LL}$, especially when
one notes that near the frequencies $\omega =
c_{\pm}\left(\theta_{\vec{q}}
\right)q$ where
both peak, the numerator of (\ref{N+22}) $\left(=
\left(c_{\pm}\left(\theta_{\vec{q}}\right)q - v_0q_{\parallel}
\right)^2\right)$, differs from
that of (\ref{N+23}) only by a $\left| \vec{q} \right|$ independent
factor of $\left( c_{\pm}\left(\theta_{\vec{q}}\right)q -
v_0q_{\parallel}
\right)^2/\rho_0\sigma
_1q^2_{\perp}$.

Given these Fourier transformed correlation functions, it is
straightforward, and instructive, to Fourier transform back to real
time.  In particular, it is simple to calculate the spatially Fourier
transformed equal time velocity correlation function:
\begin{eqnarray}
\left< v_i\left(\vec{q},t\right)v_j\left(-\vec{q},t\right)\right>
&=&P^\perp_{ij}\left(\vec{q}\right)\int^\infty_{-\infty}{d\omega
\over 2\pi} C_{TT}\left(\vec{q},\omega \right) +
L^\perp_{ij}\left(\vec{q}\right) \int^\infty_{-\infty}{d\omega
\over 2\pi} C_{LL}\left(\vec{q},\omega \right) \nonumber \\
&=&{\Delta
\over 2} \left[{P^\perp_{ij}\left(\vec{q}\right) \over
\Gamma_T\left(\vec{q}\right)} + \phi \left(\hat{q}\right)
{L^\perp_{ij}\left(\vec{q}\right)
\over
\Gamma_L\left(\vec{q}\right)}\right] \propto {1 \over q^2}
\label{N+24}
\end{eqnarray}
where the second integral over frequency has
been evaluated in the limit of $\left| \vec{q} \right| \rightarrow
0$, so that
$c \left(\theta_{\vec{q}}\right)q  \gg \Gamma_L \propto q^2$, and
the factor $\phi \left(\hat{q}\right)$ depends
{\it only} on the direction $\hat{q}$ of $\vec{q}$,
{\it not}  its magnitude, and is given by the sadly
complicated expression
\begin{eqnarray}
\phi\left(\hat{q}\right) &\equiv& {1 \over c_2
\left(\theta_{\vec{q}}
\right)q}\left[{\left(c_+\left(\theta_{\vec{q}}\right) q-v_0
q_{\parallel} \right)^2
\over c_+\left(\theta_{\vec{q}}\right) q-v_0
q_{\parallel} + \left(c_+\left(\theta_{\vec{q}}\right)
q-\gamma q_{\parallel} \right){\Gamma_{\rho}
\over \Gamma_L}}\right.\nonumber\\
&&+\left.{\left(c_-\left(\theta_{\vec{q}}\right) q-v_0
q_{\parallel} \right)^2
\over c_-\left(\theta_{\vec{q}}\right) q-v_0
q_{\parallel} + \left(c_-\left(\theta_{\vec{q}}\right)
q-\gamma q_{\parallel} \right){\Gamma_{\rho}
\over \Gamma_L}}\right]\nonumber\\
&\equiv&  {1 \over F\left(\vec{q}, \kappa, \gamma \right)}
\left[{A^2_+\left(\vec{q}, \kappa, \gamma \right)
\over A_+\left(\vec{q}, \kappa, \gamma \right) -
A_-\left(\vec{q}, \kappa, \gamma \right)
{\Gamma_{\rho}\left(\vec{q}
\right)\over \Gamma_L\left(\vec{q}
\right)}}\right. \nonumber\\
&&+ \left.{A^2_-\left(\vec{q}, \kappa, \gamma \right)
\over A_-\left(\vec{q}, \kappa, \gamma \right) -
A_+\left(\vec{q}, \kappa, \gamma \right)
{\Gamma_{\rho}\left(\vec{q}
\right)\over \Gamma_L\left(\vec{q}
\right)}}
\right]
\label{phi def}
\end{eqnarray}
where we've defined
\begin{eqnarray}
F \left(\vec{q}; \kappa, \gamma \right) \equiv
\sqrt{\left({\gamma - v_0
\over 2 v_0}
\right)^2 \kappa ^2 \left({q_{\parallel}
\over q_{\perp}}\right)^2 + 1} \, ,
\label{F def}
\end{eqnarray}
\begin{eqnarray}
A_\pm \left(\vec{q}; \kappa, \gamma \right) \equiv
\pm F\left(\vec{q}; \kappa, \gamma \right) + \left({\gamma -
v_0
\over 2 v_0}
\right) \kappa  {q_{\parallel}
\over q_{\perp}}  \, ,
\label{A def}
\end{eqnarray}
and
\begin{eqnarray}
\kappa \equiv {v_0 \over \sqrt{\sigma_1 \rho_0}} .
\label{kappa def}
\end{eqnarray}
The second equality in (\ref{phi def}) is obtained from the first
simply by cancelling common factors of $\sigma_1
\rho_0q^2_{\perp}$ out of the numerator and denominator of
various terms.

Note, and this will prove to be crucial later, that
$\phi\left(\hat{q}
\right)$ depends {\it only} on $\hat{q}$, diffusion
constants, and the dimensionless {\it ratios} $\kappa$ and $\gamma/v_0$.
This last fact
is essential for our renormalization group scaling analysis, as we
will show later.

The ${1  \over  q^2}$ divergence of (\ref{N+24}) as
$|\vec{q}|
\rightarrow 0 $ reflects the enormous long wavelength
fluctuations in this system.

These fluctuations predicted by the linearized theory are strong
enough to destroy long ranged order in $d \leq 2$.  To see this,
calculate the mean squared fluctuations in
$\vec{v}_\perp\left(\vec{r},t\right)$ at a given point $\vec{r}$,
and time $t$.  This is simply the integral of the trace of
Eqn (\ref{N+24}) over all $\vec{q}$:
\begin{eqnarray}
\left< \left| \vec{v}_\perp \left(\vec{r},t\right)\right|^2\right> &=&
\int {d^dq  \over  (2\pi)^d} \left< v_i \left(\vec{q}, t \right)
v_i
\left(-\vec{q}, -t \right)\right> \nonumber\\ &=&{\Delta
\over 2}
\left[\int {d^dq  \over  \left(2\pi
\right)^d} \left( {(d -2)
\over D_Tq^2_\perp + D_\parallel q^2_\parallel} +
{\phi\left(\hat{q}\right)
\over  D_Lq^2_\perp + D_\parallel q^2_\parallel}
\right)\right]
\label{v real lin}
\end{eqnarray}
The last integral clearly diverges in the infrared
$\left(\left|\vec{q}
\right| \rightarrow 0\right)$ for $d \leq 2$.  The divergence in the
ultraviolet $\left(\left|\vec{q} \right| \rightarrow \infty \right)$
for $d \geq 2$ is not a concern, since we don't expect our theory to
apply for $\left|\vec{q} \right|$ larger than the inverse of a
microscopic length (such as the interaction range $\ell_0$).
Presumably, at larger wavenumbers, the correlation function falls
off fast enough that the wavevector integral in Eqn (\ref{v real
lin}) converges in the ultraviolet.

Indeed, we will in subsequent calculations mimic the effect of this
putative more rapid decay of correlations as $\left|\vec{q} \right|
\rightarrow \infty$ with a sharp ultraviolet cutoff.   We will
restrict integrals over wavevectors to  hypercylindrical shell with
long ({\it very} long!) axis along the direction of flock
motion $x_\parallel$:
\begin{eqnarray}
\left|\vec{q}_{\perp} \right|< \Lambda , \quad - \infty \leq
q_{\parallel}
\leq
\infty
\label{N+26}
\end{eqnarray} with the ultraviolet cutoff $\Lambda$ of order the
inverse of a microscopic length (e.~g., $\ell_0$).

Obviously, this is quite an arbitrary choice of ultraviolet cutoff, and
any result that depends on the precise form of this cutoff will not
be accurately calculated by this prescription.  However,
{\it universal}, long wavelength properties of the flock
should be unaffected by the precise choice of cutoff, and it is on
those properties that we will focus our attention.

The infra-red divergence in Eqn (\ref{v real lin}) for $d \leq 2$
cannot be dismissed so easily, since our hydrodynamic theory
should get better as  $\left|\vec{q} \right|\rightarrow 0$.  Indeed,
in the absence of non-linear effects, this divergence is real, and
signifies the destruction of long ranged order in the linearized
model by fluctuations, even for arbitrarily small noise $\Delta$, in
spatial dimensions $d \leq 2$, and in particular {\it in} $d = 2$,
where the integral in Eqn (\ref{N+24}) diverges logarithmically in
the infra-red.  This is so since, if $\left<\left|\vec{v}_\perp\right|^2
\right>$ is arbitrarily large even for arbitrarily small $\Delta$, our
original assumption that $\vec{v}$ can be written as a mean value
$\left<\vec{v} \right>$ plus a {\it small} fluctuation
$\vec{v}_\perp$ is clearly mistaken; indeed, the divergence of
$\vec{v}_\perp$ suggests that the velocity can swing through
{\it all} possible directions, implying that $\left<\vec{v}
\right> = 0$ for $d \leq 2$.

In $d = 2$, this result is very reminiscent of the familiar
Mermin-Wagner-Hohenberg (MWH) theorem\cite{MW}, which
states that in equilibrium, a spontaneously broken continuous
symmetry is impossible in $d= 2$ spatial dimensions, precisely
because of the type of logarithmic divergence of fluctuations that
we have just found here.

In the next section, we will show that this prediction is invalidated
by non-linear effects, and, in fact, much of the scaling of
correlation functions and propagators is changed from that
predicted by the linearized theory in spatial dimensions $d \leq 4$.




\section{Non-linear Effects and Breakdown of Linear Hydrodynamics in
the Broken Symmetry State}

\subsection{Scaling analysis}

In this section we analyze the effect of the non-linearities in equations
(\ref{EOM broken}) and (\ref{cons broken}) on the  long length and
time behavior of the system, for spatial dimensions $d<4$. We will
rescale lengths, time, and the fields
$\vec{v}_{\perp}$ and $\delta\rho$ according to
\begin{eqnarray}
\vec{x}_{\perp}&\rightarrow& b\vec{x}_{\perp}\nonumber\\
x_{\parallel}&\rightarrow& b^{\zeta}x_{\parallel}\nonumber\\
t&\rightarrow& b^{z}t\nonumber\\
\vec{v}_{\perp}&\rightarrow& b^{\chi}\vec{v}_{\perp}\nonumber\\
\delta\rho&\rightarrow& b^{\chi_\rho}\delta\rho
\label{field rescale}
\end{eqnarray}
choosing the scaling exponents to keep the diffusion
constants
$D_{B, T, \rho, \parallel}$, and the strength $\Delta$
of the noise fixed. The reason for choosing to keep these particular
parameters fixed rather than, e.g., $\sigma_1$, is that these
parameters completely determine the size of the equal time
fluctuations in the linearized theory, as can be seen from Eqn.
(\ref{v real lin}).   Under the rescalings (\ref{field rescale}), the
diffusion constants
rescale according to $D_{B, T} \rightarrow b^{z-2} D_{B, T}$
and $D_{\rho, \parallel} \rightarrow b^{z-2 \zeta} D_{\rho, \parallel}$;
hence, to keep them fixed, we must choose $z = 2$ and $\zeta = 1$. The
rescaling of the random force $\vec{f}$ can then be obtained from the
form of the
$f-f$ correlations Eqn. (\ref{white noise}) and is, for this choice of $z$ and
$\zeta$:
\begin{eqnarray}
\vec{f}\rightarrow b^{-1-d/2}\vec{f}
\label{force rescale}
\end{eqnarray}
To maintain the balance between $\vec{f}$ and the
linear terms in $\vec{v}_{\perp}$ in eq. (\ref{EOM broken}), we must
rescale the
velocity field according to:
\begin{eqnarray}
\vec{v}_{\perp}\rightarrow b^{\chi}\vec{v}_{\perp}
\label{}
\end{eqnarray}
with
\begin{eqnarray}
\chi=1-d/2
\label{}
\end{eqnarray}
which is the roughness exponent for the linearized
model. That is, we expect $\vec{v}_{\perp}$ fluctuations on length scale
$L$ to scale like $L^\chi$. Therefore, the linearized hydrodynamic
equations, neglecting the nonlinear convective term and the
non-linearities in the pressure, imply that
$\vec{v}_{\perp}$ fluctuations grow without bound (like
$L^\chi$) as
$L\rightarrow\infty$ for $d\le 2$, where the above expression for
$\chi$ becomes positive. Thus, this linearized theory predicts the loss of
long range order in $d \le 2$, as we saw in Section IV by explicitly
evaluating the real space fluctuations.

Making the rescalings as described in Eqns. (\ref{field rescale}), the
equation of
motion (\ref{EOM broken}) becomes:
\begin{eqnarray}
\partial_{t} \vec{v}_{\perp} &+& b^{\gamma_v}\gamma\partial
_{\parallel}\vec{v_{\perp}} +
b^{\gamma_\lambda} \left[\lambda_1
 \left( \vec{v}_{\perp}  \cdot
\vec{\nabla}_{\perp}\right)
\vec{v}_{\perp} + \lambda_2 \left(\vec{\nabla}_{\perp} \cdot
\vec{v}_{\perp}\right)
\vec{v}_{\perp}\right]  \nonumber \\ &=& - \vec{\nabla}_{\perp}
 \left(\sum_{n=1}^\infty b^{\gamma_n}
\sigma_n (\delta \rho )^n\right)+D_B \vec{\nabla}_{\perp}
(\vec{\nabla}_{\perp}\cdot \vec{v}_{\perp})
+D_{T}\nabla^{2}_{\perp}\vec{v}_{\perp}+D_{||}\partial^{2}_{
||}\vec{v}_{\perp}+\vec{f}_{\perp}
\label{}
\end{eqnarray}
with
\begin{eqnarray}
\gamma_\lambda = \chi + 1 = 2-d/2
\label{}
\end{eqnarray}
\begin{eqnarray}
\gamma_v = z - \zeta = 1
\label{}
\end{eqnarray}
and
\begin{eqnarray}
\gamma_n = z-\chi + n\chi -1 = n + (1-n) {d\over 2} \quad .
\label{}
\end{eqnarray}
The scaling exponent $\chi_\rho$ for $\delta\rho$ is
given by $\chi_\rho =
\chi$, since the density fluctuations $\delta\rho$ are comparable in
magnitude to the $\vec{v}_{\perp}$ fluctuations. To see this, note that
the eigenmode of the linearized equations of motion that involves
$\delta\rho$ is a sound mode, with dispersion relation
$\omega = c_{\pm}\left(\theta_{\vec{q}}\right)q$. Inserting this into the
Fourier transform of the continuity equation (\ref{cons broken}), we see that
$\delta\rho \sim {\vec{q}_{\perp}\cdot
\vec{v}_{\perp}\over q_{\perp}}$. The magnitude of $\vec{q}_{\perp}$
drops out of the right hand side of this expression; hence $\delta\rho$
scales like $\left| \vec{v}_{\perp} \right|$ at long distances. Therefore, we
will choose $\chi_\rho=\chi=1 -{d \over 2}$.

The first two of these scaling exponents for the non-linearities to become
positive as the spatial dimension $d$ is decreased are
$\gamma_\lambda$ and $\gamma_2$, which both become positive for
$d<4$, indicating that the $\lambda_1\left(\vec{v}_{\perp} \cdot
\vec{\nabla}\right)\vec{v}_{\perp}, \lambda_2
\left(\vec{\nabla}_{\perp}
\cdot  \vec{v}_{\perp} \right) \vec{v}_{\perp}$ and $\sigma_2
\vec{\nabla}_{\perp} (\delta\rho^2)$ non-linearities are all relevant
perturbations for $d<4$.  So, for $d<4$, the linearized hydrodynamics will
break down.

What can we say about the behavior of Eq. (\ref{EOM broken}) and
(\ref{cons broken}) for $d<4$, when the linearized hydrodynamics no
longer holds? The standard approach for such problems is the
dynamical renormalization group. In most cases, this approach is only
practical near the upper critical dimension (in our case $d_{c} = 4$),
and yields the anomalous exponents in an expansion in $\epsilon =
d_c -d$.   This approach will obviously not be of much use in our
problem in $d = 2$, where the ostensibly small parameter in this
expansion $\epsilon = 2$.  We will nonetheless undertake this
approach in subsection B, and show that, for unfortunate technical
reasons, we learn little even near $d = 4$.  Fortunately, as we show in
subsection C, because of the various symmetries in Eq. (\ref{EOM
broken}), we can obtain the {\it exact} scaling exponents in
$d=2$.

\subsection{Renormalization Group Analysis, $d < 4$}

In this subsection, we  analyze the effect of the relevant
non-linearities $\lambda_1, \lambda_2$, and
$\sigma_2$ on the broken symmetry state in spatial dimensions
$d < 4 $.

Our tool is the dynamical renormalization group (for
details, see, e.g., the excellent description in Forster, Nelson and Stephen
\cite{FNS}). We will summarize the essential features of this
procedure here; readers interested in details are referred to \cite{FNS}.

We proceed through the iteration of the following 3 steps:
\begin{enumerate}
\item We separate the fields $\vec{v}_{\perp}$ and $\delta\rho$, and the
random forces
$\vec{f}$ into short and long wavelength components, according to:
\begin{eqnarray}
\vec{v}_{\perp >}\left(\vec{r}, t\right) = \int_> {d^d q d\omega \over (2
\pi)^d}\,  \vec{v}_{\perp}\left(\vec{q}, \omega \right) e^{i\left(\vec{q}
\cdot
\vec{r} -\omega t \right)}
\label{}
\end{eqnarray}
\begin{eqnarray}
\vec{v}_{\perp <} \left(\vec{r}, t\right) = \int_< {d^dq d\omega \over
\left(2
\pi\right)^d} \,
\vec{v}_{\perp}\left(\vec {q}, \omega \right) e^{i\left(\vec {q} \cdot
\vec{r} -\omega t \right)}
\end{eqnarray}
where $\int_>$ denotes a wavevector integral
restricted to a hypercylindrical shell $b^{-1} \Lambda < \left|
\vec{q}_{\perp}
\right| <
\Lambda$, where $\Lambda$ is an ultraviolet cutoff, and $\int_<$
likewise denotes an integral over the interior of this shell:
$\left| \vec{q}_{\perp} \right| < b^{-1} \Lambda$.  $\delta\rho$ and
$\vec{f}_\perp$ are likewise separated.

\item Average the EOM over the short wavelength fields
$\vec{v}_{\perp >}$, $\delta\rho_>$, and $\vec{f}_>$ to get new, effective
EOM for the long-wavelength fields $\vec{v}_{\perp <}$ and
$\delta\rho_<$,  with ``intermediate'' renormalized parameters
$D^I_{\perp}$, etc. This average is performed perturbatively in the
non-linearities in the EOM. The perturbation theory can be represented
graphically; the interested reader is referred to the previously
mentioned \cite{FNS} for further details on the mechanics of
this.

\item We now rescale the time, space, and the fields in the EOM
according to (\ref{field rescale}) in order to restore the original
ultraviolet cutoff
$\Lambda$ of the problem. We will choose rescaling exponents $z$,
$\zeta$ and $\chi$ to produce fixed points.
\end{enumerate}

Of course, the exponents are, in fact, completely arbitrary.  We need not
choose them to produce fixed points.  However, it is very convenient to
do so, since, as we will show in more detail later, the values of $z$,
$\zeta$, and $\chi$ that {\em do} produce fixed points are exactly the
values of the physical observable time, anisotropy, and roughness
exponents that characterize the scaling properties of various correlation
functions.

Performing this RG procedure, we find the following recursion relations:
\begin{eqnarray}
{dD_{B,T}  \over  d\ell} = \left(z - 2 +
G^{D}_{B,T}(g)
\right)D_{B,T}
\label{Rec DBT}
\end{eqnarray}
\begin{eqnarray}
{dD_{\parallel,\rho}  \over  d\ell} = \left(z-2\zeta +
G_{\parallel, \rho}
\left(\{g_i \}
\right)\right)D_{\parallel, \rho}
\label{Rec D parallel}
\end{eqnarray}
\begin{eqnarray}
{d\sigma_n  \over  d\ell} = \left( z + (n - 1) \chi - 1 +
G^\sigma_n\left( \{g_i \} \right) \right) \sigma_n
\label{Rec sigma}
\end{eqnarray}
\begin{eqnarray} {d\rho_0  \over  d\ell} = \left(z-1 \right) \rho_0
\label{Rec rho}
\end{eqnarray}
\begin{eqnarray}
{d\lambda_{1,2,\rho}  \over  d\ell} = \left(\chi-1+z+
G^\lambda_{1,2,\rho} \left( \{g_i \} \right)
\right) \lambda_{1,2,\rho}
\label{Rec lambda}
\end{eqnarray}
\begin{eqnarray}
{d\Delta  \over  d\ell} = \left(z - \zeta - 2 \chi + 1 - d
+ G_\Delta  \left( \{g_i \} \right)
\right) \Delta
\label{Rec delta}
\end{eqnarray}
\begin{eqnarray}
{dv_0 \over d\ell} = \left(z - \zeta \right) v_0
\label{v0 rec}
\end{eqnarray}
\begin{eqnarray}
{d\gamma \over d\ell} = \left(z - \zeta \right) \gamma
\label{gamma rec}
\end{eqnarray}

where we've taken $b = 1 + d\ell$, $d\ell \ll 1$, to
obtain differential recursion relations, the $G$'s represent graphical (i.\
e., perturbative) corrections, and the $\{g_i \}$'s are a set of
dimensionless coupling constants involving ratios of powers of the
dynamical parameters.
We have also dropped ``irrelevant''
terms in these recursion relations.  The coupling constant
$\lambda_\rho$ is the coefficient of the $\vec{\nabla}_{\perp} \cdot
\left(\vec{v}_{\perp}\delta \rho
\right)$ non-linearity in the $\rho$ equation of motion.  This
coupling constant is equal to
1, and must, up to trivial rescaling corrections, remain equal to 1 upon
renormalization.  This is a simple consequence of the fact that mass
conservation is {\em exact}; that is, the equation of motion for $\rho$
{\em must} remain the simple continuity equation $\partial_{t}\rho +
\vec{\nabla} \cdot \left(\rho \vec{v} \right) = 0$, except for the trivial
changes introduced by rescaling.  This implies that $G^\lambda_\rho
\left( \{g_i \} \right)  = 0$, {\em exactly}, for {\em all}
$\{g_i \}$.  We will later use this fact to obtain exact
values for the scaling exponents $\chi$, $z$ and $\zeta$ in $d = 2$.

It is worth noting that $\gamma$ is treated as an independent variable
here, only its bare value is related to the bare values of $\lambda_1$
and $v_0$ through eq. (\ref{gamma def}).

Although there is no {\it symmetry} argument forbidding renormalization
of $v_0$ and $\gamma$, simple power counting shows that there are no {\it
relevant}
graphical corrections to them; this is why no graphical corrections appear
in (\ref{v0 rec}), (\ref{gamma rec}).

As mentioned earlier, the rescaling exponents $\chi$, $z$ and $\zeta$
are arbitrary.  We chose them to produce fixed points only for
computational convenience.  However, there is {\em no} choice of
rescaling exponents that will keep {\em all} the parameters fixed.  For
instance, to keep $D_{B,T}$ and $D_{\parallel, \rho}$ fixed, we will have
to choose $z > 1$.   However, with this choice of $z$, the relations
(\ref{Rec sigma}) and (\ref{Rec rho}) show that $\sigma _1$ and
$\rho_0$ flow to infinity.

Which of the parameters, then, should we choose keep
fixed?  That is, which is most {\em convenient} to keep fixed?  The
answer to this question is provided by the renormalization group
matching formalism.  This approach enables one to use the
renormalization group to relate correlation functions in the original,
unrenormalized model at long distances and large times to the same
correlation functions in the renormalized system at shorter distances and
times.  The advantage of this approach is that long distance, large time
correlation functions are hard to calculate in $d < 4$, since, as we showed
from our earlier scaling arguments (and can also verify from the
renormalization group recursion relations (\ref{Rec DBT})-(\ref{Rec D
parallel})), these are {\em not} accurately calculable from the harmonic
theory developed in section IV, since the non-linearities
$\lambda_{1,2}$, $\lambda_\rho$, and
$\sigma_2$ have very large effects at long distances.  By mapping these
correlation functions onto those at {\em short} distances in the
renormalized equations of motion, we circumvent this problem.  Clearly,
there is a caveat here:  even at short distances, the correlation functions
in the renormalized model can {\em only} be calculated accurately in
the harmonic theory {\em if} the non-linear couplings in that
renormalized model are not too big.  This suggests that the {\em
convenient} choice of the rescaling exponents $\chi$, $z$ and $\zeta$ is
that which keeps the non-linearities $\lambda_{1,2}$, $\lambda_\rho$,
and
$\sigma_2$ fixed.

Let us illustrate these considerations explicitly for one very important
correlation function:  the equal time spatially fourier transformed
velocity-velocity autocorrelation function:
\begin{eqnarray}
C_{ij} \left(\vec{q}\right) \equiv \left< v_i \left(\vec{q}, t\right)v_j
\left(-\vec{q}, t\right)\right>
\label{RG match 1}
\end{eqnarray} This particular correlation function is important
because it gives us our best measure of the size of the velocity
fluctuations, and will ultimately determine whether or not these
fluctuations destroy the long ranged orientational order of the flock
(thereby driving its mean velocity to zero).

$C_{ij} \left(\vec{q}\right)$ is, of course, a function of the flock dynamical
parameters $D_{B,T}$, $D_{\parallel, \rho}$, $\Delta$, etc., as well as of
$\vec{q}$.  Furthermore, at small $\vec{q}$, it is difficult to calculate in
spatial dimension $d < 4$ due to the non-linear terms, for the reasons
discussed above.  So let's follow this renormalization group matching
procedure to relate $C_{ij}\left(\vec{q}; \{B^0_i\}\right) $ where
$\{B^0_i\}$ denotes the set of dynamical parameters $D^0_{B,T}$,
$D^0_{\parallel}$, $\Delta_0$, etc.  in the unrenormalized model, to the
same correlation function in the renormalized model, a
renormalization group time $\ell$ later:
\begin{eqnarray}
C_{ij} \left(\vec{q}_{\perp}, q_{\parallel}; \{B^0_i\} \right)
= b^{\left(2\chi +
\zeta + d -1\right) \ell}\,  C_{ij}\left(e^{\ell} \vec{q}_{\perp}, e^{\zeta
\ell}q_{\parallel}; \{B_i(\ell)\}\right)
\label{RG match 2}
\end{eqnarray}
where the $\{B_i(\ell)\}$ denote the renormalized
parameters.

In the discussion that follows, we will first consider the case
$\left({q_{\parallel}\over
\Lambda}\right)\ll \left({q_{\perp}\over{\Lambda}}\right)^\zeta$ . At the
conclusion of the
discussion of this special case, we will briefly indicate how the general case
can be treated to obtain the scaling laws quoted in the introduction. For
the case $\left({q_\parallel\over
\Lambda}\right)\ll \left({q_\perp\over\Lambda}\right)^\zeta$, we will
choose $\ell =
\ell_{\ast}\left(\vec{q}_{\perp}\right) = ln\left({\Lambda
\over q_{\perp}}\right)$, where $\Lambda$ is the ultraviolet cutoff, on
the right hand side, and obtain
\begin{eqnarray}
C_{ij} \left(\vec{q}_{\perp}, q_{\parallel}; \{B^0_i\}\right)
= \left({\Lambda
\over q_{\perp}} \right)^{2\chi + \zeta + d - 1} \, C_{ij} \left(\Lambda, {
q_{\parallel} \over  \left({q_{\perp}  \over  \Lambda}\right)^\zeta};
\{B_i\left(
\ell_{\ast}\left(\vec{q}_{\perp}\right)\right) \}
\right)
\label{RG match 3}
\end{eqnarray}
Now, {\em if} the original $q_{\perp}$ was small $(\ll
\Lambda)$ {\em and} we have chosen the rescaling exponents $\chi,
\zeta$, and $z$ so that the non-linearities $\lambda_1(\ell)$,
$\lambda_2(\ell)$, $\lambda_\rho(\ell)$ and $\sigma_2(\ell)$ on the
right hand side of (\ref{RG match 3}) flow, as $\ell \rightarrow \infty$, to
$O(1)$ fixed point values $\left(\lambda^{\ast}_{1,2,
\rho},
\sigma^{\ast}_2\right)$, then
$\lambda_1\left(\ell_{\ast}\right)$,
$\lambda_2\left(\ell_{\ast}\right)$,
$\lambda_\rho\left(\ell_{\ast}\right)$ and
$\sigma_2\left(\ell_{\ast}\right)$ can be replaced by
those fixed point values, since
$\ell_{\ast}$ will be large.  Because those fixed point
values are, by assumption, $O(1)$, then, up to $O(1)$
correction factors coming from these non-linearities, the
right hand side of (\ref{RG match 3}) can be evaluated in
the harmonic approximation Eqns. (\ref{N+18}), (\ref{N+21}), (\ref{N+22}).
(The correction
factors are only of $O(1)$ - i. e., not divergent - because
the right hand side of (\ref{RG match 3}) is evaluated at
large
$\vec{q} \, \left(\left|\vec{q}_{\perp}\right| = \Lambda
\right)$, where the infrared divergences associated with the strong
relevance of the non-linearities do not matter.  It is precisely {\em
because} of those infrared divergences that we could {\em not} evaluate
the left hand side of (\ref{RG match 3}) directly, but rather were forced
to go through this seemingly circuitous $R G$ matching formalism).
Making that harmonic approximation on the right hand side, we obtain:
\begin{eqnarray}
C_{ij}\left(\Lambda \, , \, {q_{\parallel} \over  \left({q_{\perp}
\over
\Lambda}\right)^\zeta};  \{\Gamma \left(\ell _{\ast}
\right) \}\right) = {\Delta_{\ast} \over  D^{\ast}_T
\Lambda^2 + D^{\ast}_{\parallel}\left({q_{\parallel} \over
\left({q_{\perp} \over
\Lambda}\right)^\zeta}\right)^2} \,  P^{\perp}_{ij}\left(\hat{q}_{\perp}
\right) \nonumber\\
+\, {\Delta_{\ast}
\over  D^{\ast}_L \Lambda^2 +
D^{\ast}_{\parallel}\left({q_{\parallel}
\over  \left({q_{\perp}  \over \Lambda}\right)^\zeta}\right)^2} \, \phi
\left(\Lambda, {q_{\parallel}
\over  \left({q_{\perp}
\over
\Lambda}\right)^\zeta} ;   \{B_i \left(\ell_{\ast}\right)
\}\right) L^{\perp}_{ij}\left(\hat{q}_{\perp}
\right)
\label{RG match 4}
\end{eqnarray}
where we have used the fact that we've chosen the
scaling exponents to make $\Delta$  and all the diffusion coefficients
$\{D_i\}$ flow to fixed points
$\Delta^{\ast}$, $\{D^{\ast}_i\}$.  We wish to show that
this expression depends on $\vec{q}$ \, {\em only}
through the scaling ratio $u
\equiv \left({q_{\parallel} \over  \left({q_{\perp} \over
\Lambda}\right)^\zeta}\right)$.  The first (transverse) term in (\ref{RG
match 4}) explicitly has this property.  The second (longitudinal term)
would also, except for the $\phi$ factor, to which we now turn.  From
Eqn. (\ref{phi def}) for $\phi$, we see that to calculate this factor we must
calculate
\begin{eqnarray}
F\left(\Lambda, {q_{\parallel} \over \left({q_{\perp} \over
\Lambda}\right)^\zeta}; \{B_i \left(\ell _{\ast} \right) \}\right) =
\sqrt{\left({\gamma \left(\ell_{\ast}\right) -
v_0 \left(\ell_{\ast}\right) \over 2 v_0 \left(\ell_{\ast}\right)}
\right)^2 \kappa^2
\left(\ell _{\ast}\right)
\left({q_{\parallel} \over \Lambda \left({q_{\perp} \over
\Lambda}\right)^\zeta} \right)^2 + 1} \quad ,
\label{F match}
\end{eqnarray}

\begin{eqnarray}
A_{\pm} \left(\Lambda, {q_{\parallel} \over \left({q_{\perp} \over
\Lambda}\right)^\zeta}; \{B_i \left(\ell _{\ast} \right) \}\right) &=&
\pm F\left(\Lambda, {q_{\parallel} \over \left({q_{\perp} \over
\Lambda}\right)^\zeta}; \{B_i \left(\ell _{\ast} \right) \}\right)
\nonumber\\
&-&{\left(\gamma \left(\ell_{\ast}\right) -
v_0 \left(\ell_{\ast}\right) \right)\over 2v_0 \left(\ell _{\ast}
\right)}\kappa
\left(\ell _{\ast}\right) \left({q_{\parallel} \over \Lambda
\left({q_{\perp} \over
\Lambda}\right)^\zeta } \right)
\label{A match}
\end{eqnarray}
and
\begin{eqnarray}
B_{\pm} \left(\Lambda, {q_{\parallel} \over \left({q_{\perp} \over
\Lambda}\right)^\zeta}; \{B_i \left(\ell _{\ast} \right) \}\right) &=&
\pm F\left(\Lambda, {q_{\parallel} \over \left({q_{\perp} \over
\Lambda}\right)^\zeta}; \{B_i \left(\ell _{\ast} \right) \}\right)
\nonumber\\
&+&{\left(\gamma \left(\ell_{\ast}\right) -
v_0 \left(\ell_{\ast}\right) \right)\over 2v_0 \left(\ell _{\ast}
\right)}\kappa
\left(\ell _{\ast}\right) \left({q_{\parallel} \over \Lambda
\left({q_{\perp} \over
\Lambda }\right)^\zeta }\right)
\label{B match}
\end{eqnarray}
all of which are clearly dependent {\it only} on the fixed point value
of the ratio ${\gamma \left(\ell_{\ast}\right) \over
v_0 \left(\ell_{\ast}\right)}$ (which is just a number of $O(1)$ since
$\gamma \left(\ell_{\ast}\right)$ and $ v_0\left(\ell_{\ast}\right)$
have the same dependence on
$\ell_{\ast}: exp
\left[\left(z -
\zeta
\right) \ell_{\ast}
\right]$, as can be seen from their recursion relations), and the combination
\begin{eqnarray}
\kappa \left(\ell _{\ast}\right) {q_{\parallel} \over \left({q_{\perp} \over
\Lambda}\right)^\zeta \Lambda} \, .
\label{comb match}
\end{eqnarray}

By combining the recursion relations  (\ref{v0 rec}),
(\ref{Rec sigma}), and (\ref{Rec rho}) into a recursion relation for
$\kappa$
\begin{eqnarray}
{d  \over  d\ell} \left(ln \kappa \right) = {d ln
v_0  \over  d\ell} - {1
\over  2} {d  \over  d\ell} \left(ln \sigma_1 + ln \rho_0 \right) = 1 -
\zeta
\label{kappa rec}
\end{eqnarray}
we find
\begin{eqnarray}
\kappa(\ell) = e^{(1-\zeta)\ell} \kappa_0
\label{kappa ell}
\end{eqnarray}
which implies that
\begin{eqnarray}
\kappa \left(\ell _{\ast}\right) = \kappa_0 \left(e^{\ell _{\ast}}\right)^{1 -
\zeta}= \kappa_0 \left({\Lambda \over q_{\perp}}\right)^{1 -
\zeta}
\label{kappa match}
\end{eqnarray}
Using this in (\ref{comb match}), we see that the combination:
\begin{eqnarray}
\kappa \left(\ell _{\ast}\right) {q_{\parallel} \over
\Lambda \left({q_{\perp}
\over \Lambda}\right)^\zeta} = \kappa_0 \, {q_{\parallel} \over
q_{\perp}} \quad .
\label{comb match 2}
\end{eqnarray}
takes on {\it precisely} the value it would take on using the {\it
unrenormalized} parameters and the {\it unrescaled} wavevector
$\vec{q}$. Hence, the same is true of $F$, $A_{\pm}$, and $B_{\pm}$.
And, therefore, the same is true of $\phi\left(\hat{q}\right)$.

Thus, we can replace $\phi (\Lambda, {q_{\parallel} \over
\left({q_{\perp} \over
\Lambda}\right)^\zeta} ; \{B_i \left(\ell _{\ast} \right) \})$ in (\ref{RG
match 4})
with $\phi\left(\hat{q}; \{B^0_i  \} \right )$, its unrenormalized value
straight from the linearized theory.  Doing so, and recalling that the
unrenormalized $\phi
\left(\hat{q}\right)$ was 0(1) for {\it all} directions $\hat{q}$ of
$\vec{q}$, we see from (\ref{RG match 4}) that the correlation function
$C_{ij}$
is largest when
\begin{eqnarray}
{q_{\parallel} \over \Lambda} \sim \left({q_{\perp}
\over \Lambda}\right)^{\zeta} \quad .
\label{q dom}
\end{eqnarray}
For $\left|\vec{q}\right|< \Lambda$, where our theory applies, (\ref {q dom})
implies that $q_{\parallel} \gg q_{\perp}$ (since $\zeta < 1$). In {\it that}
limit, $\phi\left(\hat{q}\right) \rightarrow 1$; using this in the expression
(\ref{RG match 4}) for $C_{ij}$ in the renormalized system, and using eq.
(\ref{RG match 4}) in turn in our expression (\ref{RG match 3}) for $C_{ij}$ in
the original model, we obtain, for $q_{\parallel} \gg q_{\perp}$, the
scaling law:
\begin{eqnarray}
C_{ij}\left(\vec{q} \right) = q^{-\left(2 \chi + d + \zeta
-1\right)}_{\perp} \, f_{ij}\left({\left(q_{\parallel} /\Lambda \right)
\over
\left({ q_{\perp}
\over  \Lambda }
\right)^\zeta}
\right)
\label{v scale1}
\end{eqnarray}

Note that the range $q_{||} \gg q_{\perp}$ for which this scaling law
holds includes those $\vec{q}$'s which dominate the fluctations; namely, those
with ${q_{||} \over \Lambda} \stackrel{>}{\sim} \left({q_{\perp}
\over
\Lambda}
\right)^{\zeta}$.

Integrating $C_{ij}\left(\vec{q} \right)$ over all $\vec{q}$ gives the
equal time, root-mean-squared real space fluctuation of
$\vec{v}_{\perp}$:
\begin{eqnarray}
\left< \left|\vec{v}_{\perp}\left(\vec{r}, t \right) \right|^2\right> &=&
\int{d^dq \over  \left(2 \pi \right)^d} \, C_{ii}\left(\vec{q}
\right)\nonumber\\
 &=&
\int{d^{d-1}q_{\perp}  \over  \left(2 \pi \right)^{d}} \, q^{-\left(2 \chi +
d + \zeta -1\right)}_{\perp} \int {dq_{\parallel}}\,
f_{ii}\left({\left(q_{\parallel} /\Lambda \right)
\over
\left({ q_{\perp}
\over  \Lambda }\right)^\zeta}
\right) \nonumber\\
&=& A \int {d^{d-1}q_{\perp} \over  q^{2\chi + d -
1}_\perp}
\label{vv fluc}
\end{eqnarray}
where the final proportionality was obtained by scaling
$q_{\perp}$ out of the $q_{\parallel}$ integral via the change of variables
$q_{\parallel}
\equiv \Lambda Q_{\parallel} \left({q_{\perp}  \over  \Lambda}
\right)^\zeta$ and we've defined the $q_{\perp} $-independent constant
$A \equiv \int d Q_{\parallel} f_{ij}\left(Q_{\parallel}\right)\Lambda^{2
\chi + d +
\zeta}$.  The final integral in (\ref{vv fluc}) clearly converges in the
infra-red $\left(\left|\vec{q}_{\perp} \right| \rightarrow 0 \right)$ limit
if and only if $\chi < 0$.  Furthermore, if $\chi$ is $>0$, and we impose an
infra-red cutoff $\left|\vec{q}_{\perp}  \right|>L^{-1}_{\perp}$ in
(\ref{vv fluc}),  where $L_{\perp}$ is the lateral (i.e., $\perp$
direction) spatial extent of the system, we easily obtain
\begin{eqnarray}
\left<\left| \vec{v}_{\perp} \left(\vec{v}, t \right)\right|^2 \right> =
C^{\prime} L^{2\chi}_{\perp}
\label{v real 2}
\end{eqnarray}

Indeed, the connected real-space, equal time, velocity autocorrelation
function discussed in the introduction is given by
\begin{eqnarray}
C_C\left(\vec{R}\right) &\equiv& \left<
\vec{v} \left(\vec{r}+
\vec{R}, t\right)\cdot \vec{v}\left(\vec{r}, t\right)\right> - \left|\left<
\vec{v}\left(\vec{r}, t\right)\right>\right|^2 \nonumber
\\
&=&
\left< \vec{v}_{\perp}\left(\vec{r}+
\vec{R}, t\right)\cdot \vec{v}_{\perp}\left(\vec{r},
t\right)\right>\nonumber
\\ &=&\int{d^dq \over  \left(2 \pi \right)^{d}} \, C_{ii}
\left(\vec{q}\right)e ^{i\vec{q} \cdot R} \nonumber
\\
&=& \int{d^dq \over  \left(2 \pi \right)^{d}}q^{-\left(2 \chi + d
+ \zeta -1\right)}_{\perp} f_{ii}\left({\left(q_{\parallel}/\Lambda\right)
\over
\left(q_{\perp}/\Lambda\right)^\zeta}\right) e^{i\left(\vec{q}_{\perp}
\cdot
\vec{R}_{\perp} + q_{\parallel} R_{\parallel}
\right)} \quad .
\label{v real corr1}
\end{eqnarray}
Making the changes of variable
\begin{eqnarray}
\vec{q}_{\parallel} \equiv {\vec{Q}_{\perp}  \over \left|
\vec{R}_{\perp}\right| } \quad , \, \quad q_{\perp} \equiv {Q_{\parallel}
\over
\left|
\vec{R}_{\perp}\right|^\zeta } \quad ,
\label{v real corr 2}
\end{eqnarray}
we obtain the scaling law (\ref{C conn scale}) for $C_c \left(\vec{R}\right)$
quoted in the introduction, with
\begin{eqnarray}
f_v(u) \equiv \int d^{d-1}Q_{\perp} dQ_{\parallel}
f_{ii}\left({Q_{\parallel}
\over  Q^\zeta_{\perp}}\right) e^{i\left(\vec{Q}_{\perp} \cdot
\hat{R}_{\perp} + Q_{\parallel} u
\right)}Q_{\perp}^{-\left(2 \chi + d + \zeta -1 \right)}
\label{v real corr 3}
\end{eqnarray}

This shows that the $\chi$ we obtain from the renormalization group
by the prescription we have chosen - namely, making the specific set of
parameters
$D_{B,T,\parallel}$, $\Delta$,
$\lambda _{1,2,\rho}$, and
$\sigma_2$ flow to fixed points - is precisely the physical roughness
exponent
defined by the velocity fluctuations.

To summarize:  {\it if} we chose the rescaling
exponents  $\chi$, $z$, and $\zeta$ so as to make the particular subset
of the dynamical parameters $D_{B,T,\parallel,\rho}$, $\Delta$, $\lambda
_{1,2,\rho}$, and
$\sigma_2$ flow to non-zero fixed points, then {\it those} $\chi$ and
$\zeta$ are the ones that appear in the scaling law (\ref{C conn scale}).  This
directly and simply relates the $RG$ to physically observable
correlation functions, so this is the choice we will make.

A scaling law similar to (\ref{v scale1})can be
derived, by precisely the same type of arguments, for the equal-time
density-density correlation function:
\begin{eqnarray}
 C_{\rho}(\vec{q}) =
{q^{3-d-\zeta-2\chi}_{\perp} \over  q^2} f_{\rho}
\left({q_{\parallel}\ell_0 \over
\left( q_{\perp}\ell_0\right)^{\zeta}}\right) Y (\theta_{\vec{q}})
\label{rho scale 1}
\end{eqnarray}
where, in writing this relation, we have used the fact
that the ``roughness'' exponent for $\rho$, $\chi_\rho = \chi$, the
``roughness'' exponent for $\vec{v}_{\perp}$.

The alert reader will have noticed that neither of the scaling
laws (\ref{v scale1}) and (\ref{rho scale 1}) derived so far
involve the time rescaling exponent $z$.  This is unsurprising,
since we've considered only equal time correlation functions up to
now.

To fully study the dynamics of the model, we need to consider
correlations between different {\it times}, as well as positions.
These {\it different} time correlation functions will involve $z$.

It is easiest to work with the space  {\it and time} Fourier
transform
$C_{ij}\left(\vec{q}, \omega\right)$, defined by
\begin{eqnarray}
\left<v_i \left(\vec{q}, \omega\right)  v_j \left(-\vec{q},
\omega^{\prime}
\right)\right>
\equiv \delta \left(\omega+ \omega^{\prime}\right) C_{ij}\left(\vec{q},
\omega\right)
\label{z1}
\end{eqnarray}

We will find, as we've asserted many times already in this paper,
that $C_{ij}\left(\vec{q}, \omega\right)$  does {\it not} have a simple
scaling form, unlike the equal time correlations.  Nonetheless we
can derive an expression for it in terms of functions of $\vec{q}$
which {\it do} show simple scaling behavior; namely, effective
wavevector dependent diffusion constants that diverge as
$\left|\vec{q}\right|, \omega \rightarrow 0$.

We begin this derivation by separating $C_{ij}$ into its transverse
and longitude parts:
\begin{eqnarray}
C_{ij}\left(\vec{q}, \omega \right) \equiv L^{\perp}_{ij}
C_{L}\left(\vec{q}, \omega\right) + P^{\perp}_{ij}
C_{T}\left(\vec{q}, \omega \right)
\label{z2}
\end{eqnarray}
where $L^{\perp}_{ij} \equiv {q^{\perp}_i q^{\perp}_j \over
q^2_{\perp}}$ and $P^{\perp}_{ij} = \delta_{ij}- L^{\perp}_{ij} -
\delta_{i,||} \delta_{j,||}$ are, respectively, the longitudinal and
transverse projection operators defined in section IV.  Both the
transverse and longitudinal pieces $C_{L,T}\left(\vec{q},
\omega \right)$ obey the same renormalization group
transformation
\begin{eqnarray}
C_{L,T}\left(\vec{q}_{\perp} ,  q_{\parallel} , \omega;
\left\{B^0_i\right\}\right) = e^{\left(2\chi + z + \zeta +
d-1\right)\ell}  C_{L,T}\left(e^{\ell}\vec{q}_{\perp}, e^{\zeta\ell}
q_{\parallel} , e^{z\ell}\omega;
\left\{B_i\left({\ell}\right) \right\}\right)
\label{z3}
\end{eqnarray}
where we've been careful to take into account the rescaling of the
delta function in (\ref{z1}) in deriving the argument of the
exponential in (\ref{z3}).

As for the equal-time correlation function, it is convenient here to
chose the rescaling factor $e^{\ell}$ such that
$e^{\ell}\vec{q}_{\perp}$ is right on the Brillouin zone boundary; i.e.,
$e^{\ell}\left|\vec{q}_{\perp}\right| =
\Lambda$.  Making this choice, and taking $\Lambda = 1$, we obtain
\begin{eqnarray}
C_{L,T}\left(q_{\perp}, q_{\parallel}, \omega;
\left\{B^0_i\right\}\right) = q_{\perp}^{-\left(2\chi + z + \zeta
+ d-1\right)}  C_{L,T}\left(\hat{q}_{\perp}, {q_{\parallel} \over
q^{\zeta}_{\perp}} , {\omega \over
q^{z}_{\perp}} ;
\left\{B_i\left({\ell_{\ast}}\right) \right\}\right)
\label{z4}
\end{eqnarray}
where, on the right hand side, we've defined
\begin{eqnarray}
\ell _{\ast} = ln \left({ 1
\over  \left| \vec{q}_{\perp} \right|}
\right) \quad .
\label{z5}
\end{eqnarray}
For the moment, let's focus on the longitudinal piece
$C_L\left(\vec{q}, \omega\right)$.  As we argued for the
equal-time correlation function, here too we can evaluate the right
hand side in the harmonic approximation Eqn. (\ref{N+22}).  This gives
\begin{eqnarray}
C_L \left(\vec{q}_{\perp},
q_{\parallel},
\omega;
\left\{B^0_i\right\}\right) =
{\Delta_{\ast}  (\omega-v_0(\ell _0^*) q_{\parallel}q_{\perp}^{z-\zeta}
 )^2 q^{-\left(2\chi
+ 3z +
\zeta + d - 1\right)}_{\perp} \over {\rm DEN}}
\label{z6}
\end{eqnarray}
where
\begin{eqnarray}
{\rm DEN} =
\left\{\left[{\omega\over  q^z_{\perp}} -
\omega_+\left(\ell_{\ast}\right)\right]^2
\left[{\omega \over q^z_{\perp}} - \omega_-\left(\ell_{\ast}\right)
\right]^2 \right.
\hspace{5 cm} \nonumber\\
+ \left. \left[ \left({\omega \over  q^z_{\perp}}\right)
\left(\Gamma _L \left(\ell_{\ast} \right) +  \Gamma
_{\rho}\left(\ell _{\ast}\right)\right) -
v_0\left(\ell_{\ast}\right) \left(\Gamma _L
\left(\ell_{\ast} \right) +  {\gamma \left(\ell_{\ast}\right) \over v_0
\left(\ell_{\ast}\right)} \Gamma _{\rho}\left(\ell _{\ast}\right)\right)
\left({q_{\parallel} \over q^{\zeta}_{\perp}} \right) \right]^2 \right\}
\label{z6 DEN}
\end{eqnarray}
where
\begin{eqnarray}
\omega_{\pm}\left(\ell_{\ast}\right) \equiv
\omega_{\pm}\left(\hat{q}_{\perp}, {q_{\parallel} \over
q^{\zeta}_{\perp}}; \gamma \left(\ell_{\ast}\right),
\rho_0 \left(\ell_{\ast}\right), \left\{ B_i
\left(\ell_{\ast}\right)\right\}
\right) \quad ,
\label{omega star}
\end{eqnarray}
with the sound frequencies $\omega_\pm \left(\vec{q};
\gamma, \rho_0, \left\{ B_i \right\}\right)$ obtained in the harmonic
theory:
\begin{eqnarray}
\omega_\pm \left(\vec{q}; \gamma, \rho_0,
\left\{ B_i \right\}\right)
&=& {\gamma + v_0 \over 2} \,  q \cos \theta_{\vec{q}}
\pm
\sqrt{\left({\left(\gamma - v_0 \right)   q \cos
\theta_{\vec{q}}  \over  2}\right)^2 +
\sigma_1 \rho_0 q^2 \sin^2 \theta_{\vec{q}}} \nonumber \\
&=& {\left(\gamma + v_0\right) q_{\parallel} \over 2} \pm
\sqrt{\left({\left(\gamma - v_0\right)  q_{\parallel}  \over
2}\right)^2 +
\sigma_1 \rho_0 q^2_{\perp}} ,
\label{omega pmp}
\end{eqnarray}
\begin{eqnarray}
\Gamma _L \left(\ell_{\ast} \right) \equiv D^{\ast}_L +
D^{\ast}_{\parallel}  \left({q_{\parallel} \over q^{\zeta}_{\perp}} \right)^2
\quad ,
\label{Gamma ast L}
\end{eqnarray}

\begin{eqnarray}
\Gamma _{\rho} \left(\ell_{\ast} \right) \equiv D^{\ast}_{\rho}
\left({q_{\parallel} \over q^{\zeta}_{\perp}} \right)^2
\quad ,
\label{Gamma ast rho}
\end{eqnarray}
and $D^{\ast}_L$, $D^{\ast}_{\parallel}$, and
$\Delta_{\ast}$ are the fixed point values of $D_L$,
$D_{\parallel}$, and $\Delta$, to which those parameters
will have flown for
$\ell_{\ast}$ large, as it will be for small
$q_{\perp}$.

The complication of this expression - that is, the fact that stops it from
having a simple scaling form - is that the parameters
$\gamma \left(\ell_{\ast} \right)$, $\sigma_1
\left(\ell_{\ast}
\right)$, and $\rho_0 \left(\ell_{\ast} \right)$ that
appear implicitly in (\ref{z6}) do {\it not} flow to field
point values for our ``canonical'' choice of $\chi$, $z$, and
$\zeta$, as discussed earlier.  Physically, this reflects the
fact that the scaling of the sound {\it speeds} $(\omega
\propto q)$ is different from that of their dampings
(damping rate
$\propto q^2$ in harmonic theory; we will show damping rate is
$\propto q^z_{\perp}$ here, in a moment).

To proceed, it is first useful to reorganize (\ref{z6}) slightly; by
multiplying numerator and denominator by $q^{4z}_{\perp}$:
\begin{eqnarray}
C_L \left(\vec{q}_{\perp}, q_{\parallel}, \omega;
\left\{B^0_i
\right\}\right) = \hspace{12 cm} \nonumber \\
\hspace{-2 cm} {\Delta_{\ast} (\omega-v_0(\ell_{*})
q_{\parallel}q_{\perp}^{z-\zeta})^2
q^{\left(z - \zeta - 2 \chi + 1 - d \right)}_{\perp}  \over
\left[\omega- q^z_{\perp} \omega_+
\left(\ell_{\ast}\right)\right]^2
\left[\omega-
q^z_{\perp} \omega_-
\left(\ell_{\ast}\right) \right]^2 +
\left[\omega q_{\perp}^z \left(\Gamma_{L}(\ell_{\ast})
+\Gamma_{\rho}(\ell_{\ast})\right)-\left(v_0 (\ell_{\ast})
\Gamma_{L}(\ell_{\ast})
+\gamma (\ell_{\ast})\Gamma_{\rho}(\ell_{\ast})\right)
q_{\parallel}q_{\perp}^{2z-\zeta}\right]^2}
\label{C L match}
\end{eqnarray}
Next, we solve the recursion relations for
$\gamma\left(\ell_{\ast}
\right)$, $\sigma_1 \left(\ell_{\ast}\right)$ and $\rho_0
\left(\ell_{\ast}
\right)$:
\begin{eqnarray}
v_0 \left(\ell_{\ast} \right) =
e^{(z-\zeta)\ell_{\ast}}v_0(\ell = 0) =
v_0 (0) q^{\zeta - z}_{\perp}
\label{z8}
\end{eqnarray}

\begin{eqnarray}
\gamma\left(\ell_{\ast} \right) =
e^{(z-\zeta)\ell_{\ast}}\gamma(\ell = 0) =
\gamma (0) q^{\zeta - z}_{\perp}
\label{z9}
\end{eqnarray}
\begin{eqnarray}
\sigma_1 \left(\ell_{\ast}\right)=
e^{(z-1)\ell_{\ast}}\sigma_1(\ell = 0) =
\sigma_1 (0) q^{1 - z}_{\perp}
\label{z10}
\end{eqnarray}
\begin{eqnarray}
\rho_0 \left(\ell_{\ast} \right)=
e^{(z-1)\ell_{\ast}}\rho_0(\ell = 0) =
\rho_0 (0) q^{1 - z}_{\perp}
\label{z11}
\end{eqnarray}
where in the second equality in each equation we've used
$\ell_{\ast} = ln\left({1 \over \left|\vec{q}_{\perp}
\right|}
\right)$.  Using these results and the expressions (\ref{omega star})
and (\ref{omega pmp}) for
$\omega_{\pm}\left(\ell_{\ast}\right)$, we see that $z$
and $\zeta$ drop out of the combinations
\begin{eqnarray}
q^z_{\perp}
\omega_{\pm}\left(\ell_{\ast} \right)
&=& q^z_{\perp}\left[{1 \over  2} \left(\gamma (0) +v_0(0) \right)
{q_{\parallel}  \over  q^\zeta_{\perp}}
q^{\zeta-z}_{\perp}
\pm
\sqrt{\left({ \left(\gamma (0) -v_0(0)\right)  \over  2 } {q_{\parallel}
\over  q^\zeta_{\perp}}q^{\zeta-z}_{\perp}\right)^2 +
\sigma_1\left(0\right)\rho_0 \left(0\right)q^{2(1-z)}_{\perp}}
\right]\nonumber\\
&=&{1 \over  2}
\left(\gamma + v_0 \right)q_{\parallel} \pm \sqrt{\left({ \left(\gamma (0)
-v_0(0)\right) q_{\parallel}  \over  2 }\right)^2 +
\sigma_1\left(0\right)\rho_0
\left(0\right) q^2_{\perp}} \nonumber\\
&=& \omega_\pm \left(\vec{q}_{\perp}, q_{\parallel} ;
\gamma(0),  v_0\left(0\right), \sigma_1 \left(0\right),
\rho_0
\left(0 \right) \right)
\label{z12}
\end{eqnarray}
and therefore the {\it positions} of the peaks in the full correlation
function (\ref{C L match}) are exactly those given by the harmonic theory
using the bare parameters $\sigma_1(0)$, $v_0(0)$, $\gamma(0)$, and
$\rho_0(0)$, namely,
$w_\pm \left(\vec{q}; \gamma(0), v_0(0),  \sigma_1(0),
\rho_0(0)\right)$. This is a direct consequence of the fact that there
are no (relevant) graphical renormalizations of the parameters
($\gamma, \sigma_1$, and
$\rho_0$) that determine the sound speeds (see the recursion
relations (\ref{Rec rho})-(\ref{v0 rec})) and shows that the relevant
non-linearities below
$d = 4$ do {\it not} alter the {\it positions} of the peaks in the
spatio-temporally Fourier-transformed velocity-velocity
autocorrelations.

The same cannot, however, be said for their widths.  Indeed, using
the above result for the sound speeds and Eqns. (\ref{Gamma ast L}) and
(\ref{Gamma ast rho}), we
see that
$C_L\left(\vec{q},
\omega\right)$ can be rewritten:
\begin{eqnarray}
&C_L& \left(\vec{q}, \omega \right) = \nonumber \\& &
{\Delta_{*} (\omega-v_0 (0) q_{\parallel})^2 q^{\left(z - \zeta - 2 \chi + 1 -
d\right)}_{\perp}
\over \left(\omega - c_+ \left(\theta_{\vec{q}}\right)q\right)^2
\left(\omega - c_-\left(\theta_{\vec{q}}\right) q\right)^2
+ \left[ \omega \left(
\Gamma^R_L \left(\vec{q}\right) +
\Gamma^R_{\rho}\left(\vec{q}\right)
\right) - q_{\parallel}
\left(v_0(0)
\Gamma^R_L\left(\vec{q}\right) +
\gamma(0)\Gamma^R_{\rho}\left(\vec{q}\right) \right)\right ]^2}
\label{C L scale}
\end{eqnarray}
where the sound speeds are given by the harmonic result
equation (\ref{N+9a}), and the renormalized dampings
\begin{eqnarray}
\Gamma^{R}_L\left(\vec{q}\right) = \left[D^{\ast}_{\parallel}
\left({q_{\parallel}  \over  q^\zeta_{\perp}}\right)^2 +
D^{\ast}_L\right]q^z_{\perp} \equiv q^z_{\perp} f_L
\left({q_{\parallel}
\over  q^\zeta_{\perp}}\right)
\label{Gamma L scale}
\end{eqnarray}
\begin{eqnarray}
\Gamma^R_\rho\left(\vec{q}\right) = D^{\ast}_{\rho}
\left({q_{\parallel}  \over  q^\zeta_{\perp}}\right)^2
q^z_{\perp} \equiv q^z_{\perp} f_\rho
\left({q_{\parallel}
\over  q^\zeta_{\perp}}\right)
\label{16a}
\end{eqnarray}
obey simple scaling laws.

The exact form of the scaling laws that we have obtained here (namely, e.g.,
$ f_L
\left({q_{\parallel}
\over  q^\zeta_{\perp}}\right) = \left[D^{\ast}_{\parallel}
\left({q_{\parallel}  \over  q^\zeta_{\perp}}\right)^2 +
D^{\ast}_L\right] $ ) , is not correct, because our choice of $\ell_\ast =
ln({1\over{q_\perp}})$ is only valid when $q_\perp^\zeta\gg q_{||}$.
In the opposite limit $q_\perp^\zeta\ll q_{||}$ , the fluctuations become
negligible in the renormalized problem once $D_{||} q_{||}^2$ becomes $\gg D_B
\Lambda^2$ in the renormalized problem, because at this point the linearized
approximation to the correlation functions is smaller than its largest value
at the Brillouin zone boundary. This means we can now stop the renormalization
at
$\ell_\ast$ such that $e^{\zeta\ell_\ast}q_{||} = \Lambda \times
D_B^\ast/D_{||}^\ast$ ,
which implies that $\ell_\ast = {ln({1\over{q_{||}}})\over\zeta} + O(1)$,
where the $O(1)$
factor is {\it universal}, because it depends only on the fixed point
values of the
diffusion constants.
Performing the above calculations with {\it this }
choice of $\ell_\ast$, we now obtain
\begin{eqnarray}
\Gamma^{R}_L\left(\vec{q}\right) = \left[D^{\ast}_{\parallel}
 +
D^{\ast}_L\left({q_{\perp}^2  \over
q^{2\over\zeta}_{\parallel}}\right)\right]q^{z\over\zeta}_{\parallel}
\times O(1) \equiv q^z_{\perp} f_L
\left({q_{\parallel}
\over  q^\zeta_{\perp}}\right) ,
\label{Gamma L scale||}
\end{eqnarray}
where we've now defined $f_L
\left({q_{\parallel}
\over  q^\zeta_{\perp}}\right)\equiv\left[D^{\ast}_{\parallel}
({q_{||}\over q_\perp^\zeta})^{z\over\zeta} +
D^{\ast}_L\left({q_{\perp}^\zeta  \over
q_{\parallel}}\right)^{{2-z}\over\zeta}\right]
\times O(1) $. Note that the precise form of this scaling function is
different
in this regime than that found earlier for
$q_\perp^\zeta\gg q_{||}$. Furthermore, its exact form is uncertain, due
to our
uncertainty in the O(1) factor, which, as discussed earlier, cannot be
determined
without knowing the fixed point values $D_i^\ast$ of the diffusion constants.
However, regardless of their values, we still get a scaling law with the same
power of $q_\perp$ and the same scaling {\it variable} ${q_{||}\over
q_\perp^\zeta}$
as that found earlier in the opposite limit.

In between these two limits we have to choose $\ell_\ast$ to smoothly
interpolate
between the two limits. This choice will clearly depend on the ratio
${q_{||}\over q_\perp^\zeta}$. Naively, one could imagine simply choosing
$\ell_\ast = ln (min({1\over q_\perp}, {O(1)\over q_{||}^{1\over\zeta}}))$.
A subtler choice
would take into account the $O(1)$ perturbative corrections we've
neglected, and would presumably
lead to a smooth crossover of $\ell_\ast$ between the two limits.

The moral of this discussion is three-fold:

1) we always get scaling laws of the form (\ref{Gamma L scale} ) for the
dampings,

2) the renormalized damping functions and the noise strength are always
of such a form that they depend {\it
only} on
$q_{||}$ for $q_{||}\gg q_\perp^\zeta$, and only on $q_\perp$ in the
opposite limit, and

3) we can only calculate the scaling function if we know the diffusion
constants at the
fixed point.

This last point will stop us from calculating the crossover functions in
$d=2$,
even though, as we will see, we {\it can} calculate the exponents there.

  We see from (\ref{Gamma L scale}) that
the physical significance of the exponent $z$ is that it gives the
scaling of the peak widths (in $\omega$) of $C_L\left(\vec{q},\omega\right)$
with $q_{\perp}$, while the peak {\it positions} continue to obey
the ``$ z= 1$'' scaling $\omega\propto q$.

Similar, but actually far simpler, arguments show that the
transverse correlation function $C_T\left(\vec{q},\omega\right)$
obeys
\begin{eqnarray}
C_T\left(\vec{q},\omega\right)= {f_\Delta({q_{\parallel}  \over
q^{\zeta}_{\perp}})q^{z-\zeta-2\chi+1-d}_{\perp}
\over  (\omega-\gamma q_{||})^2 + \Gamma^2_T\left(\vec{q} \right)}
\label{C T scale}
\end{eqnarray}
where $\Gamma_T\left(\vec{q}\right)$ obeys the scaling law
\begin{eqnarray}
\Gamma_T\left(\vec{q}\right) = q^z_{\perp} f_T
\left({q_{\parallel}  \over  q^{\zeta}_{\perp}}\right) ,
\label{Gamma T scale}
\end{eqnarray}
and
$f_\Delta$ is a scaling function associated with $\Delta$. For
general values of $q_{\perp}$ and $q_{||}$,  the same
scaling function  $f_{\Delta}$ should also be present in all of the other
correlation
functions as well (wherever $\Delta$ appears),  such as
eq. (\ref{C L scale}) for the longitudinal correlation function.

Likewise, the propogators of the full non-linear theory are given,
in $d < 4$, by the harmonic expressions (4.18)-(4.21), {\it
except} that $\Gamma_L$, $\Gamma_{\rho}$, and $\Gamma_T$ in those
expressions are replaced by
the anharmonic scaling laws (\ref{Gamma L scale}), (\ref{16a}) and
(\ref{Gamma T scale}).

To complete the specification of the scaling laws, we need the
asymptotic behavior of the scaling functions $f_{\Delta,L,T,\rho}(u)$.  From
(\ref{Gamma L scale}), (\ref{16a}), and the analogous result
for (\ref{Gamma T scale}), and requiring that the second point of our
tripartite moral applies,
we see that
\begin{eqnarray}
f_{\Delta}(u) \propto
\left\{\begin{array}{ll}
\mbox{constant},& u\rightarrow 0 \\
u^{z-\zeta-2\chi+1-d \over \zeta},&u \rightarrow \infty
\end{array}\right.
\label{crossover 0}
\end{eqnarray}
and
\begin{eqnarray}
f_{L,T,\rho}(u) \propto
\left\{\begin{array}{ll}
\mbox{constant},& u\rightarrow 0 \\
u^{z \over \zeta},&u \rightarrow \infty
\end{array}\right.
\label{crossover 1}
\end{eqnarray}
which implies that
\begin{eqnarray}
\Gamma_{L,T,\rho}\left(\vec{q}\right) \propto
\left\{\begin{array}{ll}
q^z_{\perp} ,& q_{\parallel} \ll q^\zeta _{\perp} \\
q^{z/ \zeta}_{\parallel},&q_{\parallel} \gg q^\zeta _{\perp}
\end{array}\right.
\label{crossover 2}
\end{eqnarray}
The simplest summary of the scaling of {\it all} correlation
functions and propagators is:  simply use the harmonic expressions
for them, {\it except} that diffusion constants $D_{T,B,\rho}$ should
be replaced by wavevector dependent quantities that diverge as
$\vec{q}
\rightarrow 0$, according to the scaling law
\begin{eqnarray}
D_{T,B,\rho}\left(\vec{q}\right) = q^{z-2}_{\perp} f_{T,B,\rho}
\left({\left({q_{\parallel}\over \Lambda}\right) \over
\left({q_{\perp} \over \Lambda} \right)^\zeta }
\right) \quad ,
\label{D perp scale}
\end{eqnarray}
the bare noise strength $\Delta$ should be replaced by
\begin{eqnarray}
\Delta \left(\vec{q}\right) = \Delta_{\ast} \left({q_{\perp}
\over \Lambda} \right) ^{z-\zeta-2\chi+1-d}f_{\Delta}
\left({\left({q_{\parallel}\over \Lambda}\right) \over
\left({q_{\perp} \over \Lambda} \right)^\zeta }
\right)
\label{Delta scale}
\end{eqnarray}
and the diffusion constant $D_{\parallel}$ should be replaced by
\begin{eqnarray}
D_{\parallel} \left(\vec{q}\right) = q^{z-2\zeta}_{\perp}
f_{\parallel}
\left({\left({q_{\parallel}\over \Lambda}\right) \over
\left( {q_{\perp} \over \Lambda} \right)^\zeta } \right)
\label{D parallel scale}
\end{eqnarray}
as can be seen by requiring that
\begin{eqnarray}
D_{\parallel} \left(\vec{q}\right)   q^2_{\parallel} =
D_{||}^{*}
\left({q_{\parallel} \over q^\zeta_{\perp}}\right)^2 q^z_{\perp}
\label{D parallel rule}
\end{eqnarray}
the right hand side being the form of the
$q_{\parallel}$ dependent term in (\ref{Gamma L scale}).

We hope the reader has not been too confused by the fact that we have
restored the ultraviolent cutoff $\Lambda \sim 1/\ell_0$ to the problem
by going back to dimensionful units where $\Lambda \neq 1$.

This completes our discussion of how the renormalization group,
and, in particular, the exponents $z$, $\chi$, and $\zeta$, relate to
physically observable correlation functions and propagators.  Now,
we turn to the problem of actually calculating those exponents.

\subsection{Exponents in $d = 2$}

To do this, we must calculate the graphical corrections in equations
(\ref{Rec DBT})-(\ref{Rec delta}).  The procedure for this, as
discussed in \cite{FNS} involves the harmonic correlation
functions and propagators and vertices representing the
non-linearities $\lambda_{1, 2, \rho}$ and $\sigma_2$.  Rather than
actually calculating these corrections, we will show that,
when $\lambda_2 = 0$, the structure of the theory is such that we
can determine the exponents $\chi$, $z$ and $\zeta$ exactly.

Consider first the $\lambda_1$ non-linearity.  Separating
$\vec{v}_{\perp}$ into transverse and longitudinal components,
\begin{eqnarray}
\vec{v}_{\perp} \equiv \vec{v}_T + \vec{v}_L
\label{LT1}
\end{eqnarray}
this can be written
\begin{eqnarray}
\left(\vec{v}_{\perp} \cdot \vec{\nabla}_{\perp}\right)\vec{v}_{\perp} =
\left(\vec{v}_T \cdot
\vec{\nabla}_{\perp}\right)\vec{v}_T +  \left(\vec{v}_L \cdot
\vec{\nabla}_{\perp}\right)\vec{v}_L +  \left(\vec{v}_T \cdot
\vec{\nabla}_{\perp}\right)\vec{v}_L +  \left(\vec{v}_L \cdot
\vec{\nabla}_{\perp}\right)\vec{v}_T \quad
\label{LT2}
\end{eqnarray}
Now consider the graphs that can be constructed from $\vec{v}_L -
\vec{v}_T$ the cross-terms
in this expression.  These will always mix transverse and longitude
propagators and
correlation functions in the internal integrals over momentum and
frequency.  But, as noted
earlier in our discussion of the harmonic theory, the peaks in the
longitudinal propagators and
correlation functions occur at different frequencies $\left(\omega=
\omega_\pm
\left(\vec{q}\right)\right)$ than those in the transverse propagators and
correlation function,
which occur at $\omega= 0$.  Furthermore, the overlap between
these peaks is negligible, since their widths ($\propto q^z_{\perp}$
with $z>1$) are much less than this offset in peak positions.  This
implies that the integral over wavevectors and frequencies of any
graph that mixes transverse and longitudinal propagators and
correlation functions will be much less (by powers of $q$) than
any similar graph containing purely transverse or purely
longitudinal propagators and correlation functions.  Hence, the
$\vec{v}_L - \vec{v}_T$ cross terms in (\ref{LT2}) are irrelevant
compared to the pure $\vec{v}_L$ and $\vec{v}_T$ terms.

Now let's consider those relevant pieces.  The Fourier transform of the
$\vec{v}_T$ piece at
wave vector $\vec{q}$  can be written in Fourier space:
\begin{eqnarray}
FT \left(\left(\vec{v}_T \cdot \vec{\nabla}_{\perp} \right)
v_{T_i}\right)_{\vec{q}} &=& i
\sum_{\vec{p}}
\vec{v}_T \left(\vec{p} \right) \left(\vec{q}_{\perp} -
\vec{p}_{\perp}\right)
v_{T_i}\left(\vec{q} - \vec{p}\right)
\label{LT3}\nonumber\\
&=& i q_{j}^{\perp} \sum_{\vec{p}} v_{T_j}
\left(\vec{p} \right)
v_{T_i}\left(\vec{q} - \vec{p}\right) \label{LT4}
\end{eqnarray}
where we have used the fact that $\vec{v}_T$ is transverse, so
$\vec{p}_{\perp} \cdot
\vec{v}_T \left(\vec{p} \right) = 0$.  So this piece of the
$\lambda_1$ vertex, which is a term in the equation for $\partial_t
v_i
\left(\vec{q}, t \right)$, is proportional to the external momentum
$\vec{q}_{\perp}$.  So is the purely longitudinal term,
as can easily be seen in real space.  Since
$\vec{v}_L$ is longitudinal, we can write
\begin{eqnarray}
\vec{v}_L = \vec{\nabla}_{\perp}\phi
\label{LT5}
\end{eqnarray}
for some scalar field $\phi$.  Now the second term in (\ref{LT2}) can be
rewritten in
terms of
$\phi$
\begin{eqnarray}
\left(\vec{v}_L \cdot  \vec{\nabla}_{\perp} \right)v_{Li} =
\left(\partial_j \phi\right)
\left(\partial_j \partial_i \phi \right) = \left(\partial_j \phi\right)
\left(\partial_i \partial_j \phi
\right) = {1  \over  2} \partial_i \left(\partial_j \phi \partial_j
\phi\right)
\label{LT6}
\end{eqnarray}
which is clearly a total derivative, whose Fourier transform is proportional to
$\vec{q}_{\perp}$.

Hence, the two {\it relevant} pieces of the $\lambda_1$ vertex are
proportional to the external momentum  $\vec{q}_{\perp}$.  Clearly,
the $\sigma_2$ term, being a total ${\perp}$ derivative, is also
proportional to
$\vec{q}_{\perp}$ in Fourier space.  Hence, when $\lambda_2
= 0$, all the remaining relevant vertices are proportional to
$\vec{q}_{\perp}$.  An immediate consequence of this is that
$\Delta$ and $D_{\parallel}$ acquire no graphical
renormalization. For $\Delta$, this can be seen by noting that any
graph that renormalizes $\Delta$ (e.g., figure 7) must contain two
external vertices each proportional to
$q_{\perp}$, and hence must be proportional to $q^2_{\perp}$.
Therefore all renormalizations of $\Delta$ must be proportional to
$q^2_{\perp}$, and hence negligible, as
$\left|\vec{q}\right|\rightarrow 0$, relative
to the bare
$\Delta$.  Likewise, any graph for the diffusion constants
must be proportional to figure 8, which must be proportional
to at least one power of $\vec{q}_{\perp}$.  Since
$D_{\parallel}$  and $D_{\rho}$ involve {\it no} powers of
$\vec{q}_{\perp}$, they {\it cannot} be renormalized graphically.

Thus, when $\lambda_2 = 0$, $\Delta$, $D_{\rho}$ and $D_{\parallel}$ get
no graphical renormalization.   That is, $G_{\parallel}$, $G_{\rho}$ and
$G_\Delta$ in (\ref{Rec D parallel}) and (\ref{Rec delta})
are, exactly, $= 0$.  Thus, the requirement that $\Delta$,
$D_{\parallel }$, and $D_{\rho}$ flow to fixed points $\left({d
D_{\parallel,\rho}
\over  d\ell}\right) = 0 = {d \Delta  \over  d\ell}$ leads to
two independent {\it exact} scaling relations between the
three independent exponents $\chi$, $z$ and $\zeta$.
Requiring ${d D_{\parallel,\rho}
\over  d\ell} = 0$ implies
\begin{eqnarray}
z = 2\zeta
\label{D parallel fix}
\end{eqnarray}
while requiring  ${d \Delta  \over  d\ell} = 0$ leads to
\begin{eqnarray}
z = \zeta + 2 \chi + d - 1 \quad .
\label{Delta fix}
\end{eqnarray}
We emphasize that we have only shown that these relations (\ref{D parallel
fix}) and
(\ref{Delta fix}) hold when $\lambda_2 = 0$.

We can obtain a third independent exact scaling relation between these
three exponents, and
thereby determine them exactly, when $\lambda_2 = 0$, by considering the
renormalization of
the non-linearities $\lambda_1$, $\lambda_\rho$, and $\sigma_2$.

We start deriving this third relation by noting that when
$\lambda_2 = 0$ {\it and}
$\lambda_1 = \lambda_\rho \equiv \lambda$, there can be {\it no}
graphical renormalization of $\lambda_1$. This is because for these
parameter values the equations of motion (\ref{EOM broken}) and (\ref{cons
broken}) have an exact symmetry which we call pseudo-Galilean invariance:
namely, they remain unchanged under the ``boost'' transformation:
\begin{eqnarray}
\vec{r}_{\perp} \rightarrow \vec{r}_{\perp}-\lambda\vec{v}_b t
\label{Galilean 1}
\end{eqnarray}
\begin{eqnarray}
\vec{v}_{\perp}
(\vec{r},t) \rightarrow \vec{v}_{\perp}(\vec{r},t)+\vec{v}_b
\label{Galilean 2}
\end{eqnarray}
where the ``boost'' velocity $\vec{v}_b$ is an arbitrary constant
vector in the
${\perp}$ plane.

This symmetry must be preserved upon renormalization with the {\it same} value
of
$\lambda$. Hence, there can be no graphical renormalization of $\lambda$, when
$\lambda_1 = \lambda_\rho$. That is, $G^\lambda_1 = 0$ when
$\lambda_1 = \lambda_\rho$. Since $G^\lambda_\rho = 0$ always, it
is clear that, if $\lambda_1 < \lambda_\rho$ initially, it will always
remain so upon renormalization.

This implies that, for flocks that start with bare $\lambda_1$ less than
the bare
$\lambda_\rho$ (which should be a finite fraction of all possible flocks),
only two types of fixed
points are possible:
\begin{description}
\item[I)]   $\lambda^{\ast} _1=\lambda^{\ast}_{\rho} = 0$,
  or
\item[II)]  $\lambda^{\ast} _{\rho} \neq 0$.
\end{description}

The first type of fixed point (I), however, is readily seen to be
unstable to $\lambda_\rho$, which must always be non-zero (and, in
fact, $=1$) initially. Hence, this fixed point is never reached, and we
must flow to a fixed point of type II. To see that fixed points of type
I are {\it unstable}, note that for such a fixed point, the only
remaining relevant non-linearity is $\sigma_2$.  But, by itself,
$\sigma_2$ can {\it not} renormalize {\it any} of the diffusion
constants
$D_B$ and $D_i$. The reason for this is that any graph (e.g., figure 8)
that renormalizes {\it any} diffusion constant must have an external
{\it velocity} leg emerging from the right. However, using only the
$\sigma_2$ vertex, which {\it only} involves $\rho$, we can {\it
only} make graphs with a $\rho$ leg emerging from the right.
Therefore, at a fixed point of type I, $G_{B, T} = 0$, exactly, in
(\ref{Rec DBT}). Thus, to find a fixed point for $D_{B, T}$, we must
choose $z = 2$. Combining this with the previous exact scaling
relations (\ref{D parallel fix}) and (\ref{Delta fix}), we find $\zeta = 1$
and $\chi = 1- {d\over 2}$. But using these values (which are
nothing but those that we found in the harmonic theory), we find
that $\lambda_\rho$ is a relevant perturbation at any fixed point of
type I:
\begin{eqnarray}
{d \lambda_\rho \over dl}=\left(2-{d \over 2}\right)
\lambda_{\rho}
\label{lambda rho vel}
\end{eqnarray}
for all spatial dimensions $d<4$. Note that (\ref{lambda rho vel}) is {\it
exact} at  any
fixed point of type I, since $G^\lambda_\rho=0$, exactly.

So fixed points of type I are unstable, and we will always
flow to a fixed point of type II. Using the recursion relation
(\ref{Rec lambda}) for $\lambda_\rho$, {\it and} the fact that
$G^\lambda_\rho = 0$ always, we immediately obtain that
$\lambda^{\ast}_\rho$ can be $\neq0$ if and only if a third
exact scaling relation is satisfied, namely:
\begin{eqnarray}
\chi=1-z.
\label{lambda fix}
\end{eqnarray}
 The three relations (\ref{D parallel fix}), (\ref{Delta fix}),  and
(\ref{lambda fix}) that hold when $\lambda_2 = 0$ can
trivially be solved, to find the exact scaling exponents in all
$d < 4$ that describe flocks with
$\lambda_2 = 0$;
\begin{eqnarray}
\zeta = {d + 1\over 5}
\label{canon 1}
\end{eqnarray}
\begin{eqnarray}
z = {2 (d + 1) \over 5}
\label{canon 2}
\end{eqnarray}
and
\begin{eqnarray}
\chi = {3 - 2 d \over 5}
\label{canon 3}
\end{eqnarray}
Note that these match continuously, at the upper critical dimension $d=4$,
onto their
harmonic values $\zeta=1$, $z=2$, and $\chi=1 - {d\over 2} = -1$, as they
should.

If $\lambda_1^{Bare}>\lambda_{\rho}^{Bare}$, then, besides the two cases that
we discussed above, there may be a third type of fixed point:

\begin{description}
\item[III)]
$\lambda_{\rho}^{*}=0$, $\lambda_{1}^{*}\neq 0$.
\end{description}

For this type of fixed point to
be stable, the exponents $\chi$ and $z$ have to satisfy:
\begin{eqnarray}
\chi +z <1
\label{case 3}
\end{eqnarray}
The above inequality, together with eq. (\ref{D parallel fix}) and
eq. (\ref{Delta fix}) give eq. (\ref{canon 1}), (\ref{canon 2}) and
(\ref{canon 3}) with the "=" signs replaced by $"<"$. For simplicity, we
will only discuss the cases with $\lambda_1^{Bare}<\lambda_{\rho}^{Bare}$,
where the exponents are given by eq. (\ref{canon 1}), (\ref{canon 2}) and
(\ref{canon 3}). However, all the qualitative results will be valid for
case III, e. g., the spontaneous symmetry broken phase will be more stable
in case III if it is stable in case II because of the inequality.   The
simplest
possible scenario is that there is only one stable fixed point, regardless of
whether
$\lambda^{Bare}_{\rho} < \lambda^{Bare}_1$
or not, and that it is type II, and has the canonical exponents (\ref{canon
1})- (\ref{canon 3}).  We consider it highly probable that this is, in
fact, the
case.
Even if it is not, (\ref{canon 1}) - (\ref{canon 3}) {\it do} hold for some
flocks
(those with $\lambda^{Bare}_{\rho}>
\lambda^{Bare}_1$).

Our derivation of these results (\ref{canon 1})-(\ref{canon 3}) depended
{\it only} on the assumption that
$\lambda^{\ast}_2=0$. Note, however, that in $d=2$, any flock is
equivalent to a flock with
$\lambda_2=0$. This is because the
$\lambda_1$ and $\lambda_2$ vertices become identical in $d=2$, where
$\vec{v}_{\perp}$ has only one component, which we'll take to be $x$.
That is, in $d = 2$,

\begin{eqnarray}
\lambda_1 \left(\vec{v}_{\perp} \cdot \vec{\nabla}_{\perp}\right)
\vec{v}_{\perp} =
\lambda_1
\hat{x} v_x
\partial_x v_x = {1\over 2} \lambda_1 \partial_x
\left(v^2_x\right)\hat{x}
\label{lambda 1 d=2}
\end{eqnarray}

\begin{eqnarray}
\lambda_2 \left(\vec{\nabla}_{\perp} \cdot \vec{v}_{\perp}\right)
\vec{v}_{\perp} =
\lambda_2 \hat{x}
\left(\partial_x v_x\right) v_x ={1\over 2} \lambda_2 \partial_x
\left(v^2_x\right)\hat{x}
\label{lambda 2 d=2}
\end{eqnarray}
so that the full $\vec{v}_{\perp}$ non-linearity becomes ${1\over
2} \left(\lambda_1 +\lambda_2\right)\partial_x \left(v^2_x\right)
\hat{x}$, which is just what we'd get if we started with a (primed)
model with
$\lambda^\prime _2=0$ and $\lambda^\prime _1=\lambda_1 +
\lambda_2$. This later model, since it has
$\lambda^\prime _2=0$, must have the ``canonical'' exponents
(\ref{canon 1})-(\ref{canon 2}) hence, so must the
$\left(\lambda_1,\lambda_2\right)$ model, which includes all
possible $d = 2$ models. So all models in $d = 2$ must have the
canonical exponents (\ref{canon 1})-(\ref{canon 3}).

Equivalently, we can derive this result by simply noting that, in $d =
2$, the full $v_{\perp} - v_{\perp}$vertex becomes ${1\over
2}\left(\lambda_1+\lambda_2\right) \partial_x \left(v^2_x\right)$, which
is a total
$x$ derivative even when $\lambda_2 \neq 0$. Furthermore, the $d = 2$
model now has the pseudo-Galilean invariance
(\ref{Galilean  1})-(\ref{Galilean  2}) when
$\lambda_1+\lambda_2=\lambda_\rho$. These two properties
($v_{\perp} - v_{\perp}$ vertex total $x$ derivative, and
pseudo-Galilean invariance at a special point) are all that we used
to derive the ``canonical'' exponent (\ref{canon 1})-(\ref{canon 3}); so
those canonical exponents
$\it{must}$ hold in $d=2$. Setting $d=2$ in (\ref{canon
1})-(\ref{canon 3}), we obtain:
\begin{eqnarray}
\zeta &=& {3 \over 5}
\label{zeta d=2 exact}
\end{eqnarray}
\begin{eqnarray}
z &=& {6 \over 5}
\label{z d=2 exact}
\end{eqnarray}
\begin{eqnarray}
\chi &=&  -{1 \over 5}
\label{chi d=2 exact}
\end{eqnarray}
Note, in particular, that $\chi < 0$. This implies, as discussed earlier,
that the flock exhibits true long ranged order.

Using the exponents (\ref{zeta d=2 exact})-(\ref{chi d=2 exact}) in the
general scaling relations, such as (5.31) and (5.33),  we obtain all of
the scaling results for
correlation functions in $d = 2$ quoted in the introduction.   Note also
that for this set of exponents
$z - \zeta - 2 \chi + 1 - d = 0$.  Hence, from equation (\ref{Delta
scale}), we see that the noise strength $\Delta$ is a constant,
independent of
$\vec{q}$, which makes sense since $\Delta$ is unrenormalized
graphically. So, in the
$d=2$ model, we can calculate all correlation functions from their harmonic
expressions, except that we replace the diffusion constants $D_{B,
T, \rho}$ with functions that diverge as $\vec{q}\rightarrow 0$
according to the scaling laws
\begin{eqnarray}
D_{B, T}\left(\vec{q}\right) = q^{-{4 \over 5}}_{\perp}
f_{B,
T}\left({{\left(q_{\parallel} \over \Lambda\right)}
\over {\left(q_{\perp} \over \Lambda \right)^{3\over 5}}}\right)
\label{d=2 D perp scale}
\end{eqnarray}
where we've used the exact  $d = 2$ exponents $z = {6
\over 5}$ and
$\zeta = {3 \over 5}$ in the general scaling law (\ref{D perp scale}).
$D_{\parallel,\rho}$, on the other hand, are, like $\Delta$, constants, since
$z=2\zeta$ (see general equation (\ref{Rec D parallel})), which also
makes sense since $D_{\parallel,\rho}$ are unrenormalized graphically.
Hence, the only replacement needed to turn the harmonic results
into the correct results for the full, non-linear theory in
$d = 2$ is (\ref{d=2 D perp scale}).
\subsection{$d > 2$}
Now we turn to $d > 2$. Here the canonical exponents need not hold if
$\lambda^{\ast}_2 \neq 0$. The obvious thing to do, therefore, is to
determine whether a small $\lambda_2$ is a relevant or irrelevant
perturbation at the $\lambda_2 = 0$ fixed point. We have attempted
to do this to leading (one-loop) order in a $4 - \epsilon$ expansion.
This involves calculating the perturbative corrections
$G^\lambda_{1,2}$ and
$G_{\sigma_2}$ in the recursion relations (\ref{Rec lambda}) - (\ref{Rec
sigma}) to one loop order, and to linear order in $\lambda_2$. Once
this is done, we can find a fixed point with
$\lambda_2=0$, and then calculate the linear renormalization group
eigenvalue of $\lambda_2$ at this fixed point. That is, we'll expand the
right hand side of the recursion relation (\ref{Rec lambda}) for
$\lambda_2$ to linear order in $\lambda_2$, obtaining:
\begin{eqnarray}
{d\lambda_2\over dl}=\gamma_2\lambda_2
\label{lambda 2 lin}
\end{eqnarray}
for $\lambda_1=\lambda^{\ast}_1$,
$\lambda_\rho=\lambda_\rho^{\ast}$,
$\sigma_2=\sigma^{\ast}_2$, and
$\lambda_2$ small. If $\gamma_2$ is $<0$, then the $\lambda_2 = 0$ fixed
point is
(at least locally) stable, and the canonical exponents (\ref{canon 1}) -
(\ref{canon 3}) will hold for all $d$.  Unfortunately, an (extremely
laborious!) calculation (involving 14 different Feynmann
graphs)
shows, after many seemingly miraculous and
unexpected cancellations between different graphs, that
$G^\lambda_1$, $G^\lambda_2$, and $G_{\sigma _2}$ are {\it exactly} zero to
one loop
order.  This implies that
\begin{eqnarray}
\chi + 1 - z = 0 (\epsilon^2)
\label{4 - epsilon 1}
\end{eqnarray}
and that, to this order at least, $\lambda_{1, 2, \rho}$ and $\sigma_2$ can
take on
{\it any} value at the fixed point.  That is, to this order, there appears
to be a fixed
``line'' (actually, a fixed 4 dimensional subspace $(\lambda_1, \lambda_2,
\lambda_{\rho}, \sigma_2)$), instead of a single fixed point.  This,
unfortunately, eliminates all of our predictive power for the
exponents. For example, keeping $D_{\parallel}$ fixed leads to
\begin{eqnarray}
z - 2\zeta = - G_{\parallel} (\lambda_1, \lambda_2, \lambda_{\rho},
\sigma_2)
\label{4 - epsilon 2} .
\end{eqnarray}
Our earlier arguments show that $G_{\parallel}$ vanishes if
$\lambda_2$ does; however, to this order $\lambda^{\ast} _2$ can
be anything; hence, so can  $G_{\parallel}$, and so we get {\it no}
information about $z$ and $\zeta$ from this relation at all.
Likewise the recursion relation for $\Delta$ leads to
\begin{eqnarray}
z - \zeta - 2 \chi + 1 - d = -G_\Delta (\lambda_2, \lambda_1, \lambda_\rho,
\sigma_2)
\label{4 - epsilon 3}
\end{eqnarray}
with the right hand side again vanishing if $\lambda_2 = 0$, but taking on
any value
you like if $\lambda_2$ can be anything, as it can, to this order.

So what actually happens for $2 < d <4$?  Unfortunately, from our one loop
calculation, we cannot say, but can only enumerate the possibilities:
\begin{description}
\item[Possibility I:]\quad  At higher order, $\lambda_2$ proves to be
irrelevant, and flows to zero at the fixed point. In this case, the canonical
exponents (\ref{canon 1}) - (\ref{canon 3}) will hold, for all flocks, in all
$d$ in the range $2 < d <4$.

\item[Possibility II:]\quad ${d\lambda_2  \over  d\ell} = 0$ to all orders
(i.e., exactly).  In this case, there is a fixed line (or, more generally,
$D$-dimensional subspace with $D
\geq 1$) with exponents that vary as continuous functions of $\lambda_2$,
which can
take on any value.  Hence, the exponents $\chi$, $z$, and $\zeta$ will be
continuously
variable functions of the parameters in the ordered phase.  This behavior is
somewhat reminiscent of that of the $d = 2$ equilibrium $X-Y$ model,
although here it is occurring for an entire range of spatial dimension $2 <
d < 4$, and, furthermore, is {\it not} associated in any way with the
absence of true long ranged orientational order, since such order is
actually present in our model.

\item[Possibility III:]\quad ${d\lambda_2  \over  d\ell} > 0$ at higher
order, and the ordered phase is controlled by a new, $\lambda^{\ast} _2
\neq 0$ fixed point.  In this case, the exponents will
again be universal, but presumably, different from the
canonical ones (\ref{canon 1}) - (\ref{canon 3}).
Unfortunately, we have no idea what they will be.
\end{description}

We should emphasize that {\it all} flocks will be described by only one
of the three possibilities enumerated above (i.e., one can't have
different possibilities realized in different flocks).  Unfortunately, we
have no idea which of the above possibilities {\it is} realized for $2 <
d < 4$.




\section{Anisotropic Model}

Not all flocks, of course, are equally likely to move in any direction in
the space
they occupy. Flocks of birds, for instance, although they occupy a $ d=3$
dimensional volume (the air), are far more likely to move horizontally than
vertically. This is presumably because gravity breaks the rotational symmetry
between the horizontal plane and vertical directions.

 One can imagine a variety of ``microscopic'' rules, like the Vicsek rule
described
earlier, that would exhibit such anisotropy. For example, one could apply a
``Vicsek'' rule in three dimensions, selecting thereby a vector $ \hat{n}$.
Instead
of moving along that vector, however, one could instead move along a vector
``compressed'' along some $(z)$ axis:
\begin{eqnarray}
\vec{n}^{\prime}=s n_z\hat{z}+\vec{n}_{\perp}
\label{comp}
\end{eqnarray}
with  $s<1$  and  $\vec{n}_{\perp}=\hat{n}-n_z\hat{z}$.  This will tend to
promote motion in the $x-y$ plane at the expense of motion in the
$z-$direction.
Alternatively, one could project all velocities into the $x-y$ plane, apply
a Vicsek
rule to them (while still sampling neighbors in three dimensions), and then
add to
this $xy$ move a random decorrelated step in the
$z$ direction\cite{Steph}.

For technical reasons that will, we hope, become obvious, we'll focus
our attention on systems which, whatever their spatial dimension $d$, have an
easy {\it plane} of motion; i.e., two components of velocity that are {\it
intrinsically} favored over the other $d-2$. We'll also assume perfect
isotropy {\it
within} this plane {\it and} within the $d-2$ dimensional ``hard''
subspace. The
case of birds flying horizontally corresponds to $d=3.$

A natural extension of our fully isotropic model (EOM) to this case is
\begin{eqnarray}
\partial_t\vec{v} + \lambda_1 \left(\vec{v} \cdot
\vec{\nabla}\right)\vec{v} + \lambda_2\left(\vec{\nabla}
\cdot \vec{v}\right)\vec{v} +
\lambda_3\vec{\nabla}\left(\left|\vec{v}\right|^2\right) =
\nonumber
\\    -\vec{\nabla}P(\rho) + \alpha\vec{v} -
\beta\left|\vec{v}\right|^2\vec{v}-\delta\alpha\vec{v}_H +
D_B\vec{\nabla}\left(\vec{\nabla}
\cdot \vec{v}\right) + D^e_T\nabla^2_e\vec{v} \nonumber \\ +
D^H_T\nabla^2_H\vec{v} + D_2\left(\vec{v}
\cdot \vec{\nabla}\right)^2\vec{v}+\vec{f}
\label{Aniso High}
\end{eqnarray}
Mass conservation, of course, still applies:
\begin{eqnarray}
\partial_t \rho + \vec{\nabla} \cdot \left(\rho\vec{v}\right) = 0
\label{Aniso cons}
\end{eqnarray}
and the pressure $P(\rho)$ will still be given by the same expansion in
$\delta\rho=\rho -\rho_o$
\begin{eqnarray}
P(\rho)=\sum^\infty_{n=1} \sigma_n\left(\delta\rho\right)^n \quad .
\label{Aniso P}
\end{eqnarray}
In equation (\ref{Aniso High}), $\vec{v}_H$ denotes the $d-2$ ``hard''
components of
$\vec{v}$,  i.e., those {\it orthogonal} to the $d=2$ easy plane.  Likewise,
$\nabla^2_e$ and $\nabla^2_H$ denote the operators
$\sum^2_{i=1} {\partial^2 \over \partial x^2_i}$ and
$\sum^d_{i=3} {\partial^2 \over \partial x^2_i}$, respectively, where
$i=1,2$ are the ``easy'' Cartesian directions, and $i = 3 \rightarrow d$
the ``hard''
ones. The term
$-\delta\alpha \left|\vec{v}_H\right|^2$, $\delta\alpha>0$ suppresses these
components relative to those in the easy plane.

Equation (\ref{Aniso High}) is not, of course, the most general anisotropic
model we could write down. For instance, one could have anisotropy in
the non-linear terms: e.g, terms like $\left(\vec{v}_e \cdot
\vec{\nabla}\right)\vec{v}_e$ could have different coefficients than
$\left(\vec{v}_H
\cdot
\vec{\nabla}\right)\vec{v}_H$. However, because $\vec{v}_H$ winds up being
``massive,'' in the sense of decaying to zero too rapidly (ie,
non-hydrodynamically) at long wavelengths and times to non-linearly affect the
hydrodynamic (long wavelength, long time) behavior of the flock (in its low
temperature phase), any additional terms in (\ref{Aniso High}) distinguishing
$\vec{v}_H$ and $\vec{v}_e$ will have no effect on the hydrodynamic behavior
in the low ``temperature'' phase. That is, (\ref{Aniso High}) already contains
enough anisotropy to generate all possible relevant, symmetry allowed terms in
the broken symmetry state. Hence, we will keep things simple and not generalize
(\ref{Aniso High}) further.

As we did for the isotropic problem, we will now break the symmetry
of this model, i.e., look for solutions to the form:
\begin{eqnarray}
\vec{v}\left(\vec{r},t\right)=\left<\vec{v}\right> +
\delta\vec{v}\left(\vec{r},t\right)
\label{Anisosymbreak}
\end{eqnarray}

Now, however, the direction of the mean velocity $\left<\vec{v}\right>$ (which
we'll chose as before, to be a static, spatially uniform solution of the
noiseless
$\left(\vec{f}=0\right)$ version of (\ref{Aniso High}) ) is not arbitrary,
but must
lie in the easy $(1,2)$ plane. To see this, let's, without loss of
generality, write
\begin {eqnarray}
\left<\vec{v}\right> = v_{oy} \hat{y} + v_{oz}\hat{z}
\label{}
\end{eqnarray}
with $v_{oy}$ and $v_{oz}$ constants, $\hat{y}$ in the easy plane and $\hat{z}$
one of the $d-2$ ``hard'' directions. To solve (\ref{Aniso High}) with
$\vec{f}=0$,
these must obey
\begin{eqnarray}
\alpha v_{oy}-\beta \left(v^2_{oy}+v^2_{oz}\right) v_{oy}=0
\label{voy fix}
\end{eqnarray}
and
\begin{eqnarray}
\left(\alpha-\delta\alpha\right) v_{oz}-\beta \left(v^2_{oy}+v^2_{oz}\right)
v_{oz}=0
\label{voz fix}
\end{eqnarray}
Subtracting $v_{oy} \times$ (\ref{voz fix}) from $v_{oz} \times$
(\ref{voy fix}) we obtain
$\delta\alpha v_{oy} v_{oz}=0$, which implies that either $v_{oy}$ or $v_{oz}$
must be zero. It is straightforward to show that the former solution is
unstable
(with two linear eigenvalues $\alpha>0$) to small $\vec{v}_e$ fluctuations,
while
the latter is stable (with $d-2$ linear eigenvalues $-\delta\alpha < 0$) to
$\vec{v}_H$ fluctuations, so the solution with $\left<\vec{v}\right>$ in
the easy
plane is the stable one. Furthermore, fluctuations in the ``hard''
directions are
``massive,'' in the sense of decaying rapidly to zero even at long
wavelengths, and
so can be neglected in the low temperature phase (just like $v_{\parallel}$
fluctuations in the isotropic case). Likewise, if we take
\begin{eqnarray}
\left<\vec{v}\right> = v_{o} \hat{y}
\label{broken}
\end{eqnarray}
fluctuations in $\delta v_y = v_y - v_{o}$, will also be massive (with linear
eigenvalue$-2\alpha$). Eliminating the massive fields $\delta v_y$ and
$\vec{v}_H$ in favor of the pressure, as we did for $\delta v_{\parallel}$
in the
isotropic case, gives
\begin{eqnarray}
\delta v_y = - D_{\rho y} \partial _y \rho
\label{vy elim}
\end{eqnarray}
\begin{eqnarray}
\vec{v}_H = - D_{\rho H}\vec{\nabla} _H \rho \quad ,
\label{vh elim}
\end{eqnarray}
where we've defined the diffusion constants
\begin{eqnarray}
D_{\rho y} \equiv {\sigma _1 \over 2\alpha}
\label{D rho y 1}
\end{eqnarray}
\begin{eqnarray}
D_{\rho H} \equiv {\sigma _1 \over \delta\alpha}
\label{D rho y 2}
\end{eqnarray}
and we've used the relation (\ref{Aniso P}) for the pressure, and dropped
all but
the leading order linear terms in $\delta \rho$, since higher powers of $\delta
\rho$ in Eqns. (\ref{vy elim}) and (\ref{vh elim}) prove to be
irrelevant.

Using the solutions (\ref{vy elim}) and (\ref{vh elim}), and taking, for
the reasons
just discussed,
\begin{eqnarray}
\vec{v} \left( \vec{r} \right) = \left(v_{o} + \delta v_y \left(\vec{r}, t
\right) \right) \hat{y} + v_x \left(\vec{r}, t \right) \hat{x} + \vec{v}_H
\left(\vec{r}, t
\right)
\label{v expand}
\end{eqnarray}
we can write a closed system of equations for $v_x\left(\vec{r}, t \right)$ and
$\delta \rho \left(\vec{r}, t \right)$:
\begin{eqnarray}
\partial _t \delta \rho + v_{o} \partial _y \delta \rho + \partial_x
\left(
\rho v_x \right) = \left(D_{\rho y} \partial ^2_y + D_{\rho H} \nabla^2_H
\right) \delta \rho
\label{Aniso EOM}
\end{eqnarray}
\begin{eqnarray}
\partial _t v_x +\gamma \partial _y v_x + {\lambda \over
2} \partial _x \left(v^2_x\right) &=&
- \sigma_1 \partial_x \left(\delta
\rho
\right)  - \sigma_2 \partial_x \left(\delta
\rho
\right)^2\nonumber \\
&+&
\left(D_{\parallel} \partial^2_y + D_x \partial^2_x + D_H \nabla^2_H
\right)v_x + f_x
\label{aniso broken r}
\end{eqnarray}
where we've defined $\lambda \equiv \lambda_1 + \lambda_2$,
and $\gamma = \lambda_1 v_0$, and dropped irrelevant terms.

Proceeding as we did in the isotropic model, we begin by linearizing these
equations, Fourier transforming them, and determining their mode structure.

The result of the first two steps is the Fourier transformed equations of
motion
\begin{eqnarray}
\left[- i \left(\omega - v_{o} q_y\right) + \Gamma _{\rho}
\left(\vec{q}\right)
\right]
\delta\rho \left(\vec{q}, \omega \right) + iq_x \rho_0 v_x \left(\vec{q},
\omega
\right) = 0
\label{FT EOM rho}
\end{eqnarray}
\begin{eqnarray}
\left[- i \left(\omega - \gamma q_y\right) + \Gamma _v
\left(\vec{q}\right)
\right] v_x
\left(\vec{q}, \omega \right) + i \sigma_1q_x\delta\rho \left(\vec{q}, \omega
\right) = f_x \left(\vec{q}, \omega \right)
\label{FT EOM v}
\end{eqnarray}
where we've defined
\begin{eqnarray}
\Gamma _\rho \left(\vec{q} \right) \equiv D_{\rho y} q^2_y + D_{\rho H}q^2_H
\label{Gamma rho}
\end{eqnarray}
\begin{eqnarray}
\Gamma _v \left(\vec{q} \right) \equiv D_{\parallel} q^2_y + D_H q^2_H +
D_x q^2_x \quad .
\label{Gamma v}
\end{eqnarray}

Again as in the isotropic model, we first determine the eigenfrequencies
$\omega
\left(\vec{q} \right)$ of these equations, finding
\begin{eqnarray}
\omega_{\pm} \left(\vec{q} \right) = c_{\pm}\left(\theta_{\vec{q}}, \phi
_{\vec{q}}
\right) q -i \epsilon_{\pm} \left( \vec{q}\right)
\label{omega aniso}
\end{eqnarray}
where the sound speeds
\begin{eqnarray}
c_{\pm} \left(\theta_{\vec{q}} ,  \phi _{\vec{q}}\right) = {1 \over 2}
\left(\gamma + v_{0}\right)  \cos \theta_{\vec{q}} \pm c_2
\left(\theta_{\vec{q}} ,
\phi _{\vec{q}}\right)
\label{canis}
\end{eqnarray}
with
\begin{eqnarray}
c_2 \left(\theta_{\vec{q}} ,  \phi _{\vec{q}}\right) \equiv \sqrt{{1
\over 4}
\left(\gamma - v_0 \right)^2 \cos ^2 \theta_{\vec{q}} +
\sigma_1 \rho _0 \sin^2 \theta_{\vec{q}}  \cos^2   \phi _{\vec{q}}}
\quad ,
\label{c 2}
\end{eqnarray}
where $\theta_{\vec{q}}$ is the polar angle between $\vec{q}$ and the $y$-axis,
and $\phi _{\vec{q}}$ is the azimuthal angle, measured relative to the
$x$-axis:
i.\ e., the angle between the {\it projection} of $\vec{q}$ orthogonal to
$y$, and
the $x$-axis.

A polar plot of this sound speed versus $\theta_{\vec{q}}$ for $ \phi
_{\vec{q}} =
0$ (i.\ e., $\vec{q}$ in the ``easy'' (i.\ e., $x-y$) plane) looks {\it
exactly} like that
for the isotropic model (figure 2).  Indeed, {\it any} slice with fixed
{$ \phi
_{\vec{q}}$ looks qualitatively like that figure, although, as $\phi _{\vec{q}}
\rightarrow {\pi \over 2}$ (i.\ e., as $\vec{q}_{\perp}$, the projection of
$\vec{q}$ orthogonal to $y$, approaches orthogonality  to the $x$-axis),
the sound velocity profile becomes two circles with their centers on the $y$
axis and both circles passing through the origin.

The dampings $\epsilon_{\pm} \left(\vec{q}\right)$ in (\ref{omega aniso}) are
$0 \left(q^2\right)$, and given by
\begin{eqnarray}
\epsilon_{\pm} \, \left(\vec{q}\right) = &\pm& {c_{\pm}
\left(\theta_{\vec{q}} ,
\phi _{\vec{q}}\right) \over 2c_2 \left(\theta_{\vec{q}} ,  \phi
_{\vec{q}}\right)}
\left(\Gamma _v \left(\vec{q}\right) +  \Gamma _{\rho} \left(\vec{q}\right)
\right)\nonumber\\
&\mp& {v_{0} \cos \left(\theta_{\vec{q}} \right) \over 2c_2
\left(\theta_{\vec{q}}
,  \phi _{\vec{q}}\right)}
\left(\Gamma _v \left(\vec{q}\right) +  {\gamma \over v_{0}}
\Gamma _{\rho}
\left(\vec{q}\right)
\right)
\label{Aniso damp}
\end{eqnarray}

Note that, unlike the isotropic problem in $d > 2$, here there are no
transverse
modes in {\it any} $d$:  we always have just two longitudinal Goldstone modes
associated with $\delta\rho$ and $v_x$.

We can now again parallel our treatment of the isotropic model and
calculate the
correlation functions and propagators.  The calculation is so similar that
we will
not repeat the details, but merely quote the results:
\begin{eqnarray}
G_{vv} \left(\vec{q}, \omega \right) = {i \left(\omega - v_{o} q_y \right) -
\Gamma_\rho \left(\vec{q}\right)
\over Den \left(\vec{q}, \omega \right) }
 \quad ,
\label{Aniso Prop vv}
\end{eqnarray}
\begin{eqnarray}
G_{v \rho} \left(\vec{q}, \omega \right)  = {i \sigma_1 q_x  \over Den
\left(\vec{q}, \omega \right) }
\quad , \label{Aniso Prop vrho}
\end{eqnarray}
\begin{eqnarray}
G_{\rho v}  \left(\vec{q}, \omega \right) = {i \rho _0 q_x  \over Den
\left(\vec{q},
\omega \right) }
\quad , \label{Aniso Prop rhov}
\end{eqnarray}
\begin{eqnarray}
G_{\rho \rho}  \left(\vec{q}, \omega \right) = {i \left(\omega -
\gamma q_y \right) - \Gamma_v
\left(\vec{q}\right)  \over Den \left(\vec{q}, \omega \right) }
\quad , \label{Aniso Prop rhorho}
\end{eqnarray}
\begin{eqnarray}
C_{vv} \left(\vec{q}, \omega \right) = {\Delta \left[ \left(\omega -  v_{o} q_y
\right)^2
+ \Gamma^2_{\rho}
\left(\vec{q}\right)\right]  \over \left| Den \left(\vec{q}, \omega
\right)\right|^2 }
\quad , \label{Aniso Corr vv}
\end{eqnarray}
\begin{eqnarray}
C_{\rho v}  \left(\vec{q}, \omega \right) = \left< \delta\rho
\left(\vec{q}, \omega
\right) v_x \left(-\vec{q}, - \omega \right)\right> =
{\Delta \sigma_1 q_x \left(\omega - v_{o} q_y - i\Gamma_\rho
\left(\vec{q}\right)\right) \over \left| Den
\left(\vec{q},
\omega
\right)\right|^2 }
\quad ,
\label{Aniso Corr rhov}
\end{eqnarray}
and
\begin{eqnarray}
C_{\rho \rho}\left(\vec{q}, \omega \right) = { \Delta \rho_0^2 q_x^2
\over \left| Den
\left(\vec{q},
\omega
\right)\right|^2} \quad ,
\label{Aniso Corr rhorho}
\end{eqnarray}
where we've defined
\begin{eqnarray}
Den \left(\vec{q},
\omega
\right) &=& \left(\omega - c_+ \left(\theta_{\vec{q}},
\phi_{\vec{q}}\right)q\right)
\left(\omega - c_- \left(\theta_{\vec{q}},
\phi_{\vec{q}}\right)q\right) \nonumber\\ &+&
i \left[\omega\left(\Gamma_{\rho}
\left(\vec{q}\right) + \Gamma_v
\left(\vec{q}\right)
\right) - q_y \left( v_{o} \Gamma_v \left(\vec{q}\right) +
\gamma
\Gamma_{\rho} \left(\vec{q}\right)
\right)
\right]
\label{Aniso Den}
\end{eqnarray}
which, of course, implies
\begin{eqnarray}
\left|Den \left(\vec{q},
\omega
\right)\right|^2 &=& \left(\omega - c_+ \left(\theta_{\vec{q}},
\phi_{\vec{q}}\right)q\right)^2
\left(\omega - c_- \left(\theta_{\vec{q}},
\phi_{\vec{q}}\right)q\right)^2\nonumber\\ &+&
\left[\omega\left(\Gamma_{\rho}
\left(\vec{q}\right) + \Gamma_v
\left(\vec{q}\right)
\right) - q_y \left(v_{o}\Gamma_v \left(\vec{q}\right) +
\gamma
\Gamma_{\rho} \left(\vec{q}\right)
\right)
\right]^2
\label{Aniso Den 2}
\end{eqnarray}
These horrific expressions actually look quite simple when plotted as a
function of
$\omega$ at fixed $\vec{q}$; indeed, such a plot of $C_{vv}$ looks
precisely like the solid line in
figure 6:  two asymmetrical peaks, centered at $\omega = c_\pm
\left(\theta_{\vec{q}},
\phi_{\vec{q}}\right) q$, with widths $\epsilon _{\pm} \left(\vec{q}\right)
\propto q^2$.

Note that, at this linear order, everything {\it scales} as it did in the
isotropic
problem:  peak positions $\propto q$, widths $\propto q^2$, and heights
$\propto {1 \over q^4}$.

Continuing to blindly follow the path we trod for the isotropic problem, we can
calculate the equal-time $v_x-v_x$ correlation function:
\begin{eqnarray}
C_{vv} \left(\vec{q}\right) &\equiv& \left< v_x \left(\vec{q}, t \right) v_x
\left(-\vec{q}, t \right)\right>\nonumber\\
&=&\int^{\infty}_{- \infty} {d\omega \over 2 \pi}C_{vv} \left(\vec{q},
\omega\right)
\nonumber\\
&=& {\Delta \over 2} \, {\phi \left(\hat{q}\right) \over \Gamma_L
\left(\vec{q}\right) }
\label{}
\end{eqnarray}
where $\phi \left(\hat{q}\right)$ depends {\it only} on the direction
$\hat{q}$ of
$\vec{q}$, and is given by
\begin{eqnarray}
\phi\left(\hat{q}\right) = {1 \over c_2 \left(\theta_{\vec{q}}, \phi
_{\vec{q}}\right) q} &&\left[{\left(c_+
\left(\theta_{\vec{q}}, \phi _{\vec{q}}\right) q - v_{o} q_y
\right)^2
\over
c_+\left(\theta_{\vec{q}}, \phi _{\vec{q}}\right) q - v_{o} q_y +
\left(c_+\left(\theta_{\vec{q}}, \phi _{\vec{q}}\right) q - \lambda_1
v_{o}
q_y\right){\Gamma_\rho \over \Gamma_L}}\right.\nonumber\\
&+& \left. {\left(c_-\left(\theta_{\vec{q}}, \phi _{\vec{q}}\right) q -
v_{o} q_y
\right)^2
\over
c_+\left(\theta_{\vec{q}}, \phi _{\vec{q}}\right) q - v_{o} q_y +
\left(c_-\left(\theta_{\vec{q}}, \phi _{vec{q}}\right) q - \lambda_1 v_{o}
q_y\right){\Gamma_\rho \over \Gamma_L}}\right]
\label{chapter 4}
\end{eqnarray}
These fluctuations again diverge like ${1 \over q^2}$ as $\left| \vec{q}\right|
\rightarrow 0$, just as in the isotropic problem.

This completes our abbreviated discussion of the linearized theory of the
anisotropic model.  The most succinct summary of this linearized theory is that
everything scales just as it did in the isotropic problem.  This implies
that the
non-linearities (i.\ e., the $\lambda$ and $\sigma_2$ terms in the equations of
motion (\ref{Aniso EOM})) become relevant in and below the same upper
critical dimension
$d_{uc} = 4$ as in the isotropic problem.  For $d < 4$, therefore, these
non-linearities will change the long-distance behavior of the anisotropic
model.
We will now treat these non-linearities using renormalization group arguments
similar to those we used for the isotropic model in $d= 2$.  Now, however, they
will work for all $d$ between $2$ and $4$.

Notice that all of the non-linearities in (\ref{Aniso EOM}) are total
$x$-derivatives, just as in the $d=2$ case for the isotropic problem. Now,
however, this is true in {\it all} spatial dimensions, not just in $d=2$.
(This, of
course, is the reason we chose to consider precisely two ``soft''
components). Thus, we will now be able to derive exact exponents in this
model for all spatial dimensions. We will not go through the arguments in
detail,
as they are virtually identical to those in the $d=2$ case for the
isotropic model,
but will simply quote the conclusions:
\begin{enumerate}
\item There are {\it no} graphical corrections to {\it any} of the diffusion
constants in (\ref{Aniso EOM}) except $D_x$.
\item The stable fixed point that controls the ordered phase {\it must} have
$\lambda^*_\rho\neq 0$ at least for $\lambda(0)<\lambda_\rho(0)$, which is a
finite fraction of all flocks, and
\item  $\Delta$ and $\lambda_\rho$ are not graphically renormalized.
\end{enumerate}
Point one suggests that, in constructing our dynamical renormalization
group for
(\ref{Aniso EOM}), we should scale the $x$-direction differently from {\it
both}
the $y$-direction and the $d-2$ hard directions. Furthermore, since {\it both}
the $y$-direction {\it and} the $d-2$ hard directions are alike in having their
associated diffusion constants unrenormalized, we should scale these directions
the {\it same} way. Therefore, in our renormalization group, we will rescale as
follows:
$x\rightarrow bx$,  $(y, \vec{x}_H)\rightarrow
b^\zeta(y,\vec{x}_H)$,
$\label{Aniso scale}$ $t\rightarrow b^zt$.
 With these rescalings, the recursion relations for $D_i$, $i\neq  x,\rho$,
$\Delta$, and
$\lambda_{\rho}$ become:
\begin{eqnarray}
{dD_i \over dl}=(z-2\zeta)D_i \quad , \quad       (i\neq x)
\, ,
\label{Aniso vec}
\end{eqnarray}
\begin{eqnarray}
{d\Delta\over dl}=[z-2\chi+(1-d)\zeta-1]\Delta
\label{Aniso Delta vec}
\end{eqnarray}
\begin{eqnarray}
{d\lambda_{\rho}\over dl}=(\chi+ z -1) \lambda_{\rho}
\label{Aniso lambda vec}
\end{eqnarray}
All three relations are exact, since none of these parameters
experiences any graphical renormalization. As in the isotropic case,
we want all of these parameters to flow to fixed points; this leads to
three exact scaling relations between the three exponents $\chi$,
$z$, and $\zeta$:
\begin{eqnarray}
  z=2\zeta
\label{Aniso Exact 1}
\end{eqnarray}
\begin{eqnarray}
  z-2\chi+(1-d)\zeta=1
\label{Aniso Exact 2}
\end{eqnarray}
\begin{eqnarray}
\chi=1- z ,
\label{Aniso Exact 3}
\end{eqnarray}
hose solution is easily found in all $d<4$:
\begin{eqnarray}
\zeta={3 \over {7-d}}
\label{Aniso exact zeta}
\end{eqnarray}
\begin{eqnarray}
z={6 \over 7-d}
\label{Zanis}
\end{eqnarray}
\begin{eqnarray}
\chi={1-d \over {7-d}}  \, .
\label{Aniso exact chi}
\end{eqnarray}
Note that these reduce to our isotropic results in $d=2$, as they should,
since the
two models are identical there. They also reduce to the harmonic values $z=2$,
$\zeta=1$, and $\chi=-1$, in $d=4$, as they should, since $4$ is the upper
critical
dimension.

In the physically interesting case of $d=3$, we obtain:
\begin{eqnarray}
\zeta={3 \over 4}
\label{zeta d=3}
\end{eqnarray}
\begin{eqnarray}
z={3 \over 2}
 \label{zd = 3}
\end{eqnarray}
\begin{eqnarray}
\chi=-{1 \over 2}
\label{chi d=3}
\end{eqnarray}
As in the isotropic case, we can use scaling arguments here to show
that the effect of the non-linearities can be fully incorporated by
simply replacing
$D_x$ everywhere it appears in the linearized expressions by the
divergent, wavevector dependent scaling form:
\begin{eqnarray}
D_{x} \left(\vec{q}\right) = q^{z-2}_x f\left[{\left({q_y \over
\Lambda}\right) \over \left({q_x \over
\Lambda}\right)^\zeta}\, , {\left({q_H\over\Lambda}\right) \over \left({q_x
\over
\Lambda}\right)^\zeta}\right] \quad .
\end{eqnarray}
Doing this leads to all of the scaling laws for this anisotropic
problem quoted in the introduction.




\section{Testing the Theory in Simulations and Experiments}

In this section, we discuss how our theory can be tested in
simulations and direct observations of real flocks. The ``real'' flocks
may include, e.g., mechanical, self-propelled ``go carts'' packed so
densely that they align with their neighbors \cite{Berk}, as well as
aggregates of genuinely living organisms.

We begin with a few
suggestions about the best boundary conditions and parameter
values for simulations or experiments, and then describe how the
correlation functions and scaling exponents $\chi$, $z$ and
$\zeta$ predicted by our theory can be measured.  The most useful
boundary conditions are ``torus'' conditions; that is, reflecting walls
in $d - 1$ directions, and periodic boundary conditions in the
remaining direction, call it $y$ (see figure 9). The advantage of
these conditions is that one knows a priori that, if the flock does
spontaneously order, its mean velocity will necessarily be in the
periodic ($y$) direction.

It might be objected that imposing such
anisotropic boundary conditions breaks the rotation invariance our
model requires, but this is not, in fact, the case. A ``bird'' deep inside
the box moves with no special direction picked out a priori; it can
{\it only} find out about the breaking of rotation invariance on the
boundary if the bulk of the flock spontaneously develops long range
order. This is precisely analogous to the way one speaks of a
ferromagnet as spontaneously breaking a continuous symmetry
even if it orders in the presence of ordered boundary conditions.

So, by imposing these boundary conditions, we know the direction of the
flock motion (the $y$ direction in the simulation), and, therefore, have
oriented the simulation axes with the axes used in our theoretical
discussion; i.e., our $\parallel$ axis equals the simulation's periodic
direction.

Alternative boundary conditions add
the additional complication of having to first determine the
direction of mean flock motion before calculating correlation
functions. This complication is even worse for a finite flock (as any
simulation must treat), since the mean direction of motion will
wander, executing essentially a random walk that will explore the
full circle in a time of order $T_{flock}=2\pi\sqrt{{N \over\Delta}}$.
Our results, which assume a {\it constant} direction of flock motion,
will only apply for time scales $t<<T_{Flock}$. Even drifts of the
mean flock direction through angles $<<2\pi$ can cause problems,
however, since most of the interesting scaling behavior is
concentrated in a narrow window of angles
$q_{\parallel} \sim q^\zeta_{\perp} \gg q_{\perp}$; i.e., near
the direction of mean flock motion. So this drift greatly complicates
the experimental analysis, and is best avoided by using the toroidal
boundary conditions just described.

Of course, it is considerably
harder to produce these boundary conditions in a real experiment.
Ants walking around a cylinder may come close, although gravity
will always break rotation invariance on a real cylinder. Perhaps the
experiment could be done on the space shuttle, or with a rapidly
spinning cylinder producing artificial gravity that swamps real
gravity, or by using neutrally buoyant organisms in a fluid.
Alternatively, one could use a ``track'' such as that shown in figure
10, and take data only from the cross-hatched region, chosen to be
in the middle of the straight section of the track, far from the
curves.

Other, more ingenious ways to pre-pick the direction of
mean motion through boundary conditions may also be dreamed up
by experimentalists more clever than we are.

We strongly caution anyone attempting to test our results, however, that
it is {\it only} through boundary conditions that one may prepick the
direction of mean motion. {\it Any} approach that prepicks this direction
in the bulk of the flock, such as giving each bird a compass, letting them
be blown by a wind, or run downhill, or follow a chemical scent, etc., will
lead to a model {\it outside} the universality class of our isotropic model,
since the starting model does not have any rotation invariance to be
spontaneously broken (unless the anisotropy leaves an ``easy plane''
in which all directions are equivalent, in which case our anisotropic
model of section VI applies).  Indeed, such flocks of ``birds with
compasses'' will be less interesting than the models we've
studied here, since the ``compass'' will introduce a ``mass'' that
makes any fluctuation away from the pre-picked direction of flock
motion decay rapidly (i.e., non-hydrodynamically)  with time. In
such a model, it's easy to show that the non-linearities are
irrelevant, and there are no interesting fluctuations left at long
distances and times.

And now a few words about parameter choices. For definiteness, we
will discuss in what follows the Vicsek model, whose parameters
are ${v_0}={S \over R_0}$, where $S$ is the distance the birds travel
on each time step and $R_0$ is the radius of the circle of neighbors, the
mean number density $\rho_0$ in units of ${1 \over R_0^d}$ where d is
the dimension of the system, and the noise strength $\Delta$, which
is the mean squared angular error. Since the interesting non-linear
effects in our model come from terms proportional to $v_0^2$, those
effects will become important at shorter length scales in a faster
moving flock. That is, in, e.g., the Vicsek model, should we chose the
dimensionless velocity as large as possible, {\it consistent} with the
flock ordering. However, if we take $v_0$ too big, i.e., $v_0 \gg1$,
then, on each time step, each bird is likely to have a completely
different set of neighbors. It is difficult to see how order can
develop in such a model. So, to take $v_0$ as big as possible without
violating $v_0 \gg 1$, we should chose
$v_0 \sim 1$. The simulations of Vicsek et\ al.\  \cite{Vicsek} took
$v_0
\ll1$, and, hence, probably never explored (in their finite flocks) the
long length scale regime in which our non-linear effects become
important.

Now to the mean density $\rho_0$, which is, of course,
just determined by the total number of birds $N$ and the volume
$V$ of the box via $\rho_0={N \over V}$. We clearly want
this to be large enough that each bird usually finds some neighbors
in its neighbor sphere:  This means we want $\rho_0 R_0^d \geq O(1)$.
However, if we make $\rho_0$ too large, each bird has so many
neighbors that a simulation is considerable slowed down, since the
``direction picking'' step of the Vicsek algorithm takes a time
proportional to the number of neighbors (because we've got to
average their directions).  Thus, for simulations, one wishes to
choose $\rho_0$ as {\it small} as possible, consistent, again, with
getting good order.

Finally, we consider the noise
$\Delta$. Here again, to see our fluctuation effects, we want
$\Delta$ as big as possible. However, if $\Delta$ is too big, the flock
won't order. Furthermore, even if $\Delta$ is small enough that the
flock {it does} order, we want also to be sure that we are well below
the critical value $\Delta_c$ of $\Delta$ at which the flock
disorders.  Otherwise, for distances smaller than the correlation
length $\xi$ associated with the order-disorder transition, the
scaling properties of the flock will be controlled by the fixed point
that controls the order-disorder transition, {\it not} the low
temperature fixed point we have studied here.

If this transition is continuous, as it appears to be in Vicsek's
simulations \cite{Vicsek}, this correlation length diverges as
$\Delta \rightarrow \Delta^-_c$.  Thus, to observe scaling behavior
we predict over as many decades of length scale as possible, we
want to choose $\Delta$ substantially less than $\Delta_c$, but as
big as possible consistent with this (to maximize fluctuation
effects).  Choosing $\Delta$ to be a little below the point at which
the mean velocity $\left<\vec{v}\right>$  starts to ``saturate''
seems like a fairly good compromise between these two
competing effects.  Similar considerations apply for choosing the
optimal $\rho_0$, and $v_0$, which we want to be as small or
big, respectively, as they can be without substantially
suppressing long ranged order.  The best choices will
probably lead to all three parameters $\rho_0$, $v_0$ and
$\Delta$ begin, in suitably dimensionless units, $0(1)$.

Having chosen the appropriate parameter values and boundary
conditions, what should an experimentalist or simulator measure to
test our theory?

We have already discussed a number of such measurements in the
introduction; namely, the spatially Fourier transformed equal-time
and spatio-temporally Fourier transformed unequal time
density-density correlation functions $C\rho \left(\vec{q}\right)$
and  $C\rho \left(\vec{q}, \omega\right)$, respectively.  Our
predictions for these are given in Eqns. (\ref{rho scale 1}) and
(\ref{z2}).

One additional correlation function that can be measured quite
easily is the mean squared {\it lateral} displacement of a bird
\begin{eqnarray}
w^2(t) \equiv \left<\left|\vec{x}^{\perp}_i (t) -
\vec{x}^{\perp}_i (0) \right|^2
\right>
\label{RW 1}
\end{eqnarray}
{\it perpendicular} to the mean direction of motion of the flock.
This can easily be measured as a function of time in a simulation or
experiment simply by labeling a set of $n$ birds in a ``strip'' near
the center of the channel with its long axis running parallel to the
mean direction of bird motion (see fig.11) and then following their
subsequent motion.  It is best to center the strip in the channel so as
to postpone the birds reaching the reflecting walls as long as
possible.  Once they {\it do} reach the walls, of course $w^2(t
\rightarrow \infty)$ saturates at ${\sim L^2_{\perp}}$, $L_{\perp}$ being
the width of the channel. We will deal in the following discussion
with times much smaller than that required for a bird at the center
of the channel to wander out to its edge.   Since the mean
$x_{\perp} -
$ position $\vec{x}^{\perp}_i$ of each bird obeys
\begin{eqnarray}
\vec{x}^{\perp}_i (t) =  \vec{x}^{\perp}_i (0) + \int^t_0
\vec{v}^{\perp}_i (t) dt
\label{RW 2}
\end{eqnarray}
where $\vec{v}^{\perp}_i (t)$ is the ${\perp}$ velocity of the $i$'th
bird  at time $t$, the mean width is given by
\begin{eqnarray}
w^2 (t) = \int^t_0 d t^{\prime} \int^t_0 dt^{\prime\prime}
\left< \vec{v}^{\perp}_i (t^{\prime}) \cdot  \vec{v}^{\perp}_i
(t^{\prime\prime})\right> \quad .
\label{RW 3}
\end{eqnarray}
Now we need to relate the velocity of the $i$'th bird to the position
and time dependent velocity field $\vec{v}_{\perp}\left(\vec{r}, t
\right)$.  This is easily done:
\begin{eqnarray}
\vec{v}^{\perp}_i (t) = \vec{v}_{\perp} \left( \vec{r}_i (t), t\right)
\label{RW 5}
\end{eqnarray}
where $\vec{r}_i (t)$ is the position of the $i$'th bird at time $t$.
This is given by
\begin{eqnarray}
\vec{r}_i (t) = \vec{r}_i(0) + \bar{v}t \hat{x}_{\parallel} +
\delta x^{\parallel}_i(t) \hat{x}_{\parallel} + \vec{\delta x}^{\perp}_i
(t)
\label{RW 6}
\end{eqnarray}
where
\begin{eqnarray}
\bar{v} \equiv {1 \over N}\left|{\sum_i \vec{v}_i}\right|
\label{RW 6a}
\end{eqnarray}
is the velocity {\it averaged over all birds}, which, as discussed
earlier, is {\it not} to be confused with the space averaged $v_0
\equiv \left| \int \vec{v}\left(\vec{r}, t \right)d^dr\right|$ that
appears in the expression for the sound speeds
$c_{\pm} \left(\theta_{\vec{q}}\right)$.  This distinction proves to
be {\it crucial} here, as we shall see in a moment.  In (\ref{RW 6}),
$\delta x^{\parallel}_i(t)$ and $\delta x^{\perp}_i(t)$ reflect the
motion of the $i$'th individual bird relative to the mean motion of
the flock (at speed $\bar{v}$).

Using (\ref{RW 6}), we see that the desired single bird
autocorrelation function in (\ref{RW 3}) is
\begin{eqnarray}
\left< \vec{v}^{\perp}_i \left(t^{\prime}\right)  \cdot
\vec{v}^{\perp}_i \left(t^{\prime\prime}\right) \right> =
\nonumber\\
\left< \vec{v}_{\perp}\left(\vec{r}_i(0) +
\left(\bar{v}t^{\prime} +
\delta x_{\parallel}\left(t^{\prime}\right)
\right)\hat{x}_{\parallel} +
\vec{\delta x}_{\perp}\left(t^{\prime}\right),t{\prime}\right)
\cdot \vec{v}_{\perp} \left(\vec{r}_i(0) +
\bar{v}t^{\prime\prime} +
\delta x_{\parallel}\left(t^{\prime\prime}\right)
\hat{x}_{\parallel} +
\vec{\delta x}_{\perp}\left(t^{\prime\prime}\right),
t^{\prime\prime} \right)\right> \nonumber \\
= C_c \left(\vec{x}_{\perp} \left(t^{\prime}\right) -
\vec{x}_{\perp}
\left(t^{\prime\prime}\right), \bar{v} | t^{\prime} -
t^{\prime\prime} | + \delta
x_{\parallel}\left(t^{\prime}\right) - \delta
x_{\parallel}\left(t^{\prime\prime}\right), t^{\prime} -
t^{\prime\prime}\right)
\label{RW 7}
\end{eqnarray}
where $C_c\left(\vec{R}, t\right)$ is the real space velocity field
auto-correlation function defined in the introduction.

We assume (and will verify a posteriori) that both $\delta x_{\parallel}$
and
$\vec{\delta x}_{\perp}$ are small enough compared to the average
motion $\bar{v}t \hat{x}_{\parallel}$ that their effect on the
velocity-velocity autocorrelation in (\ref{RW 7}) is negligible.
For now neglecting them, we see that we are left with the task of
evaluating $C_v \left(\vec{R}_{\perp} = 0, R_{\parallel} = \bar{v}t, t
\right)$.

Expressing $C_c$ in terms of its Fourier transform then gives
\begin{eqnarray}
C_c \left(\vec{R}_{\perp} = 0,   R_{\parallel} = \bar{v}t, t\right) =
\int d ^{d-1} q_{\perp} dq_{\parallel}  d\omega \, e^{i\left(\omega -
\bar{v} q_{\parallel} t
\right)} C_{ii} \left(\vec{q}, \omega \right)
\label{Cv 1}
\end{eqnarray}
Using the fact that $C_{ii} \left(\vec{q}, \omega \right)$ is peaked at
$\omega = c_{\pm} \left(\theta_{\vec{q}}\right)q$ with widths that
scale like $q^z_{\perp}f\left({q_{\parallel}\ell _0 \over
\left(q_{\perp}\ell _0 \right)^\zeta} \right)$,
the dominating peak is at $\omega=\omega_{-}$ for $v_0(0)>\gamma(0)$ or
at $\omega=\omega_{+}$ for $v_0(0)<\gamma(0)$, with heights
 that scale
like $q^{-\delta}_{\perp}g\left({q_{\parallel}\ell _0 \over
\left(q_{\perp}\ell _0 \right)^\zeta} \right)$ with $\delta = 2 \chi
+ z + \zeta + d - 1$ (see Eqn. (\ref{C L scale})). Assuming that
$v_0(0)>\gamma(0)$, it
 is straightforward to show that,
upon integrating (\ref{Cv 1}) over $\omega$, we obtain
\begin{eqnarray}
C_c = \int d^{d - 1} q_{\perp} dq_{\parallel} e^{i\left(c_-
\left(\theta_{\vec{q}}\right)q-\bar{v}q_{\parallel}\right)t}
f_-\left({q_{\parallel}\ell _0 \over
\left(q_{\perp}\ell _0 \right)^\zeta} \right)q^{z - \delta}_{\perp}
\label{Cv 2}
\end{eqnarray}

This integral is dominated, as $t \rightarrow \infty$, by
$q_{\parallel} \sim \left(q_{\perp}\ell _0 \right)^\zeta/\ell _0 \gg
q_{\perp}$; hence, $\theta_{\vec{q}} \rightarrow 0$,
and we get
\begin{eqnarray} C_c = \int d^{d - 1} q_{\perp} dq_{\parallel}
e^{i \left(v_0 - \bar{v}\right) q_{\parallel}t}
f_-\left({q_{\parallel}\ell _0 \over
\left(q_{\perp}\ell _0 \right)^\zeta} \right)q^{z -
\delta}_{\perp} \quad .
\label{Cv 3}
\end{eqnarray}
We can scale the time dependence out of this integral with the
change of variables
\begin{eqnarray}
q_{\parallel} \equiv {Q_{\parallel} \over t} \quad {\rm and} \quad
\vec{q}_{\perp} \equiv {\vec{Q}_{\perp} \over t^{\left(1/\zeta
\right)}}
\label{variable change}
\end{eqnarray}
which give
\begin{eqnarray}
C_c \propto t^{2 \chi/\zeta} \quad .
\label{Cv 4}
\end{eqnarray}
Using this in (\ref{RW 7}) for the single bird velocity autocorrelation
function, and then using {\it that} autocorrelation function in the
expression (\ref{RW 3}) for the mean squared random walk distance
gives
\begin{eqnarray}
w^2 (t) \propto \int^t_0 dt^{\prime}\int^t_0
dt^{\prime\prime}\left|t^{\prime} -
t^{\prime\prime}\right|^{2\chi/\zeta}
\label{RW 8}
\end{eqnarray}
Now, we need to distinguish 2 cases:
\begin{description}
\item[Case (1):  ${2\chi \over \zeta} > -1$ .]
In this case, which holds in $d = 2$, where $\chi = -{1 \over 5}$ and
$\zeta = {3 \over 5}$, the double integral over
$t^{\prime}$ and $t^{\prime\prime}$ is dominated, for $t \gg t_0$,
the microscopic time scale, by $t^{\prime}$, $t^{\prime\prime}$, and
$\left| t^{\prime} -t^{\prime\prime}\right|$ of order $t \gg t_0$.
Hence, our calculation of $C_c$ which used the hydrodynamic (i.e.,
long time) limiting forms of the correlation functions {\it is} correct,
and (\ref{RW 8}) holds.  Changing variables to  $T^{\prime} \equiv
{t^{\prime} \over t}$ and $T^{\prime\prime} \equiv
{t^{\prime\prime} \over t}$, we see that
\begin{eqnarray}
w^2 (t) \propto t^{2\left(1 + {\chi \over \zeta} \right)}
= t^{4/3} \quad, {2 \chi \over
\zeta} > -1
\label{RWq}
\end{eqnarray}
the last equality holding in $d = 2$.  Note that this behavior is
``hyperdiffusive'':   the mean squared displacement $w^2(t)$ grows faster
than it would in a simple random walk; i.\ e., {\it faster} than {\it
linearly} with time $t$.

\item[Case (2):  ${2\chi \over \zeta} < -1$ .]  In this case, which
certainly holds for $d > 4$ (where $\chi = 1 - {d \over 2} < - 1$ and
$\zeta = 1$), the integral over $t^{\prime\prime}$ {\it converges} as
$\left|t^{\prime} - t^{\prime\prime}\right| \rightarrow \infty$.
Hence, that integral is, in fact, dominated by $\left|t^{\prime} -
t^{\prime\prime}\right| = O(t_0)$, the microscopic time, where our
hydrodynamic result Eqn. (\ref{Cv 4}) is not valid.  Presumably, the
correct $\left|t^{\prime} - t^{\prime\prime}\right| \rightarrow
0$ limit of the single bird velocity autocorrelation (\ref{RW 3}) {\it
is} finite; and, hence, so the integral over
$t^{\prime\prime}$ in  (\ref{RW 8}) approaches a finite limit as $t
\rightarrow \infty$.

Hence, we get
\begin{eqnarray}
w^2 (t) \propto \int^t_0 d t^{\prime} \times {\rm finite \,
constant} \propto t \quad, {2 \chi \over \zeta} < -1 \quad .
\label{RW 10}
\end{eqnarray}

We now need only verify our a posteriori assumptions that
$\delta x_{\parallel}$ and $\vec{\delta x}_{\perp}$ were negligible
in the velocity-velocity autocorrelation Eqn. (\ref{RW 7}).

First consider  $\vec{x}_{\perp}$; we have just shown that the
root-mean-squared $\left|\vec{x}_{\perp}
\left(t^{\prime}\right) -
\vec{x}_{\perp} \left( t^{\prime\prime}\right)\right| \propto
\left|t^{\prime} - t^{\prime\prime}\right|^{1 + {\chi \over \zeta}}$.
>From our scaling expression (6.43), we see that $C_c \left(R_{\perp},
R_{\parallel}, t \right) \approx C_c \left(R_{\perp} = 0, R_{\parallel},
t \right)$ if $R^{\zeta}_{\perp}\ll R_{\parallel}$.  In (\ref{RW 7}),
we are interested in $R_{\perp} \propto \left|t^{\prime} -
t^{\prime\prime}\right|^{1 + {\chi \over \zeta}}$ and $R_{\parallel}
\propto \left|t^{\prime} - t^{\prime\prime}\right|$; hence, the
condition
$R^{\zeta}_{\perp}\ll R_{\parallel}$ will be satisfied as
$\left|t^{\prime} - t^{\prime\prime}\right| \rightarrow \infty$
provided $\zeta +
\chi < 1$.  Since $\zeta \leq 1$ and $\chi < 0$ for all $d \geq 2$, this
condition is satisfied for all $d \geq 2$.  For
$\delta x_{\parallel}$ we need only show that $\left|\delta
x_{\parallel}\left( t^{\prime}\right) -
\delta x_{\parallel}\left( t^{\prime\prime}\right)\right| \ll
\left|t^{\prime} - t^{\prime\prime}\right|$ as $\left|t^{\prime}
- t^{\prime\prime}\right|  \rightarrow \infty$.  This is easily shown
by using the fact, alluded to earlier, that
$\delta v_{\parallel} \left(\vec{r}, t \right)$,  the fluctuation of the
velocity {\it along} the mean direction of motion, has only short
ranged temporal correlations.  Using this fact, it is straightforward
to show that $\delta x_{\parallel} (t)$ just executes a simple
random walk; that is
\begin{eqnarray}
\sqrt{\left| \delta x _{\parallel}\left( t^{\prime}\right) - \delta
x _{\parallel}\left( t^{\prime\prime}\right)\right|^2} \propto
\sqrt{\left(t^{\prime} - t^{\prime\prime}\right)} \ll \left|t^{\prime}
- t^{\prime\prime}\right|
\label{RW blah}
\end{eqnarray}
and hence these fluctuations are negligible as well.

\quad Unfortunately, the analogous calculation for the anisotropic
model shows that this random ``transverse walk'' is much less
interesting:  the mean squared transverse displacement in the
$x$-direction (the direction in the ``easy place'' of the anisotropic
model orthogonal to the mean direction of motion, $y$) is given by
an expression very similar to (\ref{RW blah})
\begin{eqnarray}
\left< \left|x_i (t) - x_i (0) \right|^2\right> \equiv w^2 (t)  =
\int ^t_0 dt^{\prime}\int ^t_0 dt^{\prime\prime} \left<
v_{ix}\left(t^{\prime}\right)v_{ix}\left(t^{\prime\prime}\right)
\right>
\label{RW A1}
\end{eqnarray}
and a calculation so closely analogous to that just given for the
isotropic model that we shan't bother to repeat it for this case shows
that
\begin{eqnarray}
\left<
v_{ix}\left(t^{\prime}\right)v_{ix}\left(t^{\prime\prime}\right)
\right> \propto  \left|t^{\prime}
- t^{\prime\prime}\right|^{3 - d - {2 \over z}}
\label{RW A2}
\end{eqnarray}
As in our analysis of the isotropic case, here, too, the question of
whether ``simple random walk'' behavior $\left(w^2 (t) \propto
t\right)$ or ``hyperdiffusive'' behavior $\left(w^2(t) \propto
t^{\gamma}, \gamma > 1\right)$ occurs hinges entirely on whether
the exponent in (\ref{RW A2}) is greater or less than $-1$, with
hyperdiffusive behavior occurring in the former case (exponent $>
-1$) and simple random walk behavior in the latter (exponent $<
-1$).  Using our exact result (\ref{Zanis}) for $z$ in the anisotropic model
for $2 \leq d \leq 4$, we see that hyperdiffusive behavior will occur
if
\begin{eqnarray}
3 - d - {2 \over z} = {2 - 2d \over 3} >  -1
\label{hyper 1}
\end{eqnarray}
which is satisfied only for $d < 5/2$.  Unfortunately, this condition is
not satisfied for either $d = 3$ or $d = 4$.  In $d = 2$, the
anisotropic model is the same as the isotropic model, while for $d >
4$, $z = 2$ and $3 - d - {2 \over z}< -1$.  So in no case in which
the anisotropic model is different from the isotropic one is
hyperdiffusive behavior observable; rather, we expect $w^2
(t) \propto t$ for all those cases.  This negative prediction could be
checked experimentally, although its confirmation, while a
non-trivial check of our theory, would clearly be less exciting than
verification of our hyperdiffusive prediction $w^2
(t) \propto t^{4/3}$ for the isotropic $d = 2$ model.
\end{description}

Some of the numerical tests discussed in this section have been carried
out recently, and good
agreement with our prediction has been reached.\cite{usnum}




\section{Future directions}

In this paper, we have only scratched the surface of a very deep and
rich new subject.  We have deliberately focussed on the most limited
possible question:  what are the properties of a flock far from its
boundaries, and deep within its ordered state? Every move away
from these restricting simplifications opens up new questions.  To
name a few that we hope to address in the coming millennium:
\begin{enumerate}
\item  The transition from the ordered (moving) to disordered
(stationary, on average) phase of the flock.  This can be studied by
analyzing the (unstable) fixed point at which the renormalized
$\alpha$ of our original model (\ref{EOM}) is zero.  The dynamical RG
analysis of this point would be technically similar to the one we've
presented here for the low temperature phase, with a few crucial
differences:
\begin{enumerate}
\item  All components of $\vec{v}$, not just the $\perp$
components, become massless at the transition.

\item The fixed point will be isotropic, since no special directions are
picked out  by $\left<\vec{v} \right>$, since $\left<\vec{v} \right>$
still $=0$ at the transition.

\item The $\beta \left|\vec{v} \right|^2 \vec{v}$ term becomes
another relevant vertex.  We know, by power counting, that {\it at}
the transition, this vertex becomes relevant in $d = 4$.  Indeed, if we
ignore the $\lambda$ vertices, our model simply reduces to a purely
relaxational time-dependent Ginsburg Landau (TDGL) model for a
spin system with the number of components $n$ of the spin equal to
the dimension $d$ of the space those spins live in.

\end{enumerate}

We have convinced ourselves by power counting that {\it at} the
transition, for $d < 4$, the $\lambda$ vertices are a {\it relevant}
perturbation to the Gaussian critical point.  Whether they constitute
a relevant perturbation to the $4-\epsilon$ TDGL fixed point,
thereby changing its critical properties, can only be answered by a
full-blown dynamical renormalization group analysis.

Obviously, a similar analysis could also be done for the anisotropic
model.

\item The shape and cohesion of an open flock, and its fluctuations.
We have thus far focussed on flocks in closed or periodic
boundary conditions.  Real flocks are usually surrounded by open
space.  How do they stay together under these circumstances?  What
shape does the flock take?  How does this shape fluctuate, and is it
stable?

This issue is somewhat similar to the problems of the shapes of
equilibrium and growing crystals (e.g., facetting, dendritic growth).  In
those problems, it was important to first understand {\it bulk} processes
(e.g., thermal diffusion in the case of dendritic growth) before one could
address surface questions (e.g., dendritic growth).  The non-trivial
aspects of the {\it bulk} processes in flocks (e.g., anomalous diffusion)
will presumably radically alter the shapes and their fluctuations.

\item  A somewhat related question is:  what happens if birds move
at different speeds?  By ``move at different speeds'', we don't mean
simply that at any instant, different birds will be moving at
different speeds (a possibility already included in our ``soft spin''
dynamical model equation (\ref{EOM})).  Rather, we mean a model in which
some birds have a different probability distribution of speeds than
others.  (In our model, this distribution of the speed of any
given bird is the same over a sufficiently long time, and controlled
by the values of the parameters $\alpha$ and $\beta$, with
large $\alpha$ and $\beta$  leading to a distribution sharply peaked
around a mean speed $v_0 = \sqrt{{\alpha \over \beta}}$, while
small
$\alpha$ and $\beta$ lead to a broader distribution).  More generally,
one could imagine 2 (or many) different species of birds, (labeled
by $k$) all flying together, each with different mean speeds
$v^{k}_0$.  What would the {\it bulk} dynamics of such a flock
be?  Would there be large scale spatial segregation, with fast birds
moving to the front of the flock, and slow birds moving to the
back?  If so, how would such segregation affect the shape of the
flock?  Would it elongate along the mean direction of motion?
Would this elongation eventually split the flock into fast and slow
moving flocks?

\item  At the other extreme, one could consider flocks in confined
geometries; e.g., inside a circular reflecting wall in $d = 2$.  In such a
case, the time averaged velocity of the flock
$\left<\vec{v}\left(\vec{r}, t
\right)
\right>_t$ could {\it not} be spatially homogeneous but would have
to circulate around the center of the circle; i.\ e.,
$\left<\vec{v}\left(\vec{r}, t
\right)
\right>_t = f(r)\hat{\theta}$.  The spatially inhomogeneous pattern of
velocity and density that resulted could be predicted by our
continuum equations.  This problem is potentially related to the
previous one, since one way a flock containing, say, some {\it very}
fast birds and other {\it very} slow birds, could stay
together would be for the fast birds to fly in circles inside the
essentially stationary volume of space filled by the slow birds.
It would be very interesting to make the connection between our continuum
theory and the recently observed circular motion of Dictyostelium cells
in a confined geometry.\cite{bacteria}

\item One could relax the constraint on conservation of bird number,
by allowing birds to be born, and die, ``on the wing''.  Numerical
studies of such models, which may be appropriate to bacteria
colonies, where reproduction and death are rapid, as well as the
migration of, e.\ g., huge herds of caribou over thousands of miles
and many months, have already been undertaken\cite{Albano};  it
should be straightforward to modify our equations by adding a
source term to the bird number conservation equation.

\item  It is possible that phase transitions other than that from the
moving to the non-moving state occur in flocks.  For example, in
some preliminary simulations of microscopic models in which birds
try to avoid getting too close to their neighbors, rather than merely
following them, we have observed (literally by eye) what appears
to be a ``flying crystal'' phase of flocks:  the birds appear to lock
themselves onto the sites of a crystalline lattice, which then appears
to move coherently.  It would be very interesting to test numerically
whether this optical appearance reflects true long-range
translational order, by looking for a non-zero expectation value of
the translational order parameters.
\begin{eqnarray}
\rho_{\vec{G}} (t) = \left<\Sigma_i e^{i\vec{G} \cdot  \vec{r}_i(t)}/N
\right>
\label{10.1}
\end{eqnarray}
which will become non-zero in the thermodynamic $(N \rightarrow
\infty)$ limit at a set of reciprocal lattice vectors $\vec{G}$ if such
long ranged order actually develops.  It will also be {\it extremely}
interesting to include the possibility of such long ranged order in our
analytic  model, and study the interplay between this translational
order and the anomalous hydrodynamics that we've found here for
``fluid'' flocks.  Will anomalous hydrodynamics suppress the
``Mermin-Wagner'' fluctuations of {\it translational} order, just as it
does those of {\it orientational} order, and lead to true long ranged
{\it translational} order, even in $d = 2$?  Will the crystallization
suppress orientational fluctuations, and thereby slow down the
anomalous diffusion that we found in the fluid case?  And in any
case, what are the temporal fluctuations of $\rho_{\vec{G}}(t)$?

It should be noted that this problem potentially has all the richness
of liquid crystal physics:  in addition to ``crystalline'' phases, in
which the set $\{\vec{G} \}$ of reciprocal lattice vectors in
(\ref{10.1}) spans all $d$ dimensions of space, one could imagine
``smectic'' phases in which all the $\vec{G}$'s lay in the same
direction; and ``discotic'' phases in $d = 3$, in which the $\vec{G}$'s
only spanned a two-dimensional subspace of this three-dimensional
space.  The melting transitions between these phases and the
``fluid'', moving flock, as well as possible direct transitions between
them and the stationary flock phase, and between each other, would
also be of great interest.

We should point out here that these models differ considerably from
recently considered models of moving flux lattices \cite{Rad}
and transversely driven charge density waves \cite{Rad,RJ} in
that here, the direction of motion of the lattice {\it not} picked out
by an external driving force, but, rather, represents a spontaneously
broken continuous symmetry.

\item  Finally, we'd like to study the problem of the growth of order
in flocks.  This is a phenomenon we've all seen every time we walk
onto a field full of geese:  eventually, our approach startles the
geese, and they take off en masse.  Initially, they fly in random
directions, but quickly the flock orders, and flies away coherently.
The dynamics of this process is clearly in many ways similar to, e.\
g., the growth of ferromagnetic order after a rapid quench from an
initial high temperature $T_i > T_c$, the Curie temperature, to a
final temperature $T_f < T_c$, a problem that has long been studied
\cite{Goldenfeld} and proven to be very rich and intriguing.  In
flocks, where, as we've seen, even the dynamics of the {\it
completely ordered} state is {\it very} non-trivial, the {\it growth}
of order seems likely to be even richer.
\end{enumerate}

Even this list of potential future problems, representing, as it does,
probably another ten years of research for several groups, clearly
represents only a narrow selection of the possible directions in
which this embryonic field can go.  We haven't even mentioned, for
example, the intriguing problem of one-dimensional flocking, with
its applications to traffic flow (and traffic jams), a topic clearly of
interest.
This problem has recently been studied \cite{Lazlo}
and found to also show a non-trivial phase transition between
moving and non-moving states.

We expect flocking to be a fascinating and fruitful topic of research
for biologists, computer scientists, and both experimental and
theoretical physicists (at least {\it these} two!) for many years to
come.

\acknowledgements

We profusely thank T.\ Vicsek for introducing us to this problem.  We
are also grateful to A.\ Csirok, E. V..\ Albano, and A.\ L.\ Barabasi for
communicating their work to us prior to publication; to M.\ Ulm and S.\
Palmer for performing some inspirational simulations, and for equally
inspirational discussions; to J.\ Sethna and K.\ Dahmen for pointing out
the existence of the $\lambda_2$ and $\lambda_3$ terms and to P. McEuen
for suggesting the go carts.  J.\ T.\ thanks the
Aspen Center for Physics for their hospitality at
several stages of this work. J.\ T.'s
work was supported in part by the National Science Foundation, Grant
DMR-9634596.




\newpage

\begin{figure}
\caption{A snapshot of a simulated flock that has reached a
statistically steady state.  Note the enormous fluctuations in the
density. Quantitatively, the statistics of the spatial Fourier transform.
$C_{\rho}\left(\vec{q}\right)$ obtained from this picture agree with 
our quantitative prediction equation (1.10). }
\label{F1}
\end{figure}

\begin{figure}
\caption{Polar plot of the direction-dependent sound speeds
$c_{\pm} \left(\theta_{\vec{q}}\right)$, with the horizontal axis
along the direction of mean flock motion.}
\label{F2}
\end{figure}

\begin{figure}
\caption{Plot of the damping $Im\omega$ versus $q_y
\equiv
\left|\vec{q}_{\perp}\right|$ where $\vec{q}_{\perp}$ is the
projection of wavevector $\vec{q}$ {\it perpendicular} to the
direction of the mean flock velocity $\left< \vec{v}\right>$ for fixed
projection $q_x$ of  $\vec{q}$ {\it parallel} to $\left<
\vec{v}\right>$.  Note that, for small $\left|\vec{q}\right|$, the
crossover between $Im\omega \propto q^z_y$ and
$Im\omega
\propto q^{z \over \zeta}_x$ occurs only for directions of
propagation
$\hat{q}$ very nearly parallel to the mean flock velocity $\left<
\vec{v}\right>$, since
$\zeta < 1$.}
\label{F3}
\end{figure}

\begin{figure}
\caption{Plot of the spatio-temporally Fourier-transformed density
correlation function $C_\rho \left(\vec{q}, \omega\right)$ versus
$\omega$ for fixed $\vec{q}$.  It shows two sharp asymmetrical
peaks at $\omega = c_{\pm} \left(\theta_{\vec{q}}\right)q$
associated with the sound modes of the flock, where $c_{\pm}
\left(\theta_{\vec{q}}\right)$ are the sound mode speeds.  The
widths of those peaks are the second mode dampings $Im
\omega_{\pm}
\left(\theta_{\vec{q}}\right) \propto q^z_{\perp} f_{\pm}
\left({q_{\parallel} \ell_0
\over \left(q_{\perp} \ell_0\right)^{\zeta}}\right)$.}
\label{F4}
\end{figure}

\begin{figure}
\caption{Geometry of the anisotropic model.  Birds prefer to fly
in the ``easy'' $x-y$ plane.  We take their (spontaneously chosen)
direction of motion within that plane to be $y$. The in-plane
direction perpendicular to that is $x$.  In general $d$, there are
$d - 2$ ``hard'' directions $\vec{r}_H$ perpendicular to this easy
plane.  The anisotropy of {\it scaling} is between $x$ and the other
$d - 1$ directions $y,\vec{r}_H$.}
\label{F5}
\end{figure}

\begin{figure}
\caption{Plot of $C_{LL} \left(\vec{q},\omega \right)$ and $C_{TT}
\left(\vec{q},\omega \right)$ versus $\omega$ for identical fixed
$\vec{q}$.  Note the smallness of the overlap between the transverse and
longitudinal peaks.}
\label{F6}
\end{figure}

\begin{figure}
\caption{Feynmann graph renormalizing the noise correlations
when $\lambda_2 = 0$. There is a factor of the external momentum
$|\vec{q}_{\perp}| \equiv q_x$ associated with each vertex; hence this
graph
does not renormalize $\Delta$, but, rather, only changes
$O(q^2 )$ pieces of the $f-f$ correlation function.}
\label{F7}
\end{figure}

\begin{figure}
\caption{Feynmann graph for diffusion constants.  When
$\lambda_2 = 0$, this graph is proportional to at least one power
of $|\vec{q}_{\perp}|\equiv q_x$, and, so, cannot renormalize
$D_{\parallel}$
or $D_{\rho}$.}
\label{F8}
\end{figure}


\begin{figure}
\caption{Illustration of the optimal boundary conditions for
simulations and experiments to test our predictions.  The top and
bottom walls are reflecting, while periodic boundary conditions
apply at the left and right walls (i.e., a bird that flies out to the
right instantly reappears at the same height on the left).  The
mean direction of spontaneous flock motion, if any occurs, is
clearly forced to be horizontal by these boundary conditions.  In
spatial dimensions $d > 2$, one should choose reflecting boundary conditions
in $d-1$ directions, and periodic in the remaining direction, thereby
forcing $\left< \vec{v}\right>$ to point along that periodic
direction.}
\label{F10}
\end{figure}

\begin{figure}
\caption{More practical ``track'' geometry for experiments on real
flocks.  Data should only be taken from the cross-hatched region
centered on the middle of the ``straightaway''.}
\label{F11}
\end{figure}

\begin{figure}
\caption{Illustration of the experiment to measure the mean
squared lateral wandering $w^2 (t)$.  One labels all of the birds
some central stripe (of width $\ll L$, the channel width), and then
measures the evolution of their mean displacements
$\vec{x}_\perp(t)$ perpendicular to the mean direction of motion
(which mean direction is horizontal in this figure).}
\label{F12}
\end{figure}


\begin{references}

\bibitem{Vicsek}
T.\ Vicsek, Phys.\ Rev.\ Lett. {\bf 75}, 1226 (1995); A.\
Czirok, H.\ E.\ Stanley, and T.\ Vicsek, J.\ Phys.\ A {\bf 30}, 1375
(1997).

\bibitem{Partridge}
B.\ L.\ Partridge, Scientific American, 114-123 (June 1982).

\bibitem{reynolds}C. Reynolds, {Computer Graphics} {\bf 21}, 25 (1987);
J.L Deneubourg and S. Goss, {Ethology, Ecology, Evolution}
{\bf 1}, 295 (1989); A. Huth and C. Wissel, in {\em Biological
Motion}, eds. W. Alt and E. Hoffmann (Springer Verlag, 1990)p. 577-590.
We thank D. Rokhsar for calling these
references to our attention.

\bibitem{TT}
J.\ Toner and Y.\ Tu, Phys.\ Rev.\ Lett. {\bf 75}, 4326 (1995).

\bibitem{MW}
N.\ D.\ Mermin and H.\ Wagner, Phys.\ Rev.\ Lett. {\bf 17}, 1133
(1966).

\bibitem{FNS}D.\ Forster, D.\ R.\ Nelson, and M\ J.\ Stephen, Phys.\ Rev.\
A{\bf 16}, 732 (1977).

\bibitem{smec}
Smectic A liquid crystals show even stronger damping at long
wavelengths than that found here for flocks.  For a theoretical
treatment, see G.\ F.\ Mazenko, S.\ Ramaswamy, and J.\ Toner;
Phys.\ Rev.\ Lett. {\bf 49}, 51 (1982); Phys.\ Rev.\ A {\bf 28},
1618 (1983).  Experimental confirmation of this theory is given by
S.\ Bhattacharya and J.\ B.\ Ketterson, Phys.\ Rev.\ Lett.\ {\bf 49},
997 (1982).  See also the discussion in P.\ G.\ deGennes and J.\
Prost, {\it The Physics of Liquid Crystals}, 2nd ed.\ (Clarendon
Press, Oxford, 1993) pages 457-465.

\bibitem{Sethna}We thank Karen Dahmen and Jim Sethna for
pointing out the existence of this term to us (although, given all
the difficulty this term has caused us, it is unclear whether thanks
are really the appropriate response).

\bibitem{KPZ}See, e.\ g., M.\ Kardar, G.\ Parisi, and Y-C.\ Zhang, Phys.\
Rev.\ Lett.\ {\bf 56}, 889 (l986).

\bibitem{Steph}We thank Stephanie Palmer for suggesting this
alternative approach to us.

\bibitem{Berk}We thank Paul McEuen
for suggesting this realization to us.

\bibitem{usnum}Y. Tu, J. Toner and M. Ulm, preprint, and to appear in PRL.

\bibitem{bacteria}H. Levine, private communication.
\bibitem{Albano}E.\ V.\ Albano, Phys.\ Rev.\ Lett.\ {\bf 77}, 2129 (l996).


\bibitem{Rad}L.\ Balents, M.\ C.\ Marchetti, and L.\ Radzihovsky,
preprint, cond-mat/9707302, and to appear in PRE.

\bibitem{RJ}L.\ Radzihovsky and J.\ Toner, unpublished.

\bibitem{Goldenfeld}See, e.\ g., M.\ Mondello and N.\ Goldenfeld, Phys.\
Rev.\ E {\bf 47}, 2384 (1993).

\bibitem{Lazlo}A.\ Czirok, A.\ L.\ Barabasi, Preprint.

\end{references}
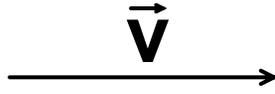
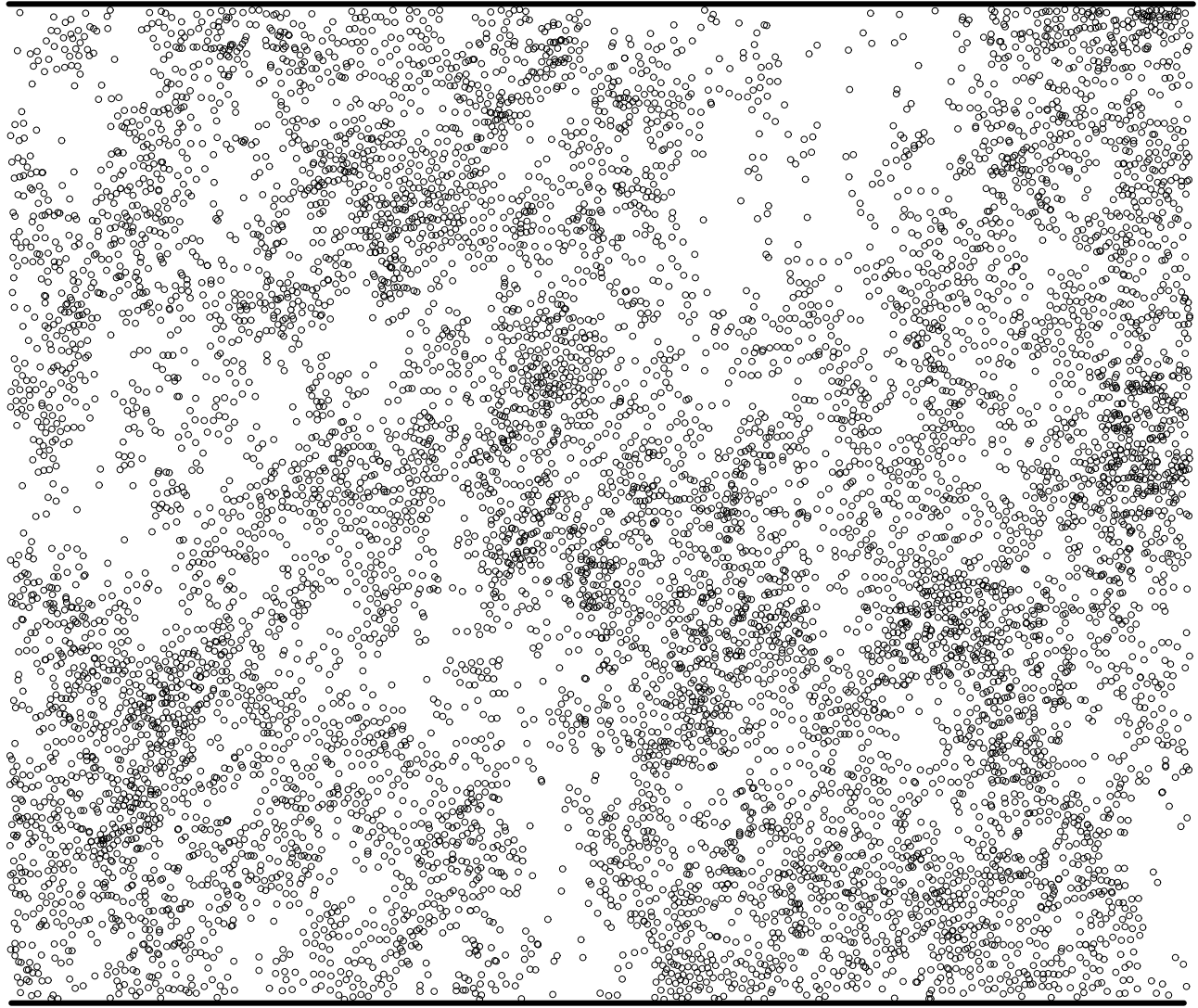
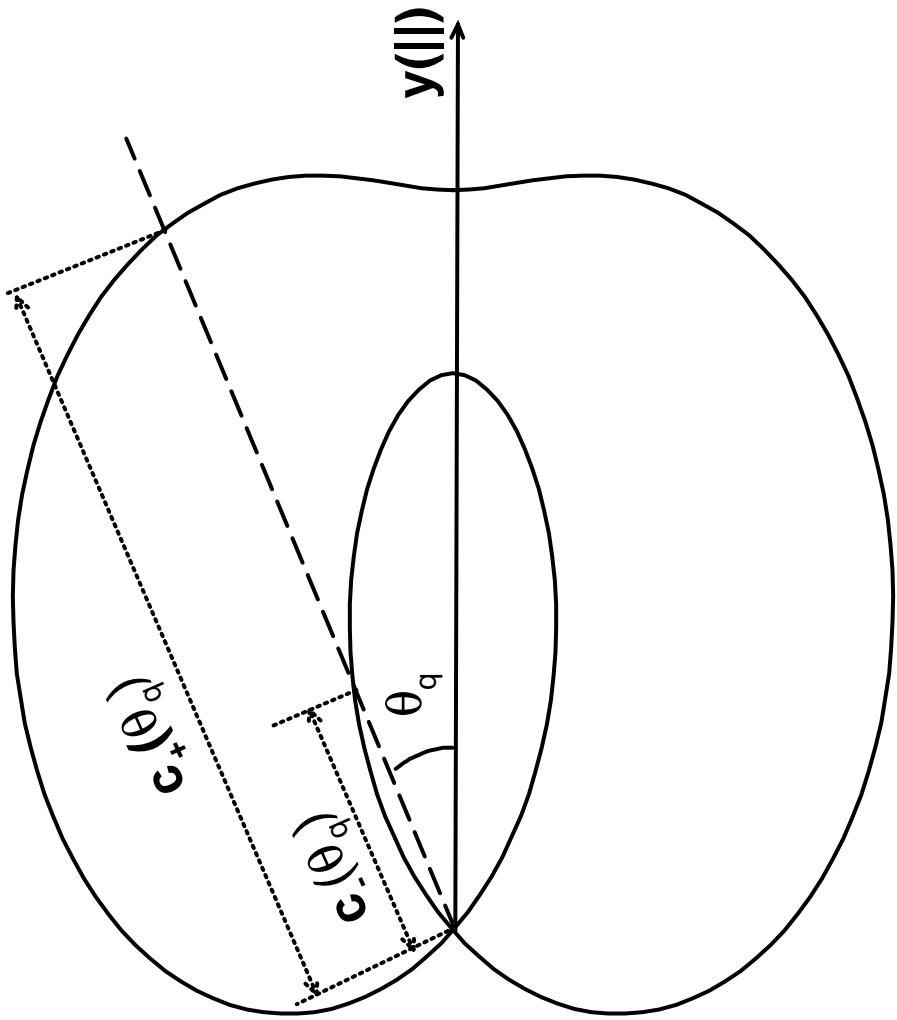
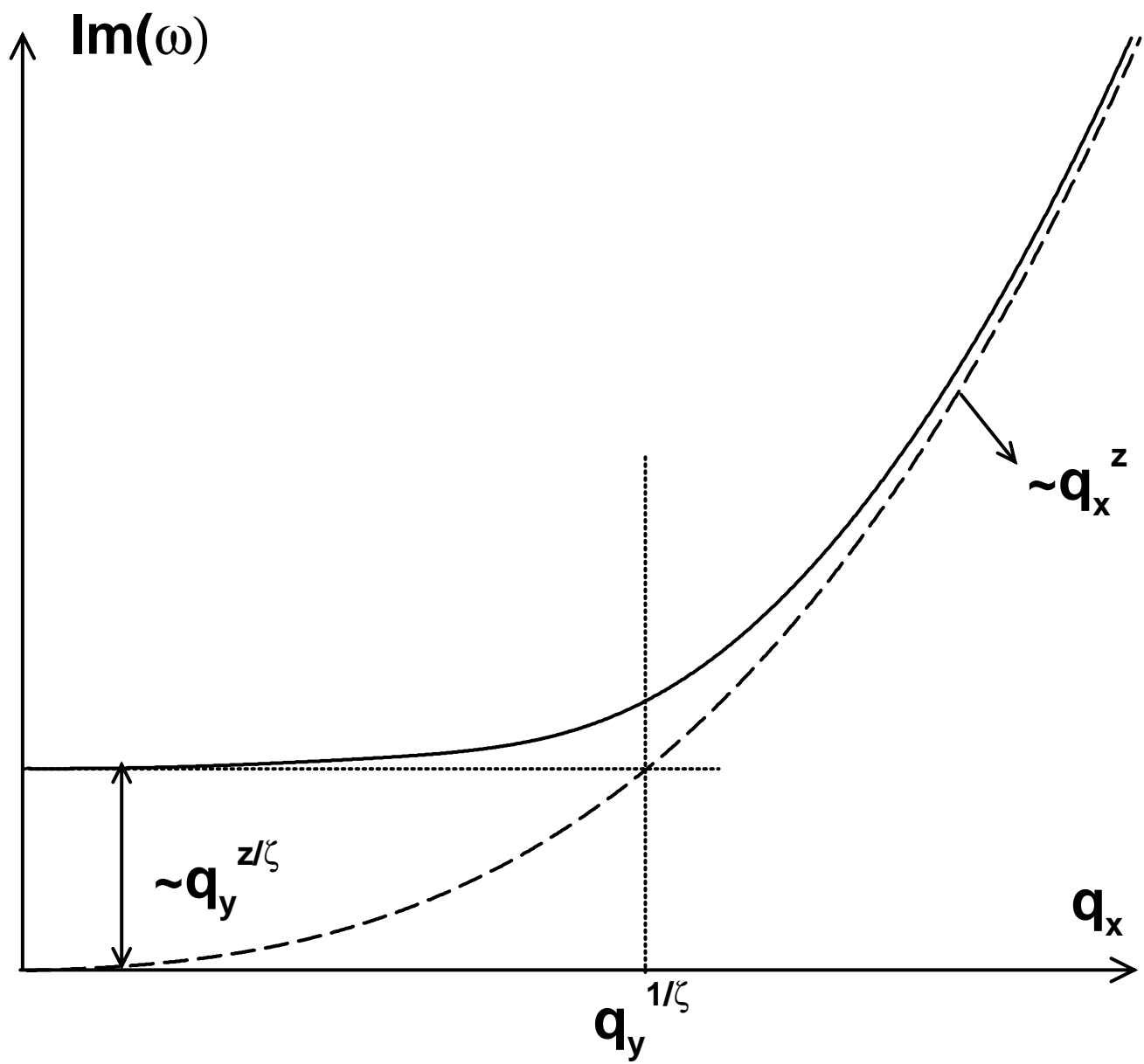
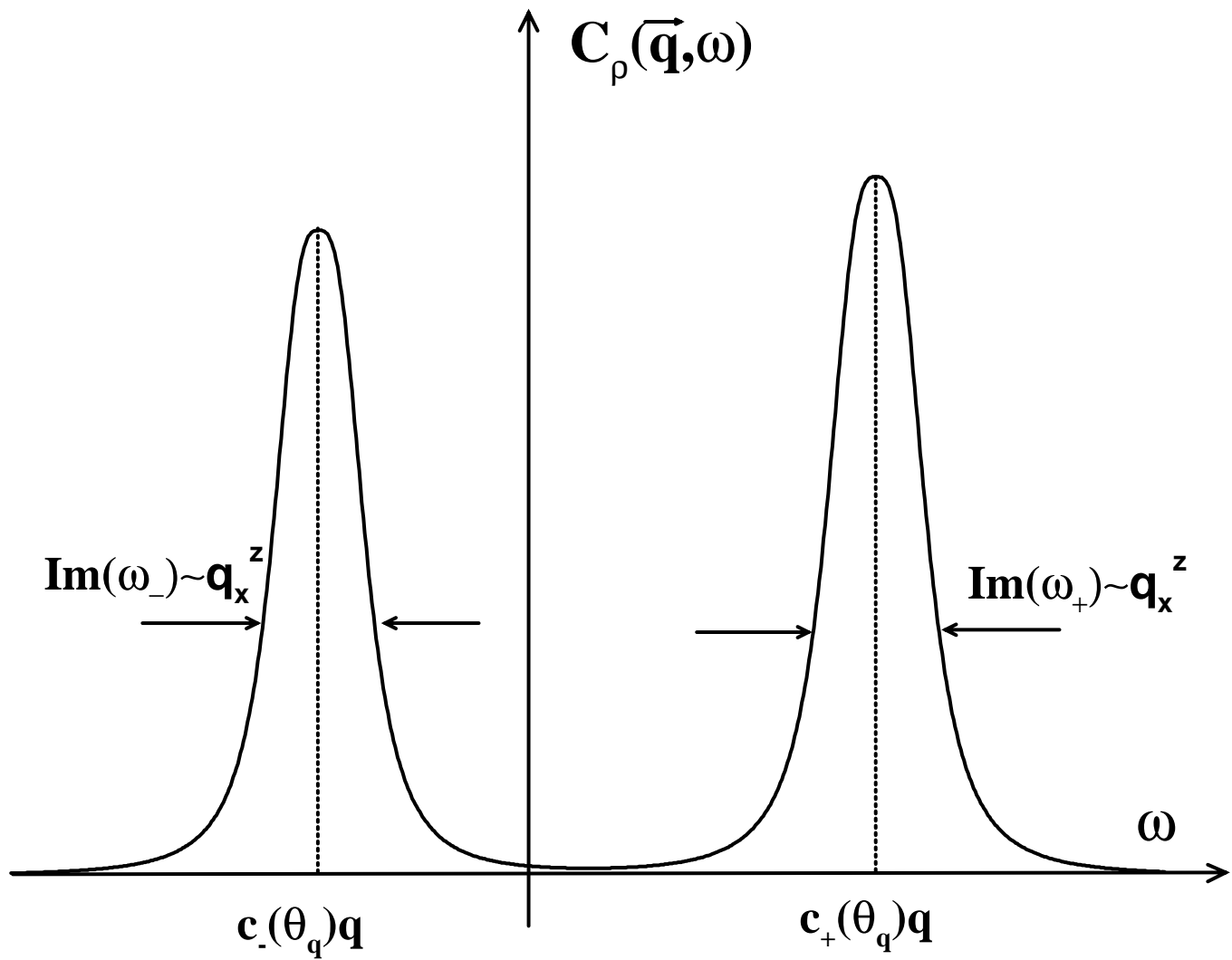
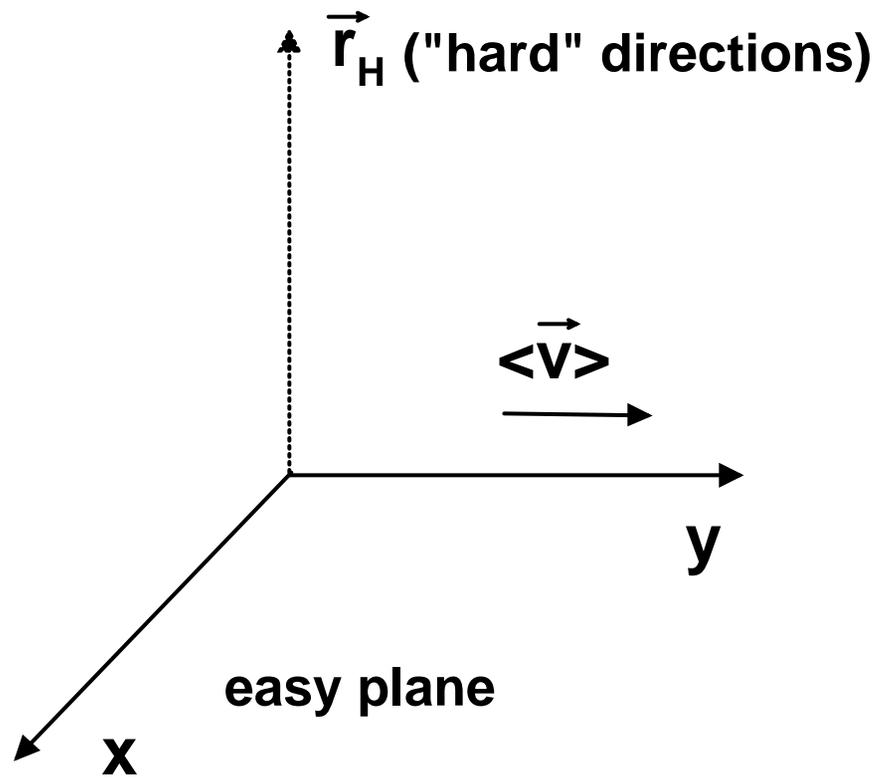
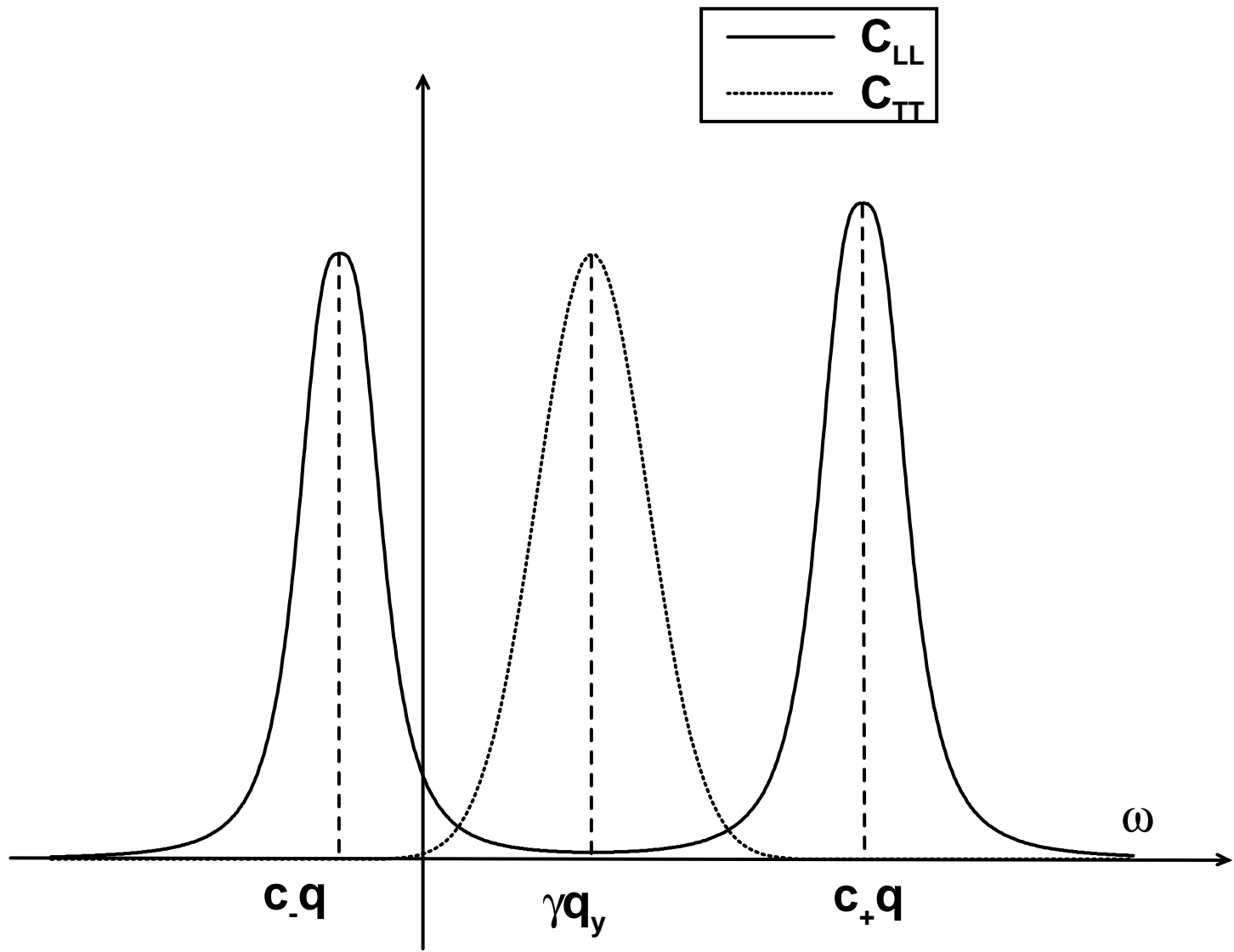
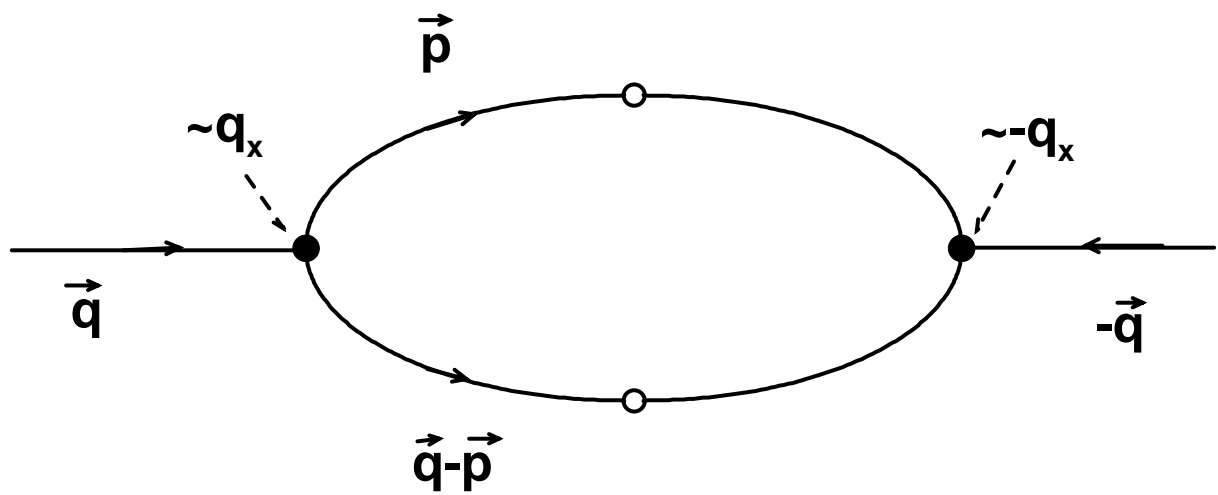
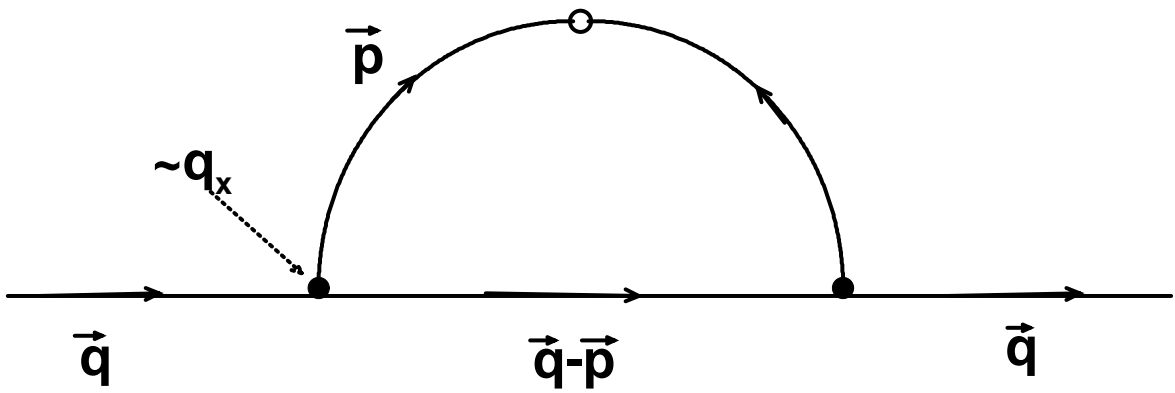
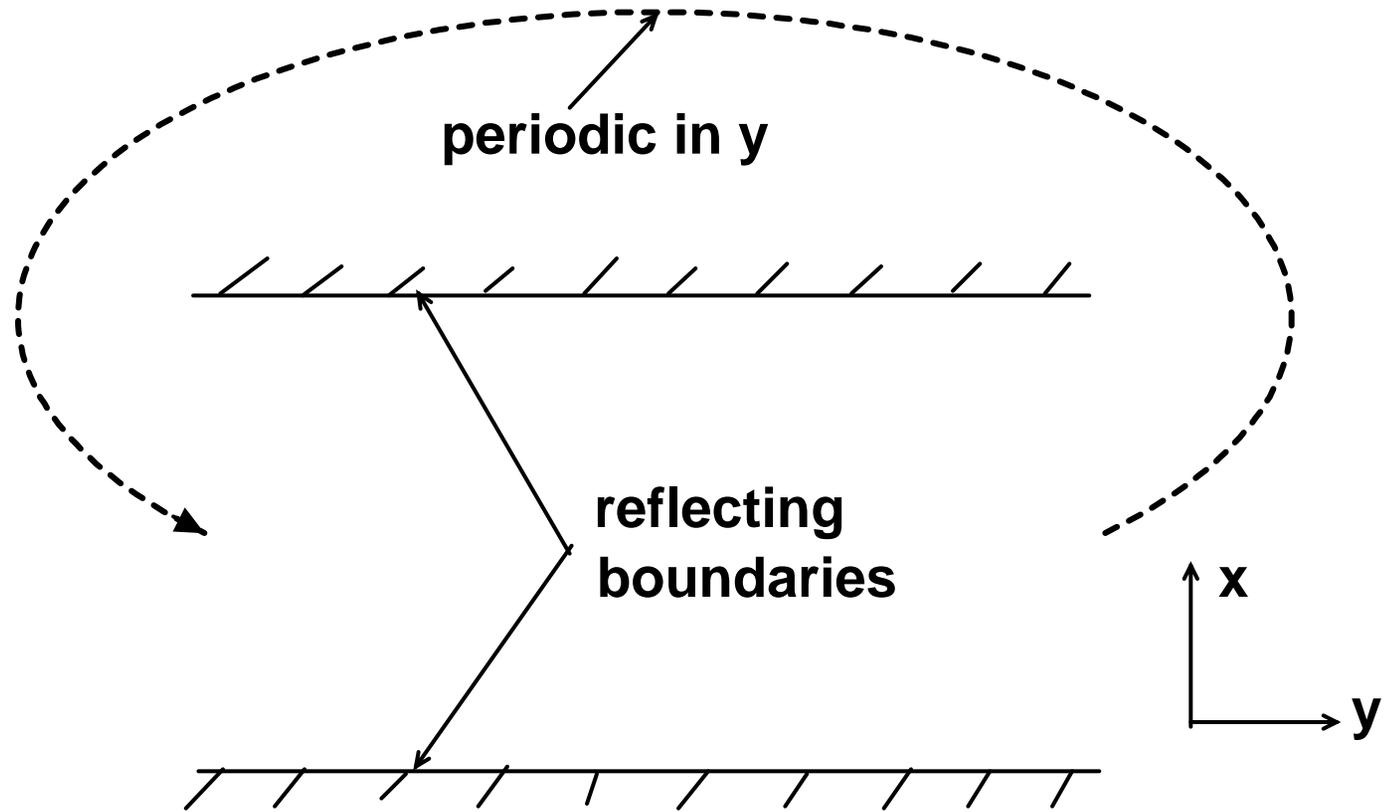
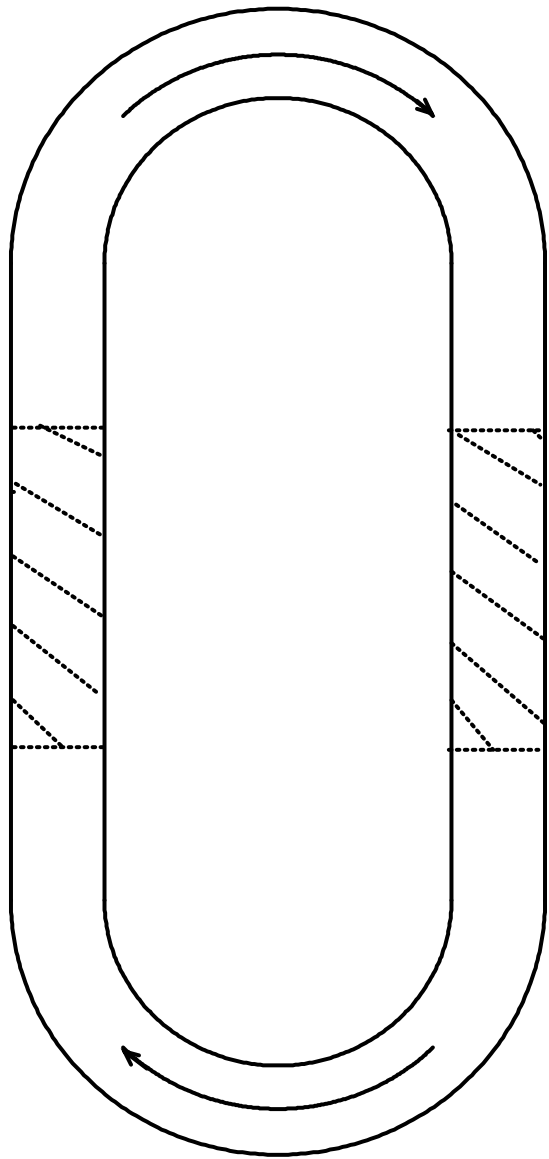
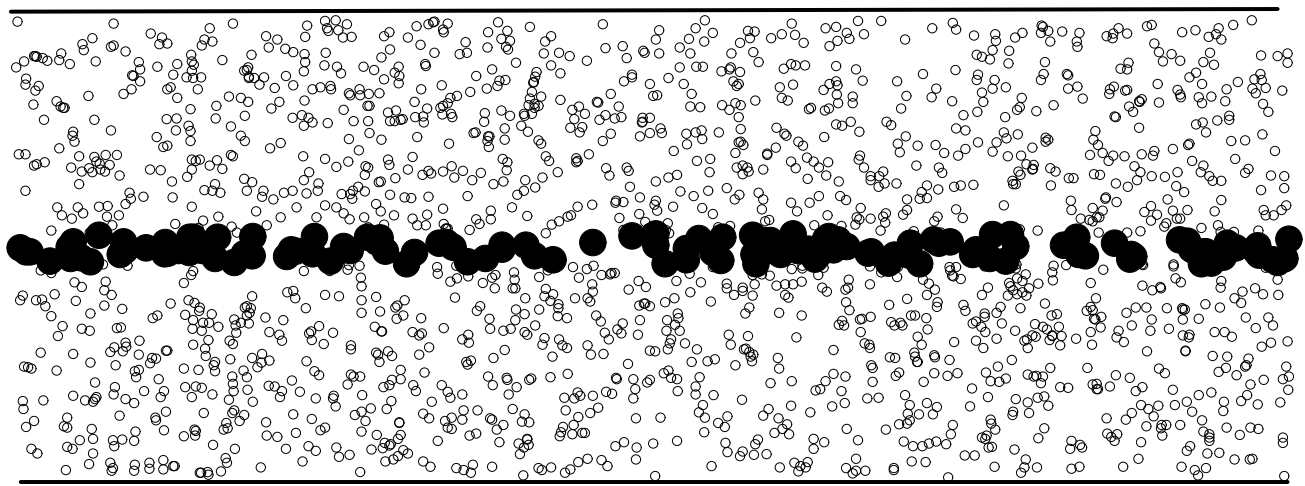
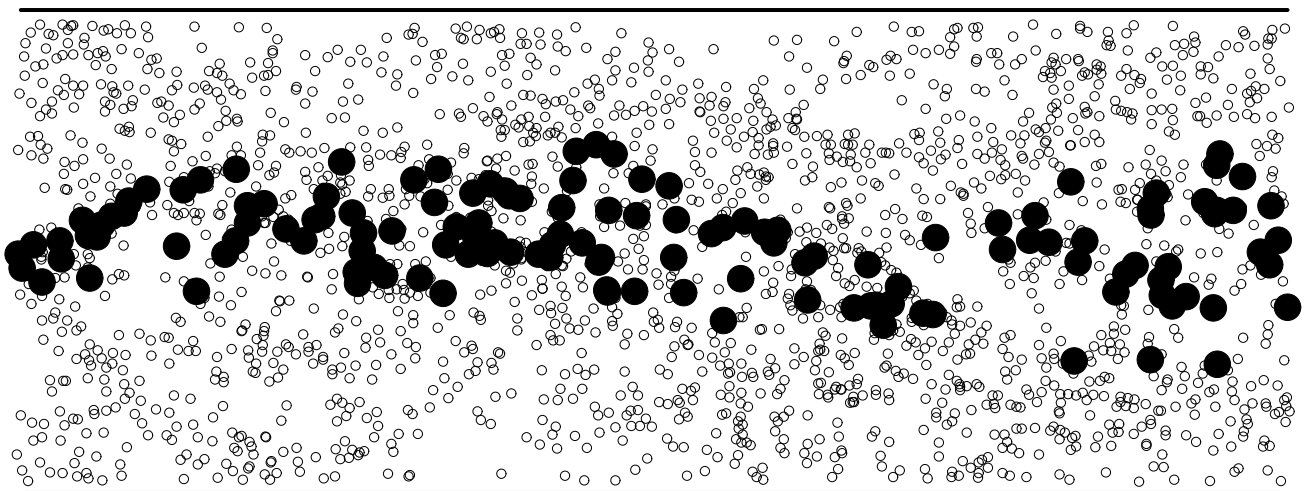
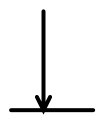
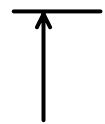
\end{document}